\documentclass[preprint,12pt]{elsarticle}

\usepackage{amsmath, amssymb}
\usepackage{graphicx}
\usepackage{booktabs}
\usepackage{siunitx}
\usepackage{hyperref}
\usepackage{multirow}
\usepackage{geometry}
\usepackage{bm}
\usepackage{amsmath,amssymb,amsfonts,mathrsfs}
\usepackage{algorithmic, booktabs}
\usepackage{graphicx}
\usepackage{graphics}
\usepackage{textcomp}
\usepackage[dvipsnames]{xcolor}
\usepackage{algorithm,algorithmic}[1]
\usepackage{setspace}
\usepackage{bbm}
\usepackage[bb=boondox]{mathalfa}

\usepackage{blindtext}
\usepackage{caption}
\usepackage{subcaption}
\usepackage{nomencl}
\usepackage{siunitx}
\usepackage[export]{adjustbox}
\def\BibTeX{{\rm B\kern-.05em{\sc i\kern-.025em b}\kern-.08em
T\kern-.1667em\lower.7ex\hbox{E}\kern-.125emX}}
\usepackage{etoolbox}
\usepackage{acronym}
\usepackage{multirow}
\usepackage{makecell}

\usepackage{nomencl}
\usepackage{multirow}
\usepackage{array}
\usepackage{pdflscape}
\usepackage{rotating}
\allowdisplaybreaks
\usepackage[percent]{overpic}
\usepackage{pict2e}
\newcommand{\tx}[1]{\text{#1}}

\usepackage{framed} % Framing content
\usepackage{multicol} % Multiple columns environment
\usepackage[framemethod=tikz]{mdframed}
\usepackage{nomencl} % Nomenclature package
\makenomenclature
\renewcommand*\nompreamble{\begin{multicols}{2}}
\renewcommand*\nompostamble{\end{multicols}}

\usepackage[most]{tcolorbox}

\newtcolorbox{myframe}{
    breakable,
    enhanced,
    colback=white,
    colframe=black,
    arc=2mm
}

\journal{Applied Energy}

\begin{document}

\begin{frontmatter}



\title{Wildfire Risk Metric Impact on Public Safety Power Shutoff Cost Savings}
\author[1]{Ryan Greenough\corref{cor1}}
\cortext[cor1]{Corresponding Author: Ryan Greenough}
\ead{rgreenou@ucsd.edu}
\author[2]{Kohei Murakami}
\ead{KMurakami@epri.com}
\author[1,3]{Jan Kleissl}
\ead{jkleissl@ucsd.edu}
\author[1]{Adil Khurram}
\ead{akhurram@ucsd.edu}

\address[1]{Department of Mechanical and Aerospace Engineering, University of California San Diego, La Jolla, CA, 92093 USA}
\address[2]{Electric Power Research Institute, Irving, TX 75062 USA}
\address[3]{Center for Energy Research, Department of Mechanical and Aerospace Engineering, University of California at San
Diego, La Jolla, CA, 92093 USA}

\begin{abstract}
Public Safety Power Shutoffs (PSPS) are a proactive strategy to mitigate wildfire risks by preemptively de-energizing power lines and redispatching generation. However, wildfire risk quantification is critical for the operational effectiveness of PSPS. Many existing PSPS formulations rely on the Wildland Fire Potential Index (WFPI) to relate wildfire risk to power system operations. However, this flammability-based wildfire risk correlates less strongly with observed wildfire ignition probabilities (OWIP) than the Large Fire Probability (WLFP). This wildfire modeling discrepancy can distort generation commitments, misinform line de-energizations, and increase real-time (RT) costs. Prior work avoided incorporating wildfire ignition probability (WIP) due to the complexity of modeling wildfire-driven failures as Bernoulli random variables, which introduces non-linear constraints. By leveraging the cross-entropy between WIP and true outages, we represent wildfire risk as the sum of each energized line’s wildfire ignition log probability (log(WIP)), rather than relying on a WFPI proxy. A cross-entropy constraint models joint line failures in a tractable manner without enumerating all failure scenarios. A stochastic day-ahead (DA) unit commitment with the PSPS framework assesses the cost impact of mapping WFPI- or WLFP-based risk metrics to WIP on the IEEE RTS 24-bus and RTS-GMLC systems. Out-of-sample results show that mapping WLFP to log(WIP) in the PSPS optimization leads to more risk-aware decisions and reduces expected and worst-case out-of-sample costs. These findings underscore the benefits of incorporating probabilistic wildfire risk metrics to improve PSPS decision-making for wildfire-resilient power systems.
\end{abstract}

\begin{graphicalabstract}
% Optional: include a small figure summarising the paper
\includegraphics[width=\linewidth]{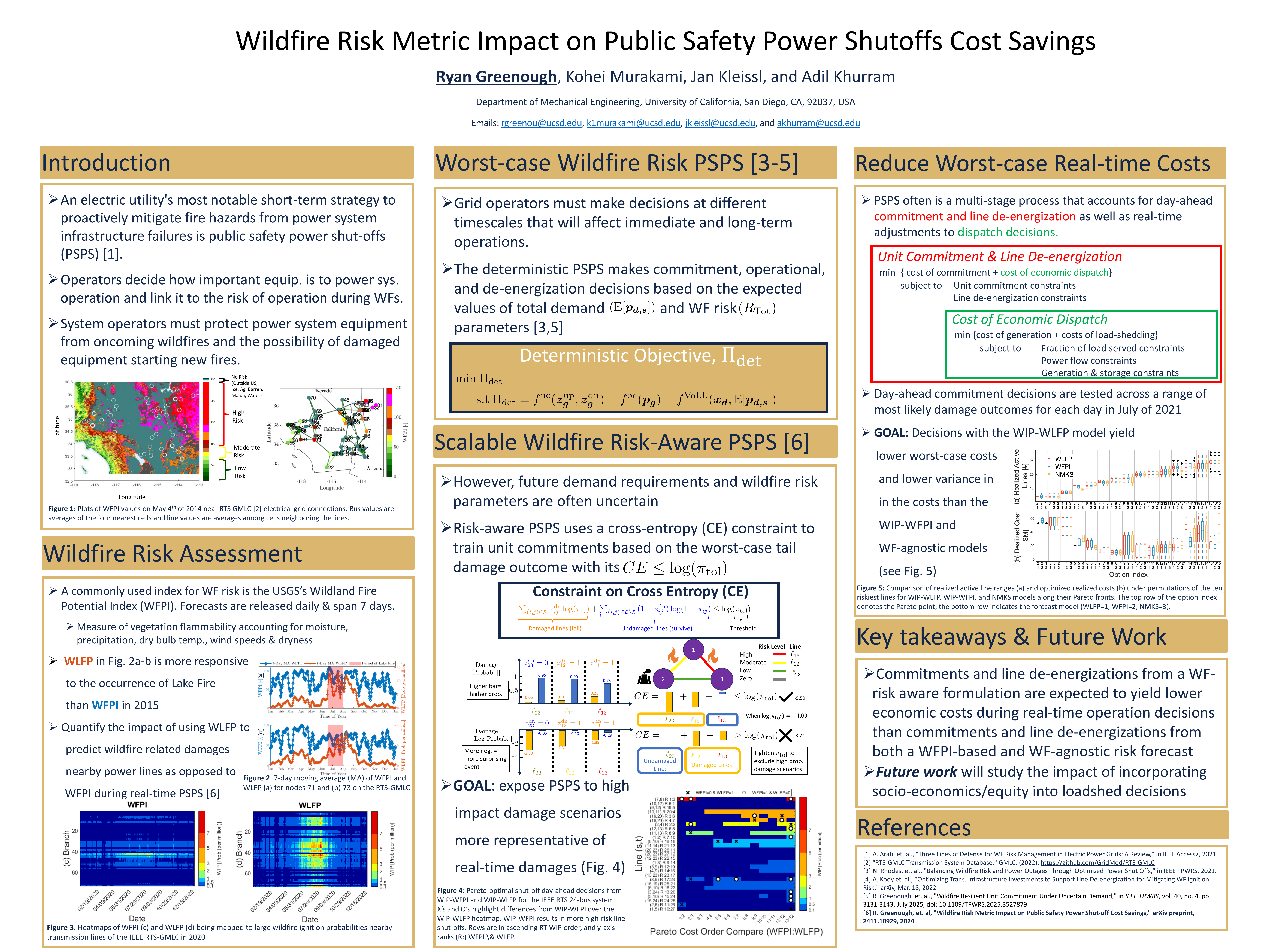}
\end{graphicalabstract}

\begin{highlights}
\item Quantifies how wildfire risk selection impacts grid costs and shutoff decisions.
\item Wildfire-risk distribution affects PSPS scenario separability and decision making.
\item New wildfire-risk model reduces maximum cost and variability.
\item Risk-aware cross-entropy constraint lowers extreme day-ahead unit commitment costs.
\item Explores trade-offs between wildfire risk reduction, cost, and load delivery.
\end{highlights}

\begin{keyword}
Wildfire mitigation \sep Power systems \sep Risk metrics \sep PSPS \sep Optimization \sep Cost analysis
\end{keyword}

\end{frontmatter}



\mbox{}
\nomenclature[]{\(\mathcal{B},\mathcal{D},\mathcal{G}\)}{Set of buses, demands, and generators}
\nomenclature[]{\(\mathcal{K}_{\psi}\subseteq  \mathcal{L}\)}{Set of damaged lines within set of transmission lines}
\nomenclature[]{\(\mathcal{P}\)}{Undirected graph of power grid}
\nomenclature[]{\(H\)}{Number of samples in DA optimization horizon}
\nomenclature[]{\(B_{ij}\)}{System susceptance  value connecting bus $i$ to bus $j$ [S*Base MVA in DC-OPF]}
\nomenclature[]{\(\tilde{\theta}_{i,t}\)}{Phasor angle at bus $i \in \mathcal{B}$ and time $t$ [rad]}
\nomenclature[]{\(\underline{p}_{g},\overline{p}_{g}\)}{Lower/upper active power limit for each gen $g \in \mathcal{G}$ [MWh]}
\nomenclature[]{\(\underline{U}_{g},\overline{U}_{g}\)}{Lower/upper power ramp limit for each gen $g \in \mathcal{G}$ [MW]}
\nomenclature[]{\(\tilde{p}_{g,t}\)}{Active generation by each generator $g \in \mathcal{G}$ at time $t$ [MWh]}
\nomenclature[]{\(\tilde{p}_{d,t}\)}{Uncertain demand at each bus $d \in \mathcal{D}$ at time $t$ [MWh]}
\nomenclature[]{\(C^{\mathrm{VoLL}}_{d}\)}{Value of Lost Load for each demand node $d \in \mathcal{D}$ [\$/MWh]}
\nomenclature[]{\(\tilde{x}_{d,t}\)}{Fraction of demand served for $d \in \mathcal{D}$ at time $t$ ($\tilde{x}_{d,t} \in [0,1]$)}
\nomenclature[]{\(\mathcal{H}\)}{Set of time-steps in the optimization horizon}
\nomenclature[]{\(R_{ij,t}\)}{Wildfire risk of a transmission line measured in WFPI [-]}
\nomenclature[]{\(z_{g,t},z_{ij,t}\)}{Active status for generators and transmission lines ($z_{g,t},z_{ij,t} \in {0,1}$)}
\nomenclature[]{\(\tilde{p}^{\text{aux}}_{g,t}\)}{Auxiliary variable for gen $g \in \mathcal{G}$ to enforce ramping limits when it is turned off. [MWh]}
\nomenclature[]{\(z^{\text{up}}_{g,t},z^{\text{dn}}_{g,t}\)}{Generator startup and shutdown binary decisions ($ z^{\text{up}}_{g,t},z^{\text{dn}}_{g,t} \in {0,1}$)}
\nomenclature[]{\(z^{\text{dn}}_{ij}\)}{Damage and subsequent de-energization of line $(i,j)$, ($z^{\text{dn}}_{ij} \in {0,1}$)}
\nomenclature[]{\(t^{\text{MinUP}}_{g},t^{\text{MinDown}}_{g}\)}{Min up and down times for each gen $g \in \mathcal{G}$ [sec]}
\nomenclature[]{\(\underline{\theta},\overline{\theta}\)}{Lower/upper limit on the phase angle difference [rad]}
\nomenclature[]{\(\tilde{p}_{ij,t}\)}{Active power flow from bus $i$ to $j$ at time $t$ [MWh]}
\nomenclature[]{\(\underline{p}_{ij},\overline{p}_{ij}\)}{Lower/upper thermal limits on line flow from $i$ to $j$ [MWh]}
\nomenclature[]{\(\omega\ \)}{Demand/WIP scenario in reduced scenario set, $\Omega$}
\nomenclature[]{\(\Omega\)}{Set of DA demand/WIP scenarios post tree reduction}
\nomenclature[]{\(\Phi\)}{Uncertain operational costs [\$]}
\nomenclature[]{\(\pi_{\omega}\)}{Probability of each demand and wildfire scenario $\omega$, ($\pi_{\omega} \in [0,1]$)}
\nomenclature[]{\(\tilde{\pi}_{ij}\)}{Probability of local fire ignition near a line $(i,j)$ ($\pi_{ij} \in [0,1]$)}
\nomenclature[]{\(C_g\)}{Variable cost of generation for each $g \in \mathcal{G}$ [\$/MWh]}
\nomenclature[]{\(C_{g}^{\mathrm{dn}},C_{g}^{\mathrm{up}}\)}{Cost to shutdown/startup a generator $g \in \mathcal{G}$ [\$]}
\nomenclature[]{\(\mathcal{A}^{c}\)}{Complement of set $\mathcal{A}$}
\nomenclature[]{\(\|\cdot\|\)}{Cardinality of Set}
\nomenclature[]{\(f^{\mathrm{uc}},f^{\mathrm{oc}},f^{\mathrm{VoLL}}\)}{Commit (uc), operating (oc) \& VoLL costs [\$]}
\nomenclature[]{\(R_{\mathrm{tol}},\pi_{\mathrm{tol}}\)}{An operator's WFPI (unitless) or WIP (in prob per million) risk tolerance level ($\pi_{\mathrm{tol}} \in [0,1]$) }
\nomenclature[]{\(\xi_{ij,\psi'}\)}{ Realized outage status of line $(i,j)$ in scenario $\psi'$ , ($\xi_{ij,\psi'}\in {0,1}$)}
\nomenclature[]{\(\mathcal{X},\mathcal{Y}\)}{Set of first, second stage variables}
\nomenclature[]{\(\tilde{\cdot}\)}{(Tilde) marking for uncertain variables}
\nomenclature[]{\(\gamma\)}{Uncertain element in demand [MWh]}
\nomenclature[]{\(\Psi\)}{Set of all permutations of line damage scenarios}
\nomenclature[]{\(\bar{\xi}_{ij,t}\)}{ Damage of line $(i,j)$ due to interdependent fire spread model (future work)  ($\bar{\xi}_{ij,t}\in {0,1}$)}
\nomenclature[]{\(\bar{\Psi}\)}{Set of all permutations of line damage scenarios when considering an interdependent fire spread model (future work) }
\nomenclature[]{\(\tilde{\xi}_{n,t}\)}{ Damage of line $(i,j)$ due to fire spread model from neighbor $n$ (future work)  ($\bar{\xi}_{ij,t}\in {0,1}$)}
\nomenclature[]{\(\mathcal{N}_{ij}\)}{ Set of all neighboring lines to line (future work)}
\nomenclature[]{\(\bar{\pi}_{ij}\)}{Probability of fire ignition near a line $(i,j)$ due to an interdependent fire spread model (future work)($\bar{\pi}_{ij} \in [0,1]$)}
\nomenclature[]{\(\pi_{n\rightarrow ij}\)}{Probability of fire ignition near a line $(i,j)$ due to a fire started at neighboring line $n$ (future work) ($\pi_{n\rightarrow ij} \in [0,1]$)}
\nomenclature[]{\(\mathcal{K}_{\bar{\psi}}\subseteq  \mathcal{L}\)}{Set of damaged lines within set of transmission lines under interdependent wildfire spread model (future work)}
\nomenclature[]{\(\hat{z}_{ij}\)}{Controllable de-energization of line $(i,j)$, ($\hat{z}_{ij} \in {0,1}$)}
\nomenclature[]{PSPS}{Public Safety Power Shutoffs}
\nomenclature[]{USGS}{United States Geological Survey}
\nomenclature[]{WFPI}{Wildland Fire Potential Index}
\nomenclature[]{NZR}{Non-zero Risk}
%\nomenclature[]{WS}{Wait-and-See}
\nomenclature[]{SMIP}{Two-stage Stochastic Mixed Integer Program}
\nomenclature[]{M-SMIP}{Multistage Stochastic Mixed Integer Program}
\nomenclature[]{WIP}{Wildfire Ignition Probability}
%\nomenclature[]{DR}{Distributionally Robust}
%\nomenclature[]{DR-PSPS}{Distributionally Robust Public Safety Power Shutoff}
%\nomenclature[]{DRO}{Distributionally Robust Optimization}
\nomenclature[]{RTS}{Reliability Test System}
\nomenclature[]{RTS-GMLC}{Reliability Test System -- Grid Modernization Lab Consortium}
\nomenclature[]{OPS}{Optimal Power Shutoff}
\nomenclature[]{S-PSPS}{Stochastic Public Safety Power Shutoff}
\nomenclature[]{FARSITE}{Fire Area Simulator}
\nomenclature[]{FlamMAP}{Flame Mapping}
\nomenclature[]{WF}{Wildfire}
%\nomenclature[]{DS}{Distribution Shaping}
\nomenclature[]{CVaR}{Conditional Value at Risk}
%\nomenclature[]{VaR}{Value at Risk}
%\nomenclature[]{ADMM}{Alternating Direction Method of Multipliers}
\nomenclature[]{MILP}{Mixed-Integer Linear Program}
\nomenclature[]{RT}{Real-time}
\nomenclature[]{RTPV}{Rooftop Photovoltaics}
\nomenclature[]{CC}{Combined Cycle}
\nomenclature[]{CT}{Combustion Turbine}
%\nomenclature[]{SVI}{Social Vulnerability Index}
\nomenclature[]{WLFP}{WFPI-based Large Fire Probability}
\nomenclature[]{VoLL}{Value of Lost Load}
%\nomenclature[]{VPI}{Value of Perfect Information}
%\nomenclature[]{REWS}{Robustified Expectation Wait-and-See}
%\nomenclature[]{REVPI}{Robustified Expectation Value of Perfect Information}
%\nomenclature[]{RERP}{Robustified Expectation Recourse Problem}
\nomenclature[]{DA}{Day-ahead}
\nomenclature[]{DC-OPF}{Direct Current Optimal Power Flow}
%\nomenclature[]{CDC}{Centers for Disease Control and Prevention}
\nomenclature[]{MIP}{Mixed-Integer Program}
\nomenclature[]{CPUC}{California Public Utilities Commission}
\nomenclature[]{WIP-WLFP}{WLFP-based Wildfire Ignition Probability}
\nomenclature[]{WIP-WFPI}{WFPI-based Wildfire Ignition Probability}
%\nomenclature[]{RA}{Risk-Averse}
%\nomenclature[]{RN}{Risk-Neutral}
\nomenclature[]{NMKS}{N-k Contingency Optimization}
\nomenclature[]{NOAA}{National Oceanic and Atmospheric Administration}
\nomenclature[]{MAE}{Mean absolute error}
\nomenclature[]{TVD}{Total variation distance}
\nomenclature[]{CE}{Cross-entropy}
\nomenclature[]{MW}{Megawatt}
\nomenclature[]{MWh}{Megawatt-hour}
\nomenclature[]{rad}{Radians}
\nomenclature[]{S}{Siemens}
\nomenclature[]{MVA}{Megavolt-amperes}
\nomenclature[]{\$1M}{Millions of Dollars}
\nomenclature[]{UC}{Unit Commitment}
\nomenclature[]{Sto.}{Stochastic}

%\printnomenclature

%\begin{table*}[t]
  %\begin{myframe}
    \printnomenclature
  %\end{myframe}
%\end{table*}
\section{Introduction}
\label{section:Introduction}

Wildfires are a significant and growing threat to California’s infrastructure, economy, and public safety. The California Department of Forestry and Fire Protection (CalFire) reports that 95\% of wildfires in California are caused by human activity, including electrical failures, campfires, debris burning, smoking, and arson~\cite{Isaacs-Thomas}. Although transmission line failures account for less than 10\% of reported wildfire ignitions in California~\cite{CPUC}, nine of California's twenty most destructive wildfires were caused by power lines~\cite{Isaacs-Thomas}. To prevent energized electrical equipment from igniting new wildfires while also ensuring reliable electrical grid operation during ongoing or potential wildfires without putting the public at risk, electric utilities execute Public Safety Power Shutoffs (PSPS)~\cite{Arab}. A PSPS is defined as the act of utilities temporarily turning off power to specific areas to reduce the risk of fires caused by electric infrastructure~\cite{CPUC}. 

A significant challenge for operators during PSPS is to coordinate equipment de-energizations to minimize load shedding, as well as prevent wildfire ignition and spread, in the face of significant uncertainty. Common planning approaches, such as the $N-1$ security-constrained approach, may not be helpful in PSPS since several power lines may intersect the region near a large wildfire. Instead, $N-k$ contingencies-based approaches are more suitable but are computationally expensive because the number of contingencies grows combinatorially with $k$, which is the number of failed components.

Wildfire forecasting models are developed in~\cite{Umunnakwe, Bayani, Bayani2} to optimize PSPS decisions. \cite{Umunnakwe} trains a neural network model for forecasting transmission line outages due to the spread of an active wildfire using features such as vegetation level, dry fuel, wind speed, etc. \cite{Bayani, Bayani2} contribute a wildfire forecasting model that approximates the non-linear PDEs describing the motion of transmission lines due to wind, which can initiate wildfires when multiple lines are in contact with each other. he deterministic formulation is introduced in \cite{Bayani}, while \cite{Bayani2} extends the framework to stochastic capacity expansion with PSPS. The forecasting machine learning model, in \cite{Bayani, Bayani2}, is trained using data from the solution of the non-linear model solved using the Runge-Kutta method, and is then used to make a transmission line wildfire risk score prediction. \cite{Bayani, Bayani2} then implement the risk score as an input to a deterministic PSPS model. \cite{Bayani} shows cost reductions from using the transmission line wildfire risk score surrogate in place of the National Oceanic and Atmospheric Administration's (NOAA's) Fire Weather Threat to power lines (e.g., transmission line wildfire risk values from 1-5) performed in \cite{Rhodes}. 

\cite{Rhodes2} demonstrated that NOAA fire warnings lack enough spatial granularity for operational PSPS optimization. Subsequent OPS formulations \cite{Kody,Kody2,Yang,Yang2} and unit commitment with OPS (UC+OPS) \cite{Greenough} instead adopted the USGS Wildland Fire Potential Index (WFPI) as the wildfire risk input. Instead, the U.S. Geological Survey's (USGS) normalized vegetation flammability index, called the Wildland Fire Potential Index (WFPI), is superior in operational models. WFPI forecasts consist of integer values ranging from 0 to 150. The USGS provides WFPI forecasts up to seven days in the future and has archived all daily forecasts since 2001.~\cite{Kody, Rhodes2, Kody2, Yang, Yang2, Greenough} use WFPI as a proxy for the impact of wildfires caused by transmission lines and define WFPI as a deterministic exogenous input to the optimization problem. Particularly, \cite{Kody, Rhodes2, Kody2} input an overly conservative WFPI forecast into their optimization model by averaging the top 10\% of all WFPI values experienced by the transmission lines over the 2019 and 2020 wildfire seasons. This approach overestimates the transmission line wildfire risk in the network compared to using the forecasted WFPI values, as done in \cite{Greenough}, and may lead to higher costs due to over-committing generating resources in anticipation of more line de-energizations.

In addition to WFPI and WLFP, a broad range of wildfire risk metrics and warning systems have been developed for fire management and situational awareness. Operational tools such as fire hazard warnings issued by the California Department of Forestry and Fire Protection \cite{CalFire}, and fire weather threat levels from the National Oceanic and Atmospheric Administration provide coarse \cite{Rhodes}, categorical assessments of wildfire danger. Complementing these are widely used fire science indices such as the Keetch-Byram Drought Index (KBDI) \cite{KBDI}, Fire Weather Index (FWI) \cite{FWI}, and Energy Release Component (ERC)\cite{ERC}, which quantify fuel dryness, weather-driven fire potential, and expected fire intensity. However, these metrics primarily characterize environmental conditions conducive to fire rather than the probability of ignition or infrastructure-specific impacts, limiting their direct applicability in power system optimization. In contrast, WFPI and WLFP provide spatially resolved measures that are more readily linked to wildfire occurrence, making them more suitable for PSPS decision-making. This work builds on these indices by mapping them to explicit wildfire ignition probabilities and embedding them within a probabilistic optimization framework that captures the joint likelihood of transmission line failures.

PSPS has been formulated as a deterministic problem in~\cite{ Umunnakwe, Bayani, Rhodes, Kody, Rhodes2, Kody2,Trakas2} and as a stochastic problem in~\cite{Bayani2, Yang, Yang2, Greenough, Greenough3, Moreira}. \cite{Umunnakwe} minimizes load shed subject to transmission line outage scenarios but does not consider the impact of energization decisions on wildfire spread and the uncertainty in demand. The Optimal Power Shutoff (OPS) model, introduced in \cite{Rhodes} and extended to contain security constraints in \cite{Rhodes2}, consists of a bi-objective function that balances maximizing the total demand served with minimizing total system transmission line wildfire risk (WFPI).\cite{Kody,Kody2} retain the deterministic OPS structure while adopting conservative upper-percentile WFPI inputs. The OPS model in \cite{Rhodes, Kody, Rhodes2, Kody2} is deterministic and does not consider commitment and operational costs. In practice, there is uncertainty in wildfire ignition and its subsequent spread, which motivates probabilistic PSPS models. \cite{Yang, Yang2} extend OPS to a two-stage and multi-stage stochastic mixed-integer program to model the operational PSPS decisions over the evolution of simulated wildfire events. \cite{Yang, Yang2} use the relative WFPI of each line to determine the likelihood of wildfire ignition events. However, \cite{Yang, Yang2} also do not consider unit commitment strategies for generators. 

In addition to WFPI forecasts, the USGS also provides an improved risk metric called Large Fire Probability (WLFP). WLFP forecasts are based on historical average probabilities of fires that spread more than 500 acres and corresponding WFPI values for those days. The USGS website provides reliability diagrams, which quantify how well WFPI and WLFP values are related to the historically observed fraction of ignited voxels by large wildfires expressed as historically observed large wildfire ignition probabilities (OWIP)\cite{WFPI_Reliability}). \cite{WFPI_Reliability} shows that more large fires occurred at the highest levels of WFPI; however, there is a large scatter in the distribution that maps large WFPI to OWIP when compared with WLFP to OWIP maps, which can affect PSPS decision-making. WLFP has not been considered as a metric for PSPS decision-making~\cite{Kody, Rhodes2, Kody2, Yang, Yang2, Greenough}.

In the author's previous work~\cite{Greenough}, a mean-Conditional Value at Risk (mean-CVaR) two-stage stochastic unit commitment problem was developed to study the impact of uncertain demand on unit commitment and line de-energization decisions. Unlike \cite{Rhodes, Kody, Rhodes2, Kody2, Yang}, \cite{Greenough} frame the optimization purely in terms of economic costs as the sum of unit commitment and operational costs instead of only the Value of Lost Load (VoLL) costs \cite{Rhodes, Kody, Rhodes2, Kody2}, or wildfire damage costs~\cite{Yang, Yang2}. 

There are two limitations of the earlier formulations. First, WFPI measures fire potential in terms of vegetation flammability and does not represent the probability of a line being damaged due to wildfires. For instance, there could be instances in the winter when dryness, fuel, and wind speeds are prime for a large fire, but the chance of that fire happening is very low. This means that a high WFPI does not correlate as strongly with the true likelihood of a wildfire ignition. Second, there are $2^L$ line de-energization scenarios, but when minimizing WFPI risk in~\cite{Kody, Rhodes2, Kody2, Yang, Yang2, Greenough}, each line de-energization decision is made independently of the other lines, which results in simply shutting off lines with the largest WFPI values. Lines with a high WFPI risk represent scenarios with a high fire potential but not necessarily a high probability of ignition. Therefore, operators could overestimate the threat of high WFPI but low WIP risk scenarios and unnecessarily de-energize lines that are critical to power delivery. This distortion of WIP prediction could operationally lead to blackouts and higher VoLL costs.  

In this work, we offer a different interpretation of wildfire risk than the WFPI-based metric originally proposed in \cite{Rhodes, Kody, Rhodes2, Kody2, Yang, Yang2, Greenough} that incorporates the probability of line damages and directly accounts for the joint likelihood of actual line failures. We model wildfire ignitions leading to line damages as Bernoulli random events and formulate a set of cross-entropy based constraints that incorporates each of the $2^L$ line failure combinations. The proposed constraint is included in an S-PSPS problem to determine line de-energization decisions, which maximize the joint likelihood of each of the line failure scenario within a desired tolerance level that represents the total system damage probability. Since the line de-energization decision variables are included in the constraints, therefore, the proposed constraint has the added benefit of trimming highly improbable scenarios instead of explicitly sampling the entire space of failure scenarios. This trimming results in a tractable formulation compared to \cite{Yang,Yang2,Greenough}.

To obtain the wildfire ignition probabilities leading to line damages, we develop a method of mapping both WFPI and WLFP values to wildfire ignition probabilities (WIP) using the OWIP~\cite{WFPI_Reliability}. We then compare the accuracy of WIP obtained from WFPI and WLFP using historical wildfire events, as well as the line de-energization decisions made in S-PSPS problems on the IEEE RTS 24-bus and RTS GMLC systems. While there is room for improving the WIP forecast, we demonstrate the value of thoroughly comparing the WIP constructions and cost outcomes from mapping two different representations of wildfire risk (WFPI and WLFP) to WIP. Since nominal WIP-WFPI and WIP-WLFP forecasts impact both the underlying damage probability model and the subsequent PSPS decisions, optimal decisions from one model may not be recoverable by adjusting the operational risk-averseness levels of another. Table \ref{table:riskComparison} and Table \ref{table:frameworkComparison} highlight the key contributions of the work in wildfire risk representations and optimization frameworks, and how it differentiates itself from existing PSPS works.

\begin{landscape}
\begin{table*}[h]
\centering
\caption{Comparison of wildfire risk representations in PSPS studies}
\begin{tabular}{l m{9.0cm} m{2.0cm} m{7.0cm}}
\hline
\textbf{Reference} &
\textbf{Risk Metric} &
\textbf{OWIP-Calibrated WIP} &
\textbf{Compares Multiple Indices} \\
\hline
Umunnakwe et al. \cite{Umunnakwe}
& ML-based risk score
& No
& No \\
Bayani et al. \cite{Bayani,Bayani2}
& Learned risk surrogate
& No
& ML surrogate vs. NOAA Fire Threat Levels \\
Rhodes et al. \cite{Rhodes,Rhodes2}
& NOAA / WFPI (operational forecast)
& No
& No \\
Kody et al. \cite{Kody,Kody2}
& WFPI (upper 10\%)
& No
& No \\
Yang et al. \cite{Yang,Yang2}
& WFPI (scaled to relative likelihood) $\rightarrow$ WIP
& No
& No \\
Greenough (prior) \cite{Greenough}
& WFPI (operational forecast)
& No
& No \\
Greenough (follow-up) \cite{Greenough3}
& WLFP $\rightarrow$ WIP (operational forecast)
& Yes
& No \\
Moreira et al. \cite{Moreira}
& Assumed failure rate adjusted with contextual info
& No
& No \\
Trakas et al. \cite{Trakas2}
& Physics-based wildfire model
& No
& No \\
\hline
\textbf{This work}
& WFPI / WLFP $\rightarrow$ WIP
& Yes
& WIP-WFPI vs. WIP-WLFP \\
\hline
\end{tabular}
\label{table:riskComparison}
\end{table*}
\end{landscape}

\begin{landscape}
\begin{table*}[h]
\centering
\caption{Comparison of optimization frameworks in PSPS studies}
\begin{tabular}{m{4.8cm} m{6.2cm} m{1.0cm} m{2.0cm} m{1.7cm} m{1.8cm} m{3.3cm}}
\hline
\textbf{Reference} &
\textbf{Framework} &
\textbf{Sto.} &
\textbf{Unit Commit.} &
\textbf{Joint Failures} &
\textbf{Scenario Filtering} &
\textbf{PSPS Objective} \\
\hline
Umunnakwe et al. \cite{Umunnakwe}
& Deterministic Economic Dispatch
& No
& No
& No
& None
& Load shed only \\
Bayani et al. \cite{Bayani,Bayani2}
& Deterministic/Stochastic OPS extension
& No/Yes
& No
& No
& None / Explicit
& Operational cost \\
Rhodes et al. \cite{Rhodes,Rhodes2}
& Deterministic OPS
& No
& No
& No
& None
& VoLL only \\
Kody et al. \cite{Kody,Kody2}
& Deterministic OPS extension
& No
& No
& No
& Explicit
& VoLL only \\
Yang et al. \cite{Yang,Yang2}
& Stochastic OPS extension
& Yes
& No
& No
& Explicit
& Operational + wildfire damage cost \\
Greenough (prior) \cite{Greenough}
& Stochastic Mean-CVaR UC-OPS
& Yes
& Yes
& No
& Explicit
& UC + operational cost \\
Greenough (follow-up) \cite{Greenough3}
& Decision-Dependent Joint DRO UC-OPS
& Yes
& Yes
& Yes
& Explicit
& UC + operational cost \\
Moreira et al. \cite{Moreira}
& Decision-Dependent DRO UC-OPS
& Yes
& Yes
& Yes
& Explicit
& UC + operational cost \\
Trakas et al. \cite{Trakas2}
& Stochastic UC
& Yes
& Yes
& Yes
& Explicit
& UC + operational cost \\
\hline
\textbf{This work}
& Stochastic UC w/ Cross-Entropy OPS
& Yes
& Yes
& Yes
& Implicit via cross entropy
& UC + operational cost \\
\hline
\end{tabular}
\label{table:frameworkComparison}
\end{table*}
\end{landscape}

The main contributions of this paper are the following: 
\begin{enumerate}
\item To the best of our knowledge, this is the first paper to illustrate and quantify the benefits of using WLFP instead of WFPI to quantify the power systems' operating risks related to wildfires. We not only illustrate that WLFP maps more accurately to OWIP than WFPI during peak wildfire season, but also demonstrate higher worst-case cost savings when WIP is mapped from WLFP forecasts instead of from WFPI forecasts.  

\item We develop a cross-entropy bound that models line failures as Bernoulli random events and incorporates WIPs as the probability of line failures due to wildfire rather than using WFPI as a proxy for exposure to potential fires. The proposed formulation is non-linear, re-formulated as a linear $\log$(WIP) constraint, and included in a two-stage day-ahead (DA) unit commitment S-PSPS problem. 

\item We use transmission line WIP forecasts to directly simulate real-time (RT) transmission line failure scenarios during PSPS events and validate the commitment decisions obtained from the S-PSPS problem. Out-of-sample RT simulations demonstrate that the WIP-based S-PSPS commitment decisions obtained using WLFP lead to lower operational costs than WFPI. 

\end{enumerate}

The paper is organized as follows. Section~\ref{section:Selection of Wildfire Risk Metric} compares WIP-WFPI and WIP-WLFP model predictions. Section~\ref{section:Modeling of Public Safety Power Shut-offs for Wildfire Risk Mitigation} introduces the PSPS strategy, including the DA and RT formulations. Section~\ref{section:Results} explains the results of the unit commitment for WFPI and WLFP on the IEEE RTS 24-bus and RTS-GMLC. Section~\ref{section:Conclusion} concludes the paper.

\section{Selection of Wildfire Risk Metric}
\label{section:Selection of Wildfire Risk Metric}
\subsection{Qualitative comparison}
\label{subsection:Qualitative comparison}

WFPI describes the ratio of live to dead fuel based on wind speed, dry bulb temperature, and rainfall~\cite{WFPI}. WFPI values range from $0$ to $150$. Values of $249-254$ mark land outside the United States, ice, agricultural land, barren land, marshland, and bodies of water; all terrain conditions with WFPI values from $249$-$254$ have zero WFPI. DA WFPI forecasts are obtained from the USGS database in deterministic OPS formulations \cite{Rhodes2,Kody,Kody2}, stochastic OPS extensions \cite{Yang}, and stochastic UC-OPS formulations \cite{Greenough}. 

WLFP estimates the probability that a new fire will burn more than 500 acres \cite{WLFP}. WLFPs are assigned per million voxels, where 1 voxel equals an area of 1 $\text{km}^2$. The WLFP forecasts are derived from historical average probabilities of large fires and are adjusted based on the WFPI of that day \cite{WLFP}. The USGS generates WFPI and WLFP forecasts for the continental US up to seven days in the future. WFPI and WLFP forecasts are validated using historical fire incidence data from \cite{Short}. 

\begin{figure}[t]
\centering
\vspace{-1em}
\begin{overpic}[width=\columnwidth]{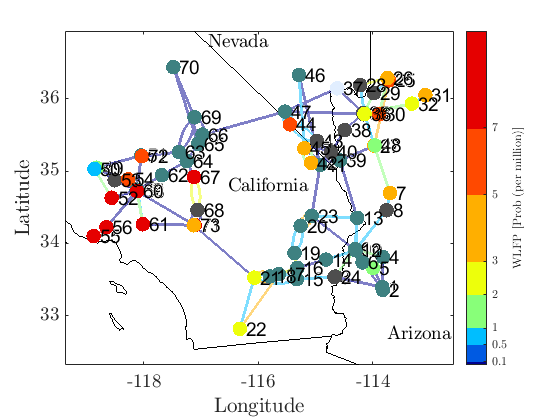}
\put(15,27){\color{Salmon} IEEE 24-bus }
\put(15,24){\color{Salmon} Region 3}
{\color{Salmon}\linethickness{0.2mm}
\polygon(15,31)(15, 47)(32, 68)(42,57)(41,32.5)}
\put(55,12){\color{Black} Region 1}{\color{Black}\linethickness{0.2mm}
\polygon(40,12)(45, 30)(55, 40)(75,45)(75,20)}
\put(60,66.5){\color{Black} Region 2}{\color{Black}\linethickness{0.2mm}
\polygon(47,60)(60, 65)(80.5, 65)(80.5,50)(55,43)}
\end{overpic}
\vspace{-1.5em}
\caption{Schematic of the RTS-GMLC with each transmission line and bus highlighted to depict its WLFP for July 31, 2020. The geographic layout is from \cite{RTSGMLC}.
Zero-risk buses are in teal, gray, and baby blue (bus 37). %1-4, 6, 8 9-20, 23,24,28,29,34,35,38-41,43,46,47,51,53,57,59,62-66,68-70 
}
%\vspace{-2em}
\label{fig:IEEERTScolored}
% Folder: H:\My Drive\MAE 207 Grid Integration Project\Project Code\Figures for Journal Paper Spring 2024 PostTSG
% File: fig2RTScolored.m
% Set: Grid='RTb';   
% Lines: Run whole script
% Figure(1)
\end{figure}

One-day-ahead forecasts (saved in .tiff format) are downloaded and combined to form timeseries of WLFP and WFPI for each of the 24 buses and along the 38 transmission line locations of the IEEE RTS 24-bus system (hereafter RTS), as well as the 73 bus and 120 transmission line locations of the RTS-GMLC system (hereafter RTS-GMLC). While all WFPI and WLFP data used in this paper are forecasts, for brevity, we drop the word ``forecast'' from hereafter. 
Each .tiff file contains a grid mesh in which each grid point has a WLFP and WFPI value and a pixel $x$-$y$ coordinate, which is translated to longitude and latitude. The four nearest neighbors from the .tiff grid mesh are averaged to form the forecast at the bus location. A collection of WFPI/WLFP forecasts is assigned to each transmission line via the Bresenham algorithm \cite{Bresenham} and to the buses at the ends of the lines. Average daily WLFP bus and line values on the RTS-GMLC in the Southwestern United States are depicted in Fig.~\ref{fig:IEEERTScolored} for illustration. Similarly, for the RTS system shown in~\cite{Greenough}, the WLFP values are obtained for all buses and lines.

\begin{figure}[t]
\vspace{-1em}
\centerline{\includegraphics[width=\columnwidth]{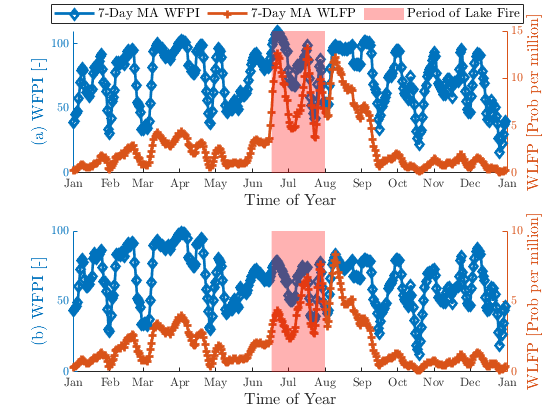}}
\vspace{-1em}
\caption{Comparison of time series of the 7-day moving average (MA) of WFPI and WLFP (a) for node 71 (23 on IEEE RTS 24-bus) and (b) node 73 on the IEEE RTS-GMLC system near San Bernardino in 2015. The transparent red region indicates the time-span of the Lake Fire.}
%\vspace{-1em}
\label{fig:WFPIvWLFP2015}
% Folder: H:\My Drive\MAE 207 Grid Integration Project\Project Code\Data for Journal Paper Spring 2024
% File: WFPIvWLFP2015.m
% Set: BUSS=73,BUSS2=71;   
% Lines: Run whole script
% Figure(2)
\end{figure}

Using the 2015 Lake Fire as a case study, we show that periods of elevated WLFP align more closely with observed wildfire ignitions than those of elevated WFPI. The Lake Fire was the largest wildfire in Southern California that year, burning 31,359 acres between June 17 and August 1 near San Bernardino, close to buses 23 (RTS) and 73 (RTS-GMLC). In Fig.~\ref{fig:WFPIvWLFP2015}, the 7-day moving average (MA) of WLFP rises sharply during this period, while WFPI shows only a modest increase or even a slight decline. For the RTS system (Fig.~\ref{fig:WFPIvWLFP2015}(a)), WLFP increases by 307\% compared to just 36\% for WFPI from June 11 to June 22. In the RTS-GMLC system (Fig.~\ref{fig:WFPIvWLFP2015}(b)), WLFP rises by 132\% during the Lake Fire, while WFPI decreases by 28\%. The top three WLFP peaks during the Lake Fire rank as the 1st, 2nd, and 22nd highest (RTS) and the 31st, 10th, and 4th highest (RTS-GMLC) MA WLFP values of 2015. All WLFP values during the Lake Fire fall above the 57th–64th percentiles. In contrast to WLFP, the top three WFPI peaks during the fire rank only as the 1st, 26th, and 105th (RTS) and the 111th, 122nd, and 115th (RTS-GMLC) highest MA WFPI values of the year, with values dipping as low as the 7th–8th percentiles. WFPI peaks in the 80th–90th percentiles appear earlier in the year (February–May), despite minimal wildfire activity. Only one large fire (1,049 acres in Riverside, April 18–24) occurred during that time \cite{Short}. These anecdotal observations of the Lake Fire suggest that WLFP tracks major wildfire events more reliably than WFPI.

The USGS provides reliability diagrams for both WFPI (Table 1 and Fig. 3 in \cite{WFPI}) and WLFP (Table 1 and Fig. 2 in \cite{WLFP}). These diagrams provide distributions of yearly historical OWIP for different ranges of WFPI or WLFP values. These OWIP capture trends in historical wildfire occurrences from 1992 to 2020 \cite{Short}. We project each WFPI/WLFP bus or line value to an OWIP distribution based on the bins provided by the reliability diagrams. The mean of that conditional OWIP (i.e., OWIP given a WFPI or WLFP range) is used to represent the WIP for that respective bus or line, which then represents the corresponding wildfire risk in the S-PSPS problem. 

\begin{figure}[t]
\centering
\subfloat{\includegraphics[width=8.2cm]{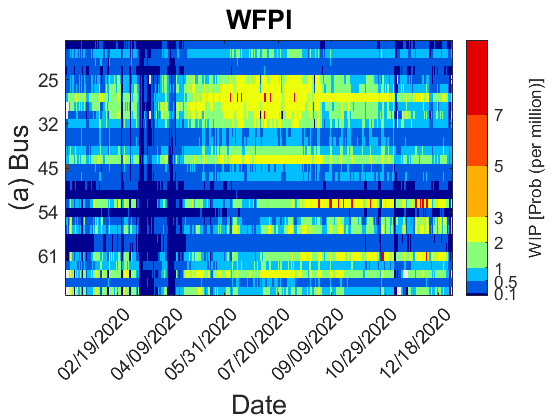}} \hfill
\subfloat{\includegraphics[width=8.2cm]{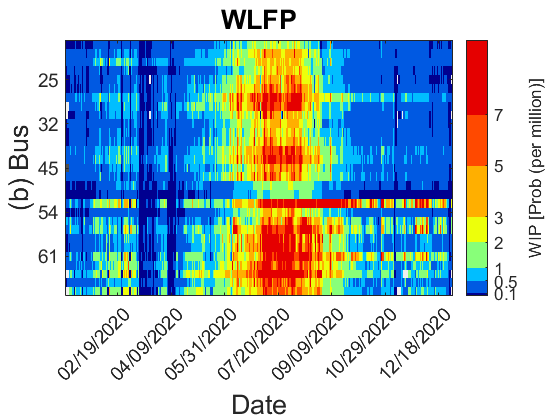}} \hfill
\subfloat{\includegraphics[width=8.2cm]{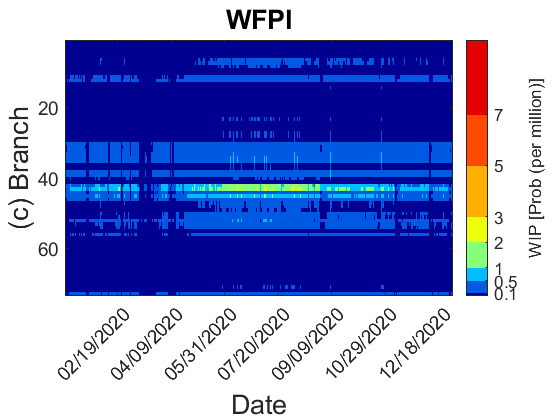}} \hfill
\subfloat{\includegraphics[width=8.2cm]{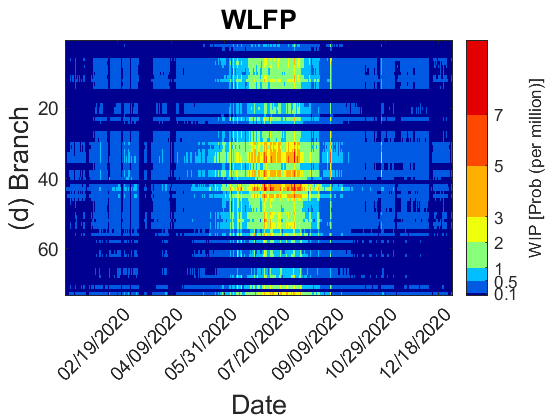}}

\vspace{-0.5em}
\caption{Time series of WIP predictions from WFPI (a, c) and WLFP (b, d) for each bus (a, b) and transmission line (or branch) (c, d) in RTS-GMLC during 2020. WLFP-based WIP aligns more closely with historical large wildfire ignitions in Southern California. Days with missing data: April 14, April 24, and July 30. Buses and lines with zero risk are excluded.}

\label{fig:WFPIvWLFP2020HeatmapRTS}

% Folder: H:\My Drive\MAE 207 Grid Integration Project\Project Code
% File: CompareWIPWLFPWFPI.m
% Set: Grid='24B';   
% Lines: Run whole script
% Figure (3)
%\vspace{-1.5em}
\end{figure}

Fig.~\ref{fig:WFPIvWLFP2020HeatmapRTS} depicts the RTS-GMLC~\cite{RTSGMLC} bus and line WIP obtained using daily average WFPI and WLFP values and reliability diagrams for the year 2020. For a tabular version of the reliability diagrams from \cite{Short}, please see Tables~\ref{table:CompareWIPWFPI}–\ref{table:CompareWIPWLFP} in the \ref{section:Extra WIP vs. OWIP Analysis for IEEE 24-bus and RTS-GMLC systems}. 
 
 The Southern California wildfire season is from late May to late September \cite{CalFire}. Again, WIP based on WFPI (hereafter WIP-WFPI) does not coincide with periods of elevated wildfire incidences during the main Summer wildfire season as consistently as WIP based on WLFP (hereafter WIP-WLFP).  

\vspace{-1em}
\subsection{Comparison of Distributions}
\label{subsection:Comparison of WIP Distributions Generated from WFPI and WLFP Forecasts to Observed WIP Distributions}

\begin{figure}[ht]
\centering
\vspace{-1.6em}
\includegraphics[width=\columnwidth]{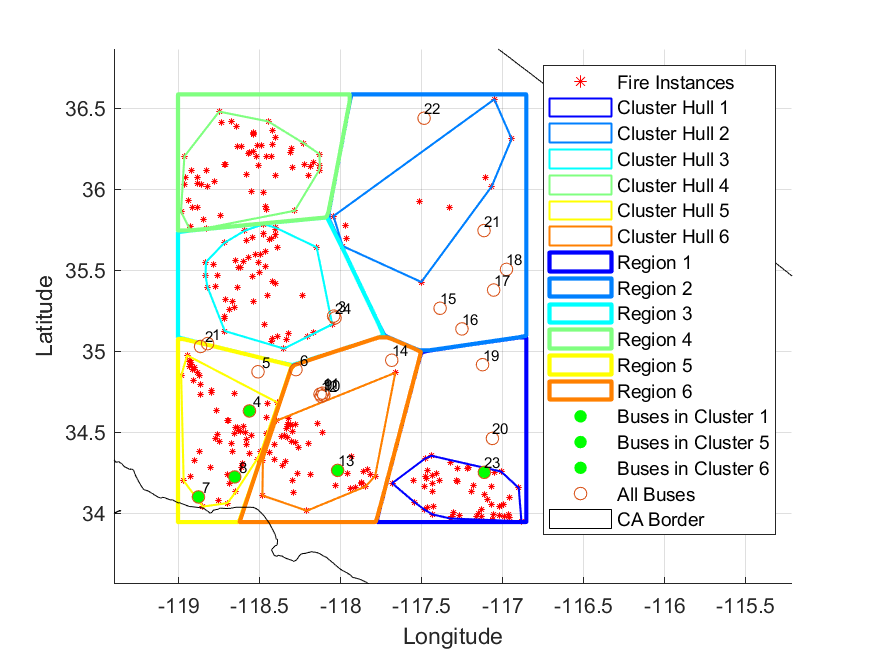}
\vspace{-1.5em}
\caption{K-Means clustering (K=6) of wildfire data from \cite{Short} to compute regional wildfire differences around the IEEE RTS 24-bus system.}
\label{fig:WFPIvWLFPDistributions}
%\vspace{-1.5em}
% Folder: H:\My Drive\MAE 207 Grid Integration Project\Project Code
% File: CompareWIPWLFPWFPI.m
% Set: Grid='24B';  
% Lines: Run whole script
% Figure (3)
\end{figure}

In this section, we further compare the WIP-WFPI and WIP-WLFP with the observed large wildfires across 20 years (from 2001 to 2020)~\cite{Short} for both RTS and RTS-GMLC systems. Consider first the RTS system. K-means clustering is performed on the longitude and latitude of all large wildfires from \cite{Short}. The bounding box shown in Fig.~\ref{fig:WFPIvWLFPDistributions} for each cluster region encompasses all RTS buses. At each RTS bus location, a 0.2-degree buffer is added to its longitude and latitude values and then used to assign the bus to a particular cluster. Out of the 6 clusters, we exclude cluster 2 (all buses contain zero risk) and cluster 4 (no buses) from subsequent analysis.

Using the start time of large wildfires in each cluster, the ground truth monthly OWIP from 2001-2020 is obtained by the number of wildfire counts normalized by the area of the convex hull of the cluster (in $\mathrm{km^2}$) times approximately 600 days (i.e., number of days in that month $\times$ 20 years). All buses in the same cluster are assigned the same OWIP, and the ground truth 20-year monthly average OWIP values are compared with the predicted monthly average WIP-WLFP and WIP-WFPI values using mean absolute error (MAE) and average cross entropy (ACE)~\cite{Goodfellow} defined below,
\begin{equation}
\mathrm{MAE}_{b}=\frac{1}{12}\sum^{12}_{m=1}|\mathrm{WIP}_{b,m}-\mathrm{OWIP}_{b,m}|, \: \forall b \in \mathcal{B}, 
\label{eq:MAE}
\end{equation}
\vspace{-1em}
\begin{align}
\mathrm{ACE}_{b}=-\frac{1}{12}\sum^{12}_{m=1}[\mathrm{OWIP}_{b,m}\log(\mathrm{WIP}_{b,m})+ %\nonumber\\
(1-\mathrm{OWIP}_{b,m})\log(1-\mathrm{WIP}_{b,m}) ],\: \forall b \in \mathcal{B}.
\label{eq:ACE}
\end{align}
where, WIP$_{b,m}$ is either equal to WIP-WLFP or WIP-WFPI for the bus $b$ and month $m$, and OWIP$_b$ is the OWIP of bus $b$ corresponding to its assigned cluster. Table \ref{table:MAE_ACE_CompactTight} shows the MAE, which is equivalent to the Total Variation Distance (TVD) between OWIP and WIP at every non-zero risk bus. 

For 8 out of the 12 buses, there is an improvement in MAE using WIP-WLFP. At buses 4, 10, 12, and 13, the elevated levels of WIP in June, July, and August contributed to a higher MAE from WIP-WLFP than WIP-WFPI. 

However, the MAE from WIP-WLFP was less than WIP-WFPI for more buses, and the total MAE across all buses is 12\% lower for WIP-WLFP. 

Table~\ref{table:MAE_ACE_CompactTight} shows the ACEs at every non-zero risk bus. ACE penalizes large distributional distances between the ground truth, OWIP, and predicted distributions WIP-WFPI or WIP-WLFP.  The ACE from WIP-WLFP is less than WIP-WFPI for all buses except Bus 13, and the total ACE across all buses is 2.3\% lower for WIP-WLFP than WIP-WFPI. The MAE and ACE results in Table \ref{table:MAE_ACE_CompactTight} show that the WIP-WLFP tends to make more accurate predictions of the WIP than WIP-WFPI. 

Similar observations were made for the RTS-GMLC system, in which K-means clustering resulted in 18 clusters (the depiction of these clusters is shown in \ref{fig:WFPIvWLFPDistributionsRTS} in \ref{section:Extra WIP vs. OWIP Analysis for IEEE 24-bus and RTS-GMLC systems}). Bus WLFP generally results in higher WIP than WFPI during wildfire season. For example, at bus 44 in cluster 4, WLFP resulted in a maximum WIP of 4.8$\times$10$^{-6}$  compared to 2.3$\times$10$^{-6}$ using WFPI. Similarly, cluster 18, which consists of buses 21 and 22, had a relatively flat WIP when using WFPI with an average of 0.3$\times$10$^{-6}$ WIP, whereas WLFP identifies elevated wildfire risk with the maximum value of 3.1$\times$10$^{-6}$ (at bus 22) during July. The corresponding plots and tables are omitted due to space limitations. However, those comparisons of the discrete monthly WIP-WFPI and WIP-WLFP distributions to the OWIP  can be seen in \ref{section:Extra WIP vs. OWIP Analysis for IEEE 24-bus and RTS-GMLC systems} for the IEEE 24-bus (Fig. \ref{fig:WFPIvWLFPDistributionsRTS}) and RTS-GMLC (Fig. \ref{fig:IEEERTSClusterAnalysis1month}).
  
\begin{table}[t]
\setlength{\tabcolsep}{3pt} 
\centering
\caption{Mean Absolute Error (MAE) [10$^{-6}$] and Average Cross Entropy (ACE) [10$^{-5}$] between observed wildfire ignition probability (OWIP) and model-based predictions using WLFP and WFPI across IEEE 24-bus clusters. The percent improvement (\% Imp.) is based on using WLFP over WFPI.}
\begin{tabular}{l@{\hskip 6pt}l@{\hskip 4pt}r@{\hskip 4pt}r@{\hskip 4pt}r@{\hskip 6pt}r@{\hskip 4pt}r@{\hskip 4pt}r}

\textbf{Cluster} & \textbf{Bus} & \multicolumn{3}{c}{\textbf{MAE}} & \multicolumn{3}{c}{\textbf{ACE}} \\
& RTS(-GMLC)& WLFP & WFPI & \% Imp. & WLFP & WFPI & \% Imp. \\
\midrule
1 & 19 (67)& 1.46 & 2.84 & 48.7 & 4.32 & 4.57 & 5.5 \\
1 & 23 (71)& 1.61 & 2.51 & 35.9 & 4.30 & 4.42 & 2.7 \\
\midrule
3 & 1  (49)& 1.32 & 1.68 & 21.6 & 2.33 & 2.55 & 8.9 \\
3 & 24 (72)& 0.75 & 1.65 & 54.4 & 2.26 & 2.45 & 7.6 \\
\midrule
5 & 2  (50)& 1.85 & 2.08 & 11.2 & 3.10 & 3.41 & 9.1 \\
5 & 4  (52)& 1.96 & 1.48 & -32.4 & 2.93 & 2.93 & 0.3 \\
5 & 7  (55)& 1.39 & 1.89 & 26.2 & 2.93 & 3.12 & 5.8 \\
5 & 8  (56)& 1.22 & 1.74 & 29.6 & 2.89 & 3.01 & 4.2 \\
\midrule
6 & 6  (54)& 0.73 & 1.27 & 42.2 & 1.96 & 2.13 & 7.9 \\
6 & 10 (58)& 1.84 & 1.24 & -48.2 & 1.99 & 2.11 & 5.4 \\
6 & 12 (60)& 1.85 & 1.24 & -49.7 & 1.99 & 2.11 & 5.4 \\
6 & 13 (61)& 1.98 & 0.85 & -132  & 1.98 & 1.94 & -2.3 \\
\midrule
\textbf{All} & -- & 18.0 & 20.5 & 12.2 & 14.8 & 15.2 & 2.3 \\
\bottomrule
\end{tabular}
\label{table:MAE_ACE_CompactTight}
%\vspace{-1em}
% % Folder: H:\My Drive\MAE 207 Grid Integration Project\Project Code
% % File: CompareWIPWLFPWFPI.m
% % Set: Grid='24B';   
% % Lines: Run whole script
\end{table}

\section{Modeling PSPS for Wildfire Risk Mitigation}
\label{section:Modeling of Public Safety Power Shut-offs for Wildfire Risk Mitigation}

\subsection{Preliminaries}
\label{subsection:Preliminaries}
Let $\mathcal{P}=(\mathcal{B}, \mathcal{L})$ be the graph describing the power grid where $\mathcal{B}=\{1, \dots, B\}$ is the set of $B$ buses in the network, and $\mathcal{L}$ is the set of $L$ edges such that two buses $i,j \in \mathcal{B}$ are connected by a transmission line if $(i,j)\in \mathcal{L}$. The sets of buses with generators and loads are collected in $\mathcal{G}$ and $\mathcal{D}$ respectively, and $\mathcal{H}=\{1, \dots, H\}$ is the set of time indices over the horizon $H$ of the optimization problem. A direct current optimal power flow (DC-OPF) is used to approximate the line power flows and bus power injections; all references to power are to its active power. At any timestep $t$, the power injected by the generator $g \in \mathcal{G}_i$ at bus $i \in \mathcal{B}$ is denoted by $p_{g,t}$ with $\bm{p}_{g}$ being the vector of all generator power injections. Similarly, $p_{d,t}$ is the load $d \in \mathcal{D}_i$ at bus $i \in \mathcal{B}$. 
Power flowing through the line $(i,j) \in \mathcal{L}$ is $p_{ij,t}$ and the phase angle at bus $i\in \mathcal{B}$ is denoted by $\theta_{i,t}$. The boldface letters represent their corresponding vectors. Finally, the binary variables are denoted by $z \in \{0,1\}$ with an appropriate subscript to capture component shutoff decisions.

\textbf{Remark:} The DC-OPF approximation is adopted due to its computational tractability and suitability for large-scale stochastic unit commitment optimization problems (MILPs). The DC-OPF approximation makes three main assumptions: linearized power flow via the small angle approximation ($\sin(\theta_i - \theta_j) \approx \theta_i- \theta_j$), neglecting reactive power (transmission lines have low resistance per reactance, r/x, ratios, which implies $G_{ij}=0$), and voltage magnitude approximations ($V_i\approx 1 \text{p.u.} :\ \forall i \in \mathcal{B}$) \cite{Dvorkin}.
Since the S-PSPS model relies on a DC-OPF approximation, it does not guarantee AC feasibility. To ensure practical implementation, scalable multi-stage stochastic PSPS models incorporating AC-OPF or relaxed AC-OPF constraints may be necessary. Additional constraints or parameter tuning may be needed in the optimization to avoid infeasibility or suboptimal solutions. However, an AC-OPF model would provide grid operators with more realistic strategies and inspire the development of algorithms capable of finding global solutions to the non-convex AC-OPF S-PSPS problem.
Furthermore, the current S-PSPS model operates under $N-k$ contingencies, whereas industry standards for maintaining supply security during transmission planning and operation typically focus on $N-1$ contingencies. Adjustments will be required to align the $N-k$ approach with accepted robust resilience strategies.
From a practical perspective, two-stage stochastic methods take significantly longer to solve than their deterministic equivalent. In addition, optimization problems, such as the unit commitment with N-k contingencies, involve combinatorial optimization. Research on ways to bundle optimal arrangements of de-enegization decisions \cite{Noyan}, package representative scenarios for demand and resource availability \cite{Gröwe-Kuska}, and construct an effective surrogate of a DC-OPF approximation \cite{Pareek} would be useful to reduce computation time for DC-OPF with $N-k$ contingencies and eventually extended to convex relaxations of the AC-OPF with $N-k$ contingencies.

\begin{figure}[ht]
\centering
\vspace{-0.6em}
\includegraphics[width=\columnwidth]{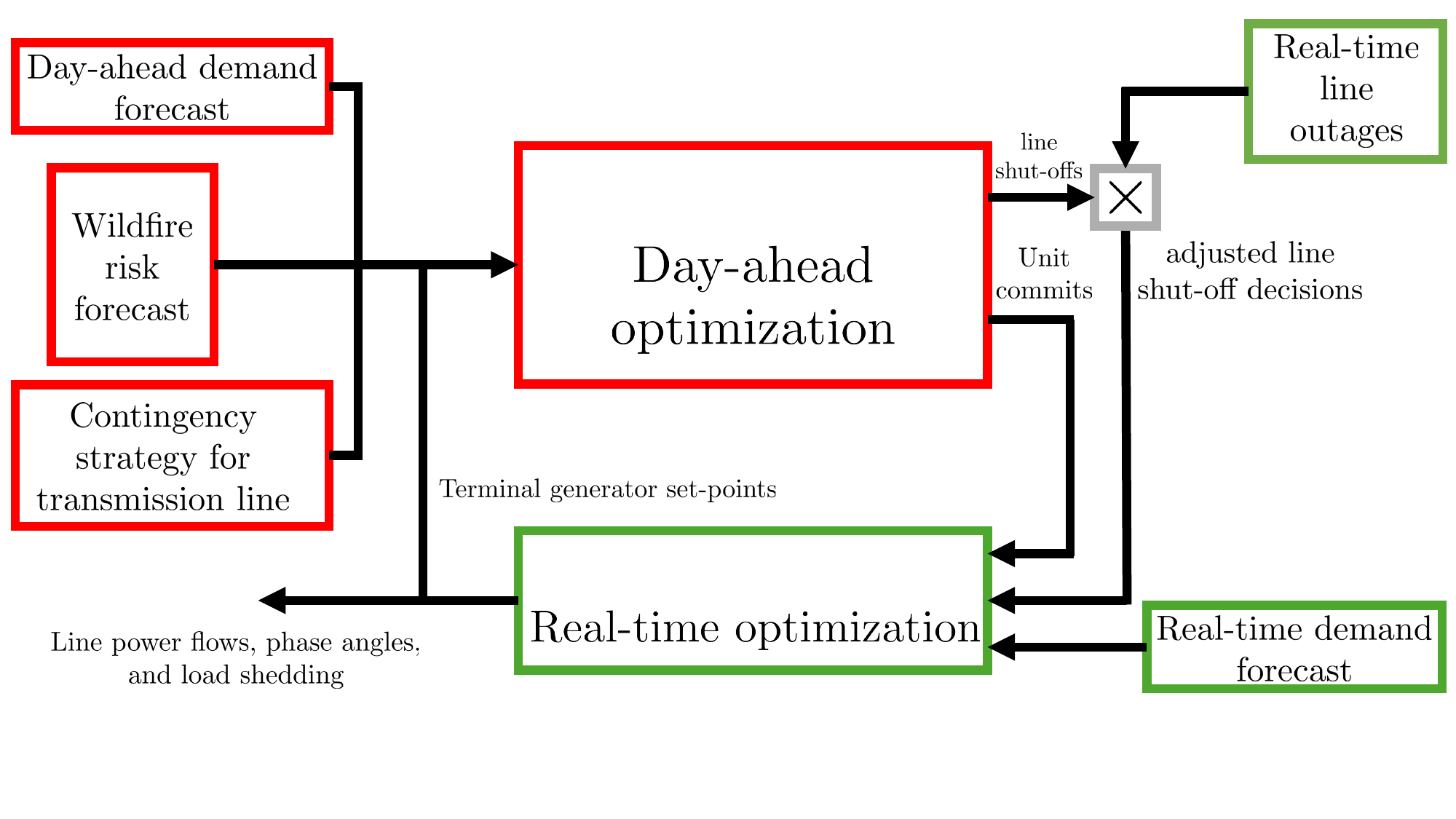}
\vspace{-4.5em}
\caption{Summary of the two stages of the DA stochastic PSPS (S-PSPS) formulation}
\label{fig:Block Diagrams for Wildfire Journal Paper 2}
%\vspace{-1.5em}
% Folder: C:\Users\rgree\OneDrive - UC San Diego\Wildfire Risk Impact on PSPS Cost Savings Data
% File: Block Diagrams for Wildfire Journal Paper 2.pdf
\end{figure}

A flowchart illustrating the methodology and workflow of the two-stage stochastic program is given in Fig. \ref{fig:Block Diagrams for Wildfire Journal Paper 2}.

\subsection{Day-ahead Stochastic PSPS formulation}
\label{subsection: Deterministic Formulation}
\begin{figure}[ht]
\centering
\vspace{-0.6em}
\includegraphics[width=\columnwidth]{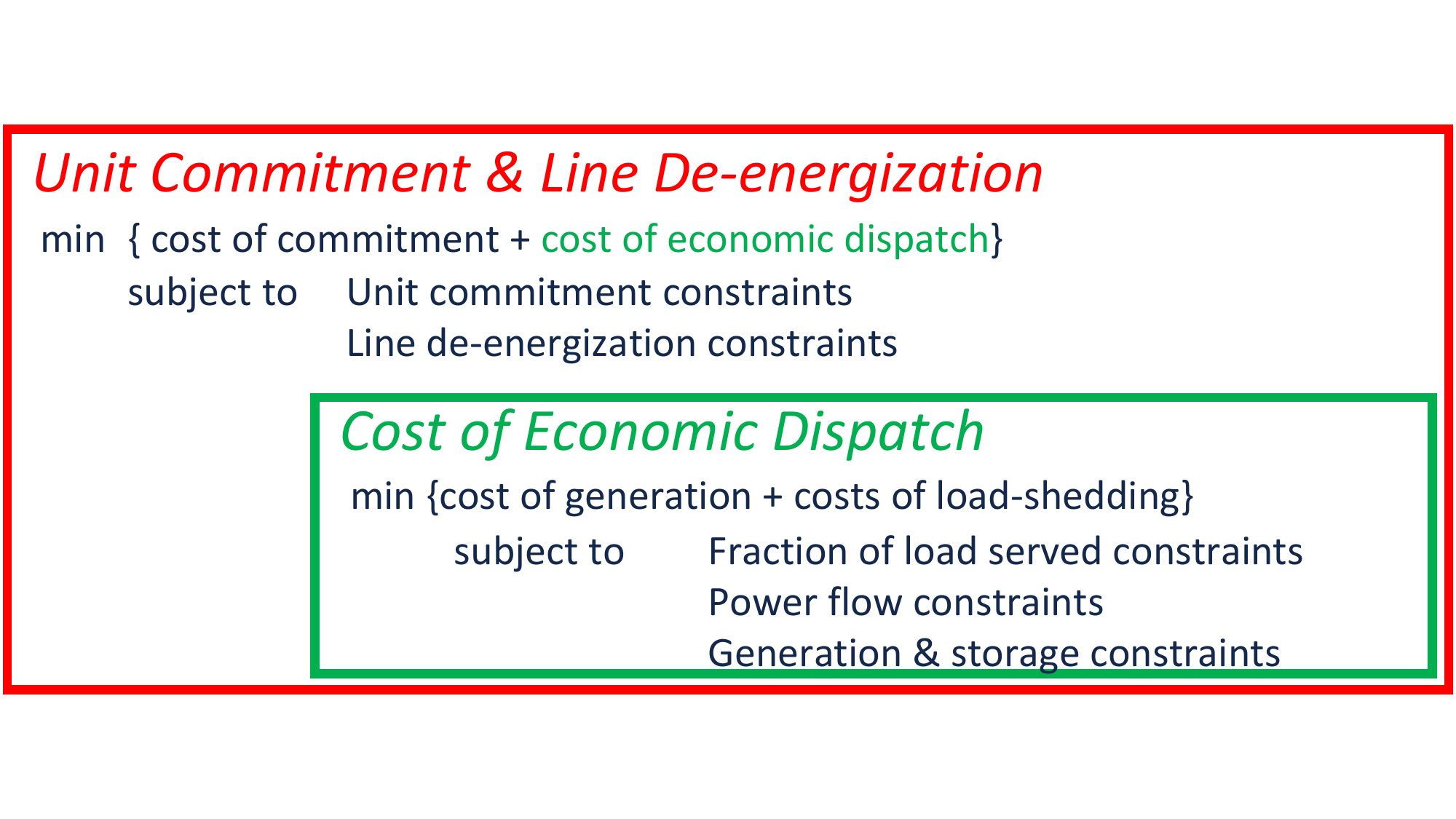}
\vspace{-4.5em}
\caption{Summary of the two stages of the DA stochastic PSPS (S-PSPS) formulation}
\label{fig:Two_Stage_Formulation_WRMIPSPSCS}
%\vspace{-1.5em}
% Folder: Ryan - UC San Diego
% File: Two_Stage_Formulation_WRMIPSPSCS.pptx
\end{figure}

We consider the following two-stage DA stochastic PSPS (S-PSPS) formulation with uncertainty in line damages due to wildfire ($\xi$) and demand ($\gamma$),
\begin{align}
&\min_{\bm{\mathcal{X}}, \bm{\mathcal{Y}}} \qquad f^{\tx{uc}}(\bm{z}_g^{\tx{up}}, \bm{z}_g^{\tx{dn}}) + \mathbb{E}_{\xi,\gamma}\left[\Phi(\tilde{\bm{p}}_g, \tilde{\bm{p}}_d, \tilde{\bm{x}}_d)\right], \\
&\text { s.t.} \quad \text{Line contingency constraint \eqref{eq:logWLFP_Nmk}}, \nonumber\\
&z_{ij,t} \geq z_{ij,t+1}, \label{eq:Damaged}\\
&z^{\tx{dn}}_{ij,t+1}=z_{ij,t}-z_{ij,t+1}, \label{eq:zdn}\\
&z_{g,t} \geq \sum_{t^{\prime} \geq t-t^{\text{MinUp}}_{g}}^{t} z^{\text{up}}_{g, t^{\prime}}, \label{eq:UC1} \\
&1-z_{g,t} \geq\sum_{t^{\prime} \geq t-t^{\text{MinDn}}_{g}} z^{\text{dn}}_{g, t^{\prime}}, \label{eq:UC2}\\
&z_{g, t+1}-z_{g, t}= z^{\text{up}}_{g, t+1}-z^{\text{dn}}_{g, t+1}, \label{eq:UC3}\\
&z_{g,t} \underline{p}_{g} \le \tilde{p}_{g,t} \le z_{g,t} \overline{p}_{g}, \label{eq:Pg}\\
&\tilde{p}^{\text{aux}}_{g,t}=\tilde{p}_{g,t}-\overline{p}_{g} z_{g, t}, \label{eq:RampAUX} \\
&\underline{U}_{g} \leq \tilde{p}^{\text{aux}}_{g, t}-\tilde{p}^{\text{aux}}_{g,t-1} \leq \overline{U}_{g,} \label{eq:RampNEW}\\
&\tilde{p}_{ij, t} \leq -B_{ij}\left(\tilde{\theta}_{i,t}-\tilde{\theta}_{j,t}+\overline{\theta}\left(1-z_{ij,t}\right)\right), \label{eq:MaxPowerFlow} \\
&\tilde{p}_{ij,t} \geq -B_{ij}(\tilde{\theta}_{i,t}-\tilde{\theta}_{j,t}+\underline{\theta}\left(1-z_{ij,t}\right)), \label{eq:MinPowerFlow}\\
&\underline{p}_{ij} \, z_{ij,t} \leq \tilde{p}_{ij,t} \leq \overline{p}_{ij}\, z_{ij,t}, \label{eq:ThermalLimit}\\
&\sum_{g \in \mathcal{G}_{i}} \tilde{p}_{g,t}+\sum_{(i,j) \in \mathcal{L}_i} \tilde{p}_{ij, t} = \sum_{d \in \mathcal{D}_{i}} \tilde{x}_{d,t} \tilde{p}_{d,t},\label{eq:PowerBalance}\\
&\forall t \in \mathcal{H}, i\in\mathcal{B}, (i,j) \in \mathcal{L}, \nonumber 
\end{align}
where the second stage variables dependent upon the demand ($\gamma$) and line damage ($\xi$) uncertainties are written with tilde,
\begin{align*}
    \bm{\mathcal{X}} &= \{\bm{z}_{g}, \bm{z}_{g}^{\tx{up}}, \bm{z}_{g}^{\tx{dn}}, \bm{z}_{ij}, \bm{z}_{ij}^{\tx{dn}}\}, \\
    \bm{\mathcal{Y}} &= \{\tilde{\bm{p}}_{g}, \tilde{\bm{p}}^{\tx{aux}}_{g}, \tilde{\bm{p}}_{ij}, \tilde{\bm{\theta}}_{i}, \tilde{\bm{p}}_{d}, \tilde{\bm{x}}_{d}\}.
\end{align*}
An overview of the two-stage stochastic program is summarized in Fig. \ref{fig:Two_Stage_Formulation_WRMIPSPSCS}.

Let $p_{g} = (p_{g,1}, \dots, p_{g,H})^\top$ be the vector of $p_{g,t}$ for generator $g$, then $\bm{p}_{\bm{g}}$ for all generators are denoted as $\bm{p}_{\bm{g}} = (p_{1}, \dots, p_{G})^\top$. The variables $\bm{p}_{\bm{d}}$, $\bm{x}_{\bm{d}}$, $\bm{z_g}^{\text{up}}$,  $\bm{z_g}^{\text{dn}}$, and $\bm{z_{ij}}$ are defined in a similar manner.
In~\eqref{eq:Damaged}, it is assumed that any damaged line remains off for the remainder of the optimization period. ~\eqref{eq:zdn} assumes that when line $(i,j)$ is damaged at time $t+1$,($z^{\text{dn}}_{ij,t+1}=1$), the line status switches from on ($z^{\text{dn}}_{ij,t}=1$) to off ($z^{\text{dn}}_{ij,t+1}=0$).
The unit commitment constraints \eqref{eq:UC1}-\eqref{eq:UC2} are used to enforce minimum up time ($t^{\text{MinUp}}_{g}$) and down time ($t^{\text{MinDn}}_{g}$) of generators given a generator $g$'s on-status ($z_{g,t}$). Similarly, the constraint~\eqref{eq:UC3} guarantees consistency between the binary variables $z^{\text{up}}_{g,t}$ and $z^{\text{dn}}_{g,t}$. In all simulations, it is assumed that all generators are initially off, $z_{g,0}=0, \:\forall g\in\mathcal{G}$.
The uncertain power of generator $g$ ($\tilde{p}_{g,t}$) is limited between its minimum ($\underline{p}_{g}$) and maximum ($\overline{p}_{g}$) power generation limits in~\eqref{eq:Pg}.
Similarly,~\eqref{eq:RampAUX}-\eqref{eq:RampNEW} implement the generator ramp rate limit between minimum ($\underline{U}_{g}$) and maximum ($\overline{U}_{g}$) limits. The auxiliary variable {$\tilde{p}^{\text{aux}}_{g, t}$} is introduced to prevent ramp violations during the startup process.
To model the uncertain power flow through the transmission lines for all $t\in \mathcal{H}$ $(i,j) \in \mathcal{L}$, the DC-OPF approximation is used in~\eqref{eq:MaxPowerFlow}-\eqref{eq:ThermalLimit} with minimum ($\underline{p}_{ij,t}$) and maximum ($\overline{p}_{ij,t}$) line thermal limits. Constraints~\eqref{eq:MaxPowerFlow} and~\eqref{eq:MinPowerFlow} limit power flow $\tilde{p}_{ij,t}$ between maximum and minimum limits, where $B_{ij}$ is the susceptance of the line. Since we measure power injections ($\tilde{p}_{g,t}$) in MWh, demand ($\tilde{p}_{d,t}$) in MWh, and power flows ($\tilde{p}_{ij,t}$) in MWh and phase angles ($\theta_{ij,t}$) in radians, the susceptance in Siemens must be multiplied by the transmission grid's base MVA value (100 MVA in the RTS and RTS-GMLC systems).
When $z_{ij,t} = 1$,~\eqref{eq:MaxPowerFlow} and \eqref{eq:MinPowerFlow} hold with equality. When $z_{ij,t} = 0$, the power flow across transmission line $(i,j)$ at time $t$ is zero, and the uncertain phase angle $\tilde{\theta}_{i,t}$ is limited between its maximum ($\overline{\theta}$) and minimum ($\underline{\theta}$) limits. Constraint~\eqref{eq:ThermalLimit} enforces the thermal limits, $\underline{p}_{ij,t}$ and $\overline{p}_{ij,t}$, on power flow, $\tilde{p}_{ij,t}$. The decision variable, $z_{ij,t}$, is used in constraint \eqref{eq:ThermalLimit} to ensure that if $\tilde{p}_{ij,t}$ = 0, the line from bus $i$ to $j$ is de-energized.
Finally, the uncertain bus power balance at each $t\in \mathcal{H}$ and every bus $i \in \mathcal{B}$ is given by the equality constraint~\eqref{eq:PowerBalance}. In~\eqref{eq:PowerBalance}, operators can optimize the fraction of the uncertain load served via $\tilde{x}_{d,t} \in [0,1]$ at each demand $d$ and time step $t$; the full load served is measured by $\tilde{p}_{d,t}$.

The objective function consists of the first stage unit commitment costs ($f^{\text{uc}}$), and second stage costs ($\Phi$) that further consist of operational costs ($f^{\text{oc}}$), and value of lost load cost ($f^{\text{VoLL}}$) defined as,
\begin{align}
    &\Phi[(\tilde{\bm{p}}_g, \tilde{\bm{p}}_d, \tilde{\bm{x}}_d)] = f^{\text{oc}}\!(\tilde{\bm{p}}_{g}) + f^{\text{VoLL}}(\tilde{\bm{x}}_{d}, \tilde{\bm{p}}_{d})\label{obj:PSPS},\\
    &{f^{\text{uc}}}(\bm{z^{\text{up}}_{g}},\bm{z^{\text{dn}}_{g}}) = \mathop \sum \limits_{t \in \mathcal{H}} (\mathop \sum \limits_{g \in \mathcal{G}} {C_g^{\text{up}}{z^{\text{up}}_{g,t}} + C_g^{\text{dn}}{z^{\text{dn}}_{g,t}}} ), \label{eq:STOUC}\\ 
    &f^{\text{oc}}(\tilde{\bm{p}}_{g}) = \mathop \sum \limits_{t \in \mathcal{H}} (\sum \limits_{g \in \mathcal{G}} {C_g \tilde{p}_{g,t}} ), \label{eq:STOOC} \\
    &{f^{\text{VoLL}}}(\tilde{\bm{x}}_{d}, \tilde{\bm{p}}_{d}) = \mathop \sum \limits_{t \in \mathcal{H}} (\mathop \sum \limits_{d \in \mathcal{{D}}} C_{d}^{\text{VoLL}}\left(1-\tilde{x}_{d,t}\right)\tilde{p}_{d,t}). \label{eq:STOVoLL}
\end{align}
In~\eqref{obj:PSPS}-\eqref{eq:STOOC}, $C_g^{\text{up}}$, $C_g^{\text{dn}}$, and $C_g$ are the generator start-up cost, shut-down cost, and marginal cost, respectively. In~\eqref{eq:STOVoLL}, $C_{d}^{\text{VoLL}}$ is Value of Lost Load cost for each demand. Variable costs $C_g$ and $C_{d}^{\text{VoLL}}$ are measured in \$/MWh while commitment costs $C_g^{\text{up}}$ and $C_g^{\text{dn}}$ in \$. The proposed model for the line contingencies is discussed next.

\subsection{Proposed WIP-based probabilistic model}

\label{subsection:Constraint for Enacting No Less Than k Damaged Lines with WLFP}
We first present the model developed in the author's earlier work~\cite{Greenough} that directly used the deterministic flammability risk score WFPI. Let $R_{ij,t}$ be the WFPI score of line $(i,j)$, then the following constraint is used to obtain a line shutoff strategy,
\begin{align} 
\sum_{\left(i,j \right) \in \mathcal{L}} \hat{z}_{ij,t}R_{ij,t} \leq R_{\text{tol}},\quad \forall t \in \mathcal{H}, 
 \label{eq:WFPI_Nmk}
\end{align}
where $R_{\text{tol}}$ is a parameter that limits the total system wildfire risk within an operator-selected tolerance level. $R_{\text{tol}}$ guarantees a certain level of security for the system while also penalizing the operation of lines within regions of higher WFPI. However,~\eqref{eq:WFPI_Nmk} only considers risk weighted line shutoff decisions ($\hat{z}_{ij,t}$) and does not explicitly consider wildfire ignitions leading to line damages ($\xi$). In \cite{Greenough}, the wildfire risk was modeled deterministically, where an NZR line is either de-energized ($\hat{z}_{ij,t}=0$) or assumed to be active in the near term and then fails and ignites a wildfire after the optimization horizon ($\hat{z}_{ij,t}=1$). This conservative assumption reduces the problem to a single WF scenario where all active NZR lines fail. In contrast, the stochastic formulation captures multiple outcomes (e.g., line survival or failure), providing a better balance between wildfire mitigation and reliability. Although deterministic models are more tractable, they overestimate risk. Stochastic PSPS policies achieve significant cost savings \cite{Greenough3}. For tractability reasons, we avoid modeling wildfire damages as an endogenous uncertainty, as done in \cite{Greenough3}, and instead model wildfire damages as exogenous uncertainties and line statuses, $z_{ij,t}$, as uncontrollable. In this paper, we develop a probabilistic formulation that explicitly includes line damage scenarios. 

Let $\tilde{\xi}_{ij,t} \in \{0,1\}$ be the event that a transmission line experiences damage from a nearby wildfire. $\tilde{\xi}_{ij,t}$ is a Bernoulli random variable with probability $\mathbbm{P}_{\xi}[\tilde{\xi}_{ij,t} = 1] = \tilde{\pi}_{ij,t}$, where $\tilde{\pi}_{ij,t}$ is either equal to WIP-WFPI or WIP-WLFP. We can construct a binary vector of line damages $\tilde{\bm{\xi}}_{ij} = (\tilde{\xi}_{ij,t})_{(i,j) \in \mathcal{L}}$ of dimension $L$ along with its probability distribution $\mathbbm{P}_{\xi}[\bm{\xi}_{ij,t} = \bm{\xi}]$. There are a total of $2^L$ combinations of line damage scenarios obtained as permutations of the entries in $\tilde{\bm{\xi}}_{ij,t}$ that are collected in the set $\Psi$. 

For any $\psi \in \Psi$, define $\mathcal{K}_{\psi}$ as the set of all line indices $(i,j)$ that are damaged due to wildfire, that is, $\tilde{\xi}_{ij,t} = 1$, and $\mathcal{K}^{\tx{c}}_{\psi} = \mathcal{L} \setminus \mathcal{K}_{\psi}$ as its complement, then we can obtain the joint probability of the event $\psi$ as,
\begin{align} 
\mathbb{P}_{\xi} \left[\psi\right]= \mathop \prod \limits_{(i,j) \in \mathcal{K}_{\psi}}  \tilde{\pi}_{ij,t} \mathop \prod \limits_{(i,j) \in \mathcal{K}^{\tx{c}}_{\psi}}(1 - \tilde{\pi}_{ij,t} ).\label{eq:Nmklinefailures}
\end{align}
Compared to~\eqref{eq:WFPI_Nmk} which only aims to minimize the total system wildfire risk by avoiding high flammability regions, the proposed formulation~\eqref{eq:Nmklinefailures} using WIP explicitly considers the likelihood of all possible combinations $2^{L}$ of line damages. For more info on the justification of the modeling assumption of independent wildfire-induced line failures, see \ref{section:Independence of Wildfire Events}. Historical wildfire perimeter overlays and event timelines (Figs.~\ref{fig:histWF24B}–\ref{fig:WFgantt24B}) show that most wildfire events affect a single transmission element, with multi-line impacts being rare, spatially localized, and short-lived. 

The impact of line damaged by wildfire is incorporated in~\eqref{eq:Nmklinefailures} by including $z_{ij,t}$ and $z_{ij,t}^{\textrm{dn}}$ as given below, 
\begin{align} 
\mathop \prod \limits_{(i,j) \in \mathcal{K}_{\psi}}  \left(\tilde{\pi}_{ij,t}\right)^{z^{\tx{dn}}_{ij,t}}\mathop \prod \limits_{(i,j) \in \mathcal{K}^{\tx{c}}_{\psi}}(1 - \tilde{\pi}_{ij,t})^{z_{ij,t}} \leq {{\pi _{\tx{tol}}}}, \label{eq:WLFP_Nmk}
\end{align}
where $\pi_{\tx{tol}}$ is an upper bound on the probabilities of line damage scenarios. For simplicity, the same tolerance level is used across all wildfire scenarios in $\Psi$. This constraint is non-linear, which is reformulated as a tractable, linear constraint by taking the $\log$ of both sides of the inequality and forming a cross-entropy (defined in~\eqref{eq:ACE}) based constraint:
\begin{align} 
\mathop \sum \limits_{\left({i,j} \right) \in \mathcal{K}_{\psi}} {z^{\tx{dn}}_{ij,t}}\log\left({{\tilde{\pi} _{ij,t}}} \right) + \mathop \sum \limits_{\left({i,j} \right) \in \mathcal{K^{\tx{c}}}_{\psi}} {z_{ij,t}} {\log}\left({1 - {\tilde{\pi} _{ij,t}}} \right) %\nonumber \\
\leq {\log}\left({{\pi _{\tx{tol}}}} \right),\quad \forall \psi \in \Psi, \, \forall t \in \mathcal{H}. \label{eq:logWLFP_Nmk}\end{align}

We measure cross-entropy because minimizing cross-entropy is equivalent to minimizing the KL-divergence between the WIP-based joint distribution defined by $\tilde{\pi}_{ij,t}$ and a fixed reference distribution in which all events have equal probability.

\begin{figure}[ht]
\centering
\vspace{-0.5em}
\includegraphics[width=\columnwidth]{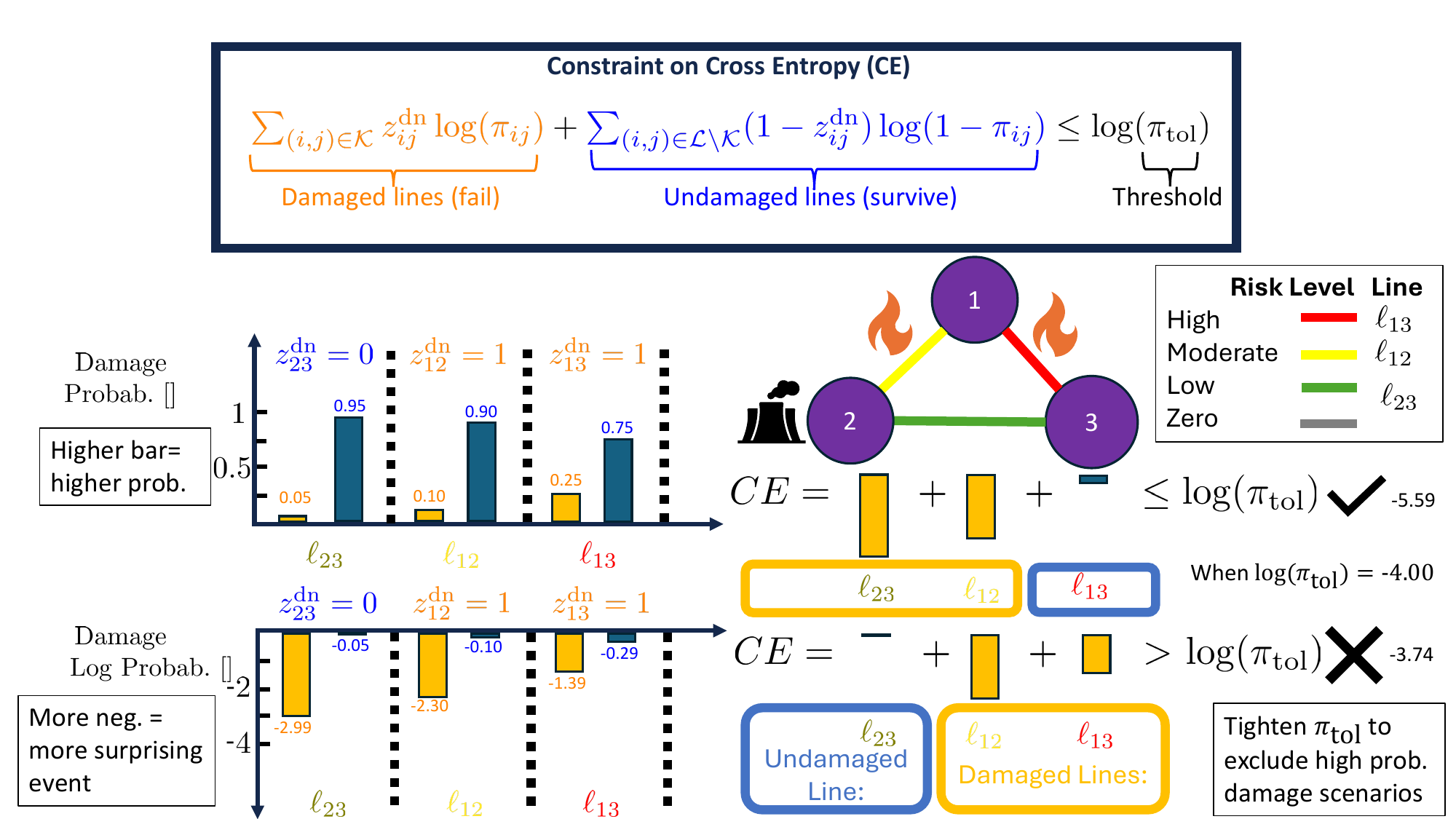}
\vspace{-1.5em}
\caption{Calculation of the cross entropy of a specific joint failure event of damages to $\ell_{12}$ and $\ell_{13}$ and survival of $\ell_{23}$ on a toy 3-bus system.The CE constraint filters out high-probability damage scenarios, while allowing a collection of low-probability or "surprising" events.}
\label{fig:Cross_Entropy_Constraint_WRMIPSPSCS}
%\vspace{-1.5em}
% Folder: Ryan - UC San Diego
% File: Cross_Entropy_Constraint_WRMIPSPSCS.pptx
\end{figure}

Including the constraint~\eqref{eq:logWLFP_Nmk} in the S-PSPS problem enables selecting only sufficiently low probability damage events. 
%selecting a line shutoff strategy $z_{ij,t}$, that avoids high wildfire probability scenarios by shutting off those lines first that are more at risk of damage due to wildfire due to the $\pi_{\tx{tol}}$ bound. 
For example, the 3-bus example in Fig. \ref{fig:Cross_Entropy_Constraint_WRMIPSPSCS} illustrates this behavior, in which lines $\ell_{23}$, $\ell_{12}$, and $\ell_{13}$ are low, moderate, and high-risk lines. As the WIP approaches zero, the $\log(1-\pi_{ij})\approx 0$. Active lines ($z_{ij}=1$ or $z^{\tx{dn}}_{ij}=0$ ) with low WIP, such as $\ell_{12}$, contribute little to the cumulative log(WIP); however, damage to $\ell_{12}$ contributes a large negative number (due to its low WIP). Large negative numbers signal that the damage event ($z^{\tx{dn}}_{12}=0$) is rare. In Fig. \ref{fig:Cross_Entropy_Constraint_WRMIPSPSCS}, constraint \eqref{eq:logWLFP_Nmk} is used to filter out highly probable damage events, such as $\{z^{\tx{dn}}_{23},z^{\tx{dn}}_{12},z^{\tx{dn}}_{13}\}=\{0,1,1\}$, but maintain $\{z^{\tx{dn}}_{23},z^{\tx{dn}}_{12},z^{\tx{dn}}_{13}\}=\{1,1,0\}$ .

To minimize operating and VoLL costs, the optimization problem tends to maximize the number of active lines at each time step, $\sum_{(i,j) \in\mathcal{L}}z_{ij,t}$. Furthermore, the optimization no longer merely avoids regions of high flammability, as in~\eqref{eq:WFPI_Nmk}. The bound $\pi_{\tx{tol}}$ forces a filtering out of operating conditions associated with relatively likely multi-line wildfire damage scenarios, while also minimizing total generator commitments costs, operating costs, and VoLL costs. 

Since the line shutoff decisions are embedded directly into constraint ~\eqref{eq:logWLFP_Nmk}, we end up with a reduced set of line damage scenarios. This set of scenarios is implicitly trimmed due to infeasibility within the log-probability bound. The resulting set can be significantly smaller than $2^L$, making it easier to solve the S-PSPS problem. Finally, the conversion of WIP to log(WIP) avoids the numerical problems of generating joint probabilities from products of rare ($10^{-6}$) marginal probabilities or WIPs. 

\subsubsection{Considering Interdependent Failures in the WLFP based probabilistic model for Future Work}
\label{subsubsection:Considering Interdependent Failures in the WLFP based probabilistic model for Future Work}

The S-PSPS uses a joint probability model from equation \eqref{eq:WLFP_Nmk} based on the independence of wildfire events to maintain the tractability of the S-PSPS. However, wildfire damage probabilities may be due to local fires and fire spread from neighboring lines. In \eqref{piijbar}, the resulting spread-adjusted probabilities preserve the tractable linearity of the cross-entropy reformulation while indirectly accounting for interdependent wildfire exposure among neighboring transmission lines \eqref{eq:logWLFP_Nmk_spread}. The derivation for the cross-entropy constraint can be seen in \ref{section:Conditional Cross-entropy-based Constraint to Capture Interdependent Failures for Future Work}. The adjusted conditional cross-entropy constraint for joint damage events is 

\begin{align}
\bar{\pi} _{ij,t}=1-(1-\tilde{\pi}_{ij})(\Pi_{n\in N_{ij}}(1-\pi_{n \rightarrow ij})) \label{piijbar} \\
\mathop \sum \limits_{\left({i,j} \right) \in \mathcal{K}_{\psi}} {z^{\tx{dn}}_{ij,t}}\log\left({{\bar{\pi} _{ij,t}}} \right) + \mathop \sum \limits_{\left({i,j} \right) \in \mathcal{K^{\tx{c}}}_{\psi}} {z_{ij,t}} {\log}\left({1 - {\bar{\pi} _{ij,t}}} \right) %\nonumber \\
\leq {\log}\left({{\pi _{\tx{tol}}}} \right),\quad \forall \psi \in \Psi, \, \forall t \in \mathcal{H}. \label{eq:logWLFP_Nmk_spread} 
\end{align}
$\pi_{n \rightarrow ij}$ are the spread probabilities that a fire is started at a line $(i,j)$ due to an ignition started at its neighboring line $n \in \mathcal{N}_{ij}$. This probability can be roughly derived from the USGS WFPI-based Fire Spread Probabilities (WFSP) (the probability a fire spreads at least 500 acres) or from a higher resolution fire spread model. In general, our modular PSPS model is agnostic to the source of WIP and choice of wildfire spread dynamics simulators. Integration of detailed fire spread simulators (e.g., FARSITE \cite{Finney}, FLAMMAP \cite{Finney2}) into $\tilde{\pi}_{ij}$ and $\pi_{n\rightarrow ij}$ is encouraged. Parameters for the fire spread dynamics models could also be generated offline via standard fire spread heat rate equations used in \cite{Trakas2}.

\subsubsection{Limitations of the Current Wildfire Forecasting Model Left for Future Work} \label{subsubsection:Limitations of the Current Wildfire Forecasting Model Left for Future Work}

It is crucial to develop accurate fire ignition and spread models that reflect the unique dynamics of power infrastructure-driven fires. These models may differ significantly from models of exogenous fires caused by lightning or human activity. Publicly available wildfire risk indices may not provide sufficient context to predict these events. Therefore, collaboration with utilities is essential, as they can supply historical ignition locations and insight into equipment failures. With failure data, academic researchers can refine S-PSPS optimization models to assess past PSPS decisions and explore strategies for improvement or prevention. For example, we could use properties, such as line conductor type (bare vs insulated) and vegetation clearance, and equipment aging, to develop a joint probability model including WIP, conductor type, vegetation clearance, and equipment aging. 

Additionally, this work assumes PSPS de-energizations fully protect power system infrastructure from both exogenous and endogenous fires. However, this may not hold in practice. This work also assumes wildfire impacts are uniformly distributed across the fire’s evolution and within the fire's boundary. However, in reality, variations in critical infrastructure and wildfire damage exist within that same region.

\subsection{Out-of-Sample Test via Real-Time Outage Simulation}
\label{subsubsection:Second Stage Formulation}
Solving the two-stage DA S-PSPS problem provides the first stage commitment and shutoff decisions ($\bm{\mathcal{X}}$). In RT, the system operator can use an economic dispatch problem to adjust second stage variables ($\bm{\mathcal{Y}}$), including generator output and demand served, based on the realized line damages. To evaluate the robustness of the first stage decisions, the economic dispatch problem is validated against realized samples of demand and line damages. Since there are many lines with zero or negligible WFPI/WLFP values, we generate line damage scenarios from the subset of $L^\prime < L$ most risky lines. This helps to reduce less likely and redundant line damage scenarios and focus on more critical line damage scenarios. 

Let $\Psi^{\prime} \subset \Psi$ be the reduced set of line damage scenarios, then the corresponding probability distribution can be constructed using~\eqref{eq:Nmklinefailures}. In Section~\ref{section:Results} we construct 1,024 line failure scenarios by sampling from $\Psi^{\prime}$ with $L^{\prime}=10$ according to the probability distribution constructed using~\eqref{eq:Nmklinefailures}. The modeling assumption ($L'=10$) is discussed further with Figures
\ref{fig:WFPIvWLFPLoadLossComparisonRTSR1}-\ref{fig:WFPIvWLFPLoadLossComparisonRTSR1VaryL} and Table \ref{table:pp9_convergence} in \ref{section:Depth of Outage Analysis RTS-GMLC}. The RT optimization problem becomes,
\begin{align}
&\min_{\bm{\mathcal{Y}}_{\psi^{\prime}}} \: {\Phi }(\bm{p}_{g,\psi^{\prime}}, \bm{p}_{d,\psi^{\prime}}, \bm{x}_{d,\psi^{\prime}})  + f^{\tx{og}}(\bm{p}_{i,\psi^{\prime}}^{\tx{og}}) \label{eq:RTobj}
\\
&\text { s.t. } \quad \text{Generator Capacity Bounds with $\bm{z}_{g,\psi'}$: } \,\eqref{eq:Pg} \nonumber \\
&\text{Generator Ramping Constraints with $\bm{z}_{g,\psi^{\prime}}$: } 
 \, \eqref{eq:RampAUX}-\eqref{eq:RampNEW} \nonumber\\
&\text{Optimal Power Flow with $\bm{z}_{\bm{ij}} \wedge \bm{\xi}_{\bm{ij},\psi^{\prime}}$:} \, \eqref{eq:MaxPowerFlow}-\eqref{eq:ThermalLimit} \nonumber\\
&\text{Real-time Demand Balance Constraints with $\bm{p}_{\bm{d},\psi^{\prime}}$: }\, \eqref{eq:PowerBalance} \nonumber \\
& \hspace{0.5 cm} \:\; i \in \mathcal{B} ,\: \forall t \in \mathcal{H},\: \forall \psi^{\prime} \in \Psi^{\prime}, \nonumber
\end{align}
where the subscript $\psi^{\prime}$ is added to distinguish between the RT variables from the first stage DA variables. 

The over-generation term $f^{\tx{og}}$ defined below with variable $p^{\tx{og}}_{i,\psi^{\prime}}$ is included to avoid generator operational infeasibility, which can happen when the fixed generator commitments from the first stage are exposed to certain outage permutations. The over-generation term is often zero. 
\begin{align}
    &f^{\text{og}}(\bm{p}^{\text{og}}_{\bm{i},\psi^{\prime}}) = \mathop \sum \limits_{t \in \mathcal{H}} (\sum \limits_{i \in \mathcal{B}} {C_{i}^{\text{og}} p^{\text{og}}_{i,t,\psi^{\prime}}} ). \label{eq:STOOG}
\end{align}

\section{Results}
\label{section:Results}

\subsection{Data and Test Case Description} \label{subsection:Data and Test Case Description}

The grid models used for analyses are the modified IEEE RTS 24 bus system and the full RTS-GMLC system, as shown in Fig.~\ref{fig:IEEERTScolored}. Simulations are performed over a full day, and decisions are made hourly (i.e., 24, 1-hour time steps). The optimization horizon of interest is one day. To determine representative scenarios for the S-PSPS optimization, total demand and cumulative bus level wildfire risk (WFPI or WLFP) scenarios are generated via a tree reduction algorithm from \cite{Gröwe-Kuska}. In the author's previous work, the scenarios were generated based on total demand only, while the DA forecast of WFPI values was obtained independently of demand. Conversely, this work generates scenarios from the joint probability distribution of total demand and WFPI/WLFP, which are then mapped to WIP to be used in constraint~\eqref{eq:logWLFP_Nmk}.

\subsubsection{Scenario generation from joint distribution of demand and WFPI/WLFP}
\label{subsubsection:Scenario generation from joint distribution of demand and WFPI/WLFP}
The tree scenario reduction method takes inputs of the four months (e.g., 121 days during peak wildfire season from June 12th to October 11th) of hourly total system load forecasts and of the cumulative bus WFPI/WLFP forecasts (across all buses) for the grid region. We create a joint distribution of total demand and cumulative bus WFPI/WLFP rather than total demand and cumulative line WFPI/WLFP, since line WFPI/WLFPs are an average risk over multiple climate zones, while bus WFPI/WLFPs represent more local wildfire risk measurements that can be related to the local demands at each bus. WFPI and WLFP are assumed to be constant throughout each day. The outputs of the scenario tree are the $M$ most likely 24-hour total demand curves and their respective daily cumulative WFPI/WLFPs across all buses. The $M$ total demand scenarios are proportionally mapped to each of the static loads given in the RTS 24 bus system or RTS-GMLC system to ensure consistency between the reduced scenarios and RTS network loads. A nearest neighbor (NN) search for each scenario is used to find which historical day's cumulative bus WFPI/WLFP is closest to each respective scenario's cumulative bus WFPI/WLFP. The NN search ensures that the spatial distribution of wildfire risk across buses is realistic and consistent with historical observations. The transmission line WFPI/WLFP values of those $M$ historical days from the NN search are the transmission line WFPI/WLFP values for each respective scenario. WFPI/WLFP values are finally converted to actionable WIP for S-PSPS using the method described in~Section~\ref{subsection:Qualitative comparison}. The resulting individual bus demand ($\bm{p}_{d,\omega}$) and transmission line WIP-WFPI or WIP-WLFP denoted as $\bm{\pi}_{ij,\omega}$ for each scenario $\omega$ are collected in the set $\Omega$ for solving the S-PSPS problem, where $\pi_{\omega}$ is the scenario probability.

\begin{figure}[t]
\centering
\subfloat{\includegraphics[width=8.2cm]{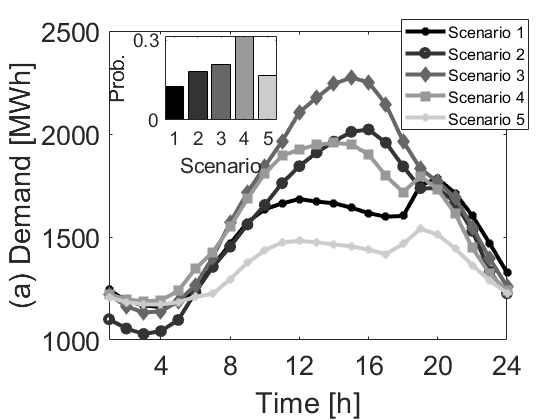}} \hfill
\subfloat{\includegraphics[width=8.2cm]{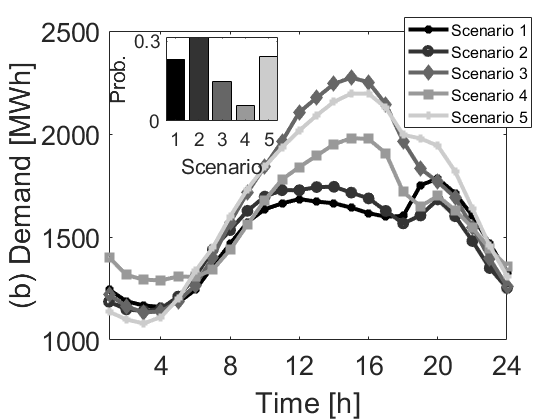}} \hfill
\subfloat{\includegraphics[width=8.2cm]{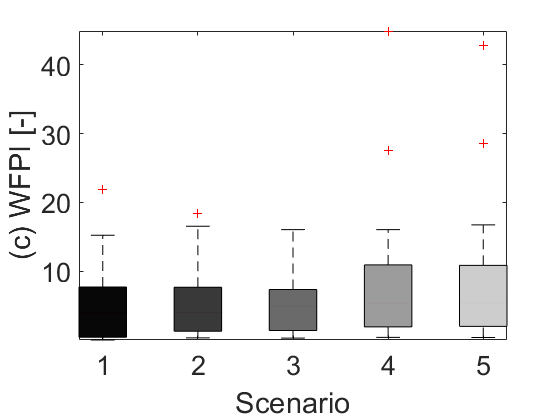}} \hfill
\subfloat{\includegraphics[width=8.2cm]{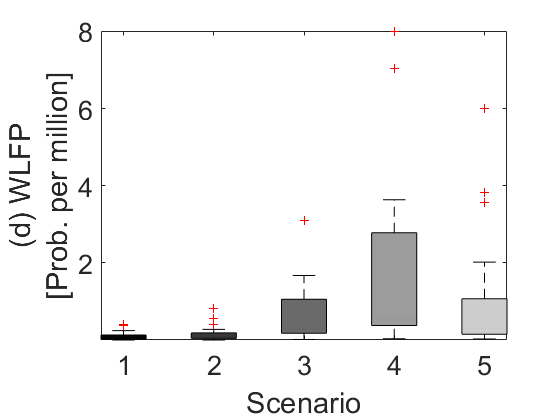}}
\caption{A sample of five scenarios (for illustration) for the RTS derived from the tree reduction based on either WFPI (a) or WLFP (b) and load data from 2020. Box plots of the line WFPI (c) and line WLFP (d) for each scenario are also shown.
}
\label{fig:Load_Profiles_IEEE24_WFPI_WLFP}
%\vspace{-1.5em}
% Folder: H:\My Drive\MAE 207 Grid Integration Project\Project Code\Figures for Journal Paper Spring 2024
% File: fig3Load_Profiles_IEEE24.m
% Lines: Run whole script
% Figure(1)-(4)
\end{figure}

As an example, Figs.~\ref{fig:Load_Profiles_IEEE24_WFPI_WLFP}(a) and \ref{fig:Load_Profiles_IEEE24_WFPI_WLFP}(b) provide the total demand timeseries for the 24~hour optimization horizon when $M=5$, along with the probabilities of each respective scenario~\cite{Gröwe-Kuska}. The reduced scenarios are intended to capture the joint relationship between electric grid demand and wildfire ignition probability throughout the year. In general, California electricity demand exhibits strong seasonal trends, with peak demand occurring during the summer months that also coincide with peak wildfire season. Elevated wildfire risk can also occur during lower-demand Fall and Winter periods associated with Santa Ana wind events, although these operating conditions occur less frequently.
The results further highlight an important distinction between WLFP- and WFPI-based wildfire risk representations. WLFP values exhibit stronger seasonal clustering with electricity demand, producing more distinct operating scenarios across both the demand and wildfire risk dimensions. In contrast, WFPI values tend to remain more uniform throughout much of the wildfire season, causing the clustering process to be driven more strongly by demand variations than wildfire risk variations. Additional clustering results and reduced scenario statistics are provided in ~\ref{section:Intuition for Scenario generation from joint distribution of demand and WFPI/WLFP}.
The transmission line WFPI distribution is relatively uniform across all scenarios compared to WLFP, similar to what was seen in Figs~\ref{fig:WFPIvWLFP2020HeatmapRTS}(c) and~\ref{fig:WFPIvWLFP2020HeatmapRTS}(d) respectively, which results in increased RT costs for WFPI as discussed in the next section.

\subsubsection{Selection of Appropriate Scenario Set Size}
The proper selection of the scenario set size is a trade-off between how similar the reduced scenario probability distribution is to the sample average approximation of demand given all historical daily scenarios and the solve time of the subsequent optimization \cite{Heitsch}. The sample average approximation assumes that each daily demand curve is given the same weight $\frac{1}{N}$ where $N$ is the number of scenarios.

\begin{figure}[t]
\centering

\begin{overpic}[width=\columnwidth, grid=false]{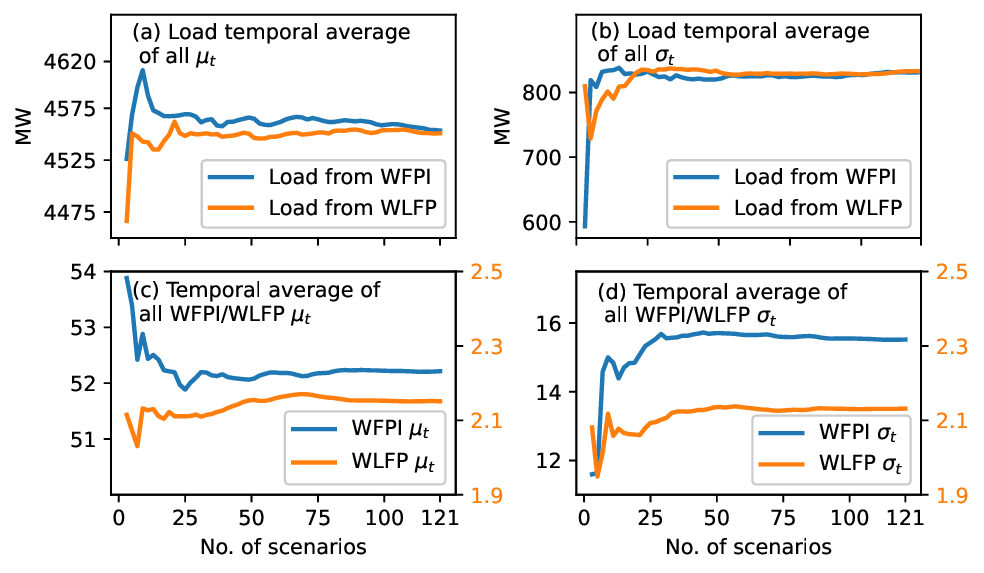}

\put(4,20){\rotatebox{90}{{\sffamily[-]}}}
\put(52,20){\rotatebox{90}{{\sffamily[-]}}}
\end{overpic}

\caption{Temporal averages of (a) expected demand and (b) demand standard deviation across $M$ scenarios. (c) Expected WFPI (blue) \& WLFP (orange). (d) Standard deviations of WFPI \& WLFP. Scenario sets use odd integers from 1 to 121.}

\label{fig:Average_MU_STD_of_Demand_Over_Time_for_Different_Scenarios_LOADWFWFPIWLFPcombined}
%\vspace{-1.5em}
% Folder: C:\Users\Ryan\OneDrive - UC San Diego\Wildfire Risk Impact on PSPS Cost Savings Data
% File: Scenario_Reduction_Validation_Demand_WF_combineplots.m
% Lines: Run full script
% Figure(1)

\end{figure}

In Fig. \ref{fig:Average_MU_STD_of_Demand_Over_Time_for_Different_Scenarios_LOADWFWFPIWLFPcombined}, we show that in the RTS system, as the number of scenarios increases, the fluctuations in the temporal averages of mean ($\mu_{t}$) and standard deviation ($\sigma_{t}$) decrease for the cases of both load and WFPI/WLFP. 

If the temporal average of the mean and standard deviation of a set with M reduced scenarios is $\hat{\mu}^M_t=\frac{1}{24}\sum_t\mu_t$ and $\hat{\sigma}^M_t=\frac{1}{24}\sum_t\sigma_t$, then the percentage error is calculated as $\frac{|\hat{\mu}^M_t-\hat{\mu}^{121}_t|}{\hat{\mu}^{121}_t}$ and $\frac{|\hat{\sigma}^M_t-\hat{\sigma}^{121}_t|}{\hat{\sigma}^{121}_t}$. After about $M=51$ samples, the temporal average of the mean and standard deviation converges to less than 0.5\% error for demand and wildfire risk compared to the sample average approximation among all 121 samples.

\subsection{Trends in Generator Commitment and Operational Costs}
\label{subsection:Trends in Commitment and Day-ahead Operational Costs}

The Value of Lost Load ($C_{d}^{\text{VoLL}}$) and the cost of over generation ($C_{i}^{\text{og}}$) are set to 5,000 \$/MWh for all demands and buses \cite{Trakas2,Mohagheghi,Farzin}. Since the value of losing 1~MWh is at least one order of magnitude higher than the cost to produce 1 MWh, the VoLL is typically the largest contributor to the total economic costs. 

Solving the full S-PSPS problem with $M=51$ scenarios and second stage variables with $2^L$ constraints~\eqref{eq:logWLFP_Nmk} for each scenario is computationally expensive. Simulating variation in both demand and WIP hides the extent to which optimal PSPS decisions are affected by wildfire risk selection. 

The sensitivity of S-PSPS to demand variability was explored in the author's earlier work~\cite{Greenough}. In this work, uncertainties in the demand and  WIP distribution are fixed to their expected demand and WIP values from scenario distributions that have stabilized in their sample mean and standard deviation. Using distributions with more than $M=51$ samples leads to a negligible impact on optimal PSPS decisions. These stabilized distributions for demand and WIP are used to isolate the impact of WIP on optimal shutoff decisions. The expected nodal demand and line WIP are calculated as, 
\begin{align*}
     \mathbbm{E}[\tilde{\bm{p}}_{d}] = \sum_{\omega \in \Omega} \pi_{\omega} \bm{p}_{d,\omega}, \quad \mathbbm{E}[\tilde{\bm{\pi}}_{ij}] = \sum_{\omega \in \Omega} \pi_{\omega}\bm{\pi}_{ij,\omega}.
\end{align*}

While the cross-entropy constraint \eqref{eq:logWLFP_Nmk} is deterministic in each line's WIP, the stochasticity is retained in the random realizations of line damages. 

Finally, to avoid any differences in realized costs or shutoff decisions stemming from the DA demand forecast, we assume that both the DA demand forecast and the realized RT demand forecast are the same demand profile from July 31st, 2020. However, the RT line outage scenarios are generated as mentioned in Section~\ref{subsubsection:Second Stage Formulation}. The effects of the joint uncertainty in demand and WIP on S-PSPS decisions are reserved for future work.
The resulting DA S-PSPS problem can be written as, 

\begin{align}
&\min_{\bm{\mathcal{X}},\bm{\mathcal{Y}}} \qquad f^{\tx{uc}}(\bm{z}_g^{\tx{up}}, \bm{z}_g^{\tx{dn}}) + \Phi(\tilde{\bm{p}}_g, \mathbbm{E}[\tilde{\bm{p}}_d], \tilde{\bm{x}}_d), \\
&\text { s.t.} \quad \tx{\eqref{eq:Damaged}-\eqref{eq:ThermalLimit}},  \nonumber\\
&\sum_{g \in \mathcal{G}_{i}} \tilde{p}_{g,t}+\sum_{(i,j) \in \mathcal{L}_i} \tilde{p}_{ij, t} = \sum_{d \in \mathcal{D}_{i}} \tilde{x}_{d,t} \mathbbm{E}[\tilde{p}_{d,t}],\label{eq:PowerBalance2}\\
&\mathop \sum \limits_{\left({i,j} \right) \in \mathcal{K}_{\psi}} {z^{\tx{dn}}_{ij,t}}\log\left({{\mathbbm{E}[\tilde{\pi}_{ij,t}]}} \right) + \mathop \sum \limits_{\left({i,j} \right) \in \mathcal{K^{\tx{c}}}_{\psi}} {z_{ij,t}} {\log}\left(1 - {{\mathbbm{E}[\tilde{\pi}_{ij,t}]}} \right) %\nonumber\\
%&\hspace{15em} 
\leq {\log}\left({{\pi _{\tx{tol}}}} \right),\\
&\forall t \in \mathcal{H}, i\in\mathcal{B}, (i,j) \in \mathcal{L}, \forall \psi \in \Psi. \nonumber 
\end{align}

\begin{figure}[t]
\centering
% \vspace{-1em}
\hspace{-0.8em}
\subfloat{\includegraphics[width=8.2cm]{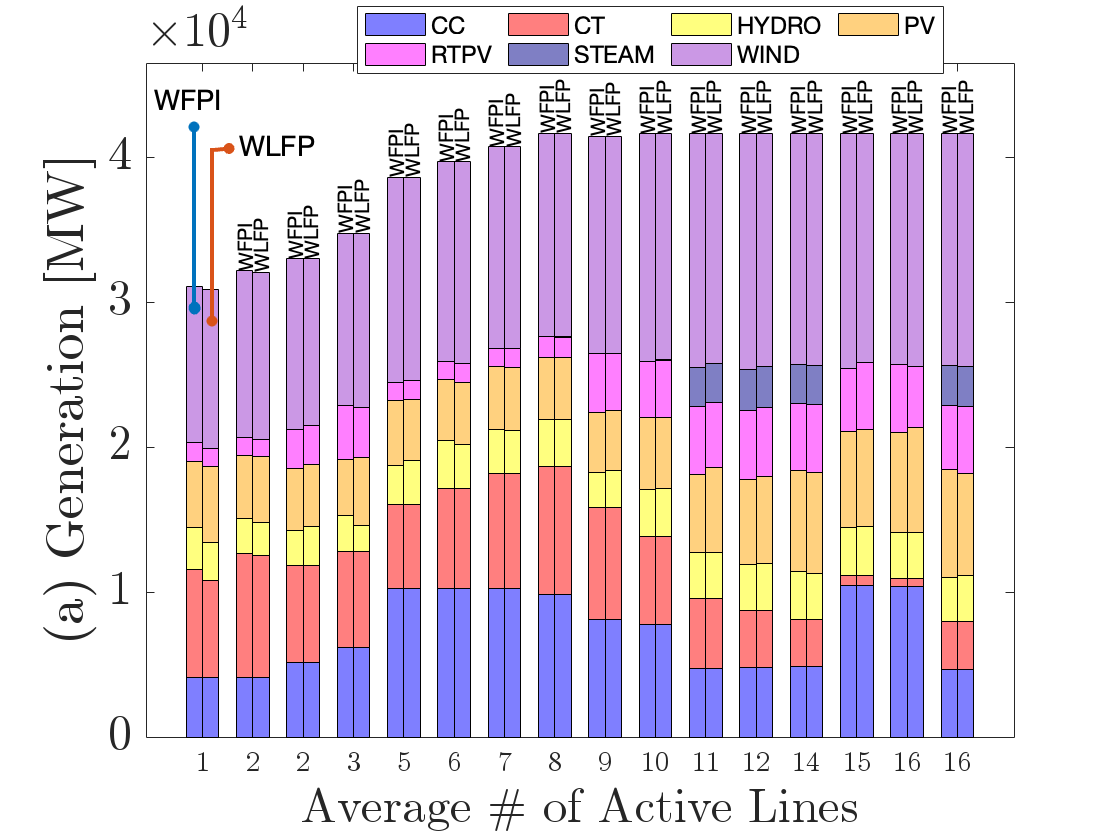}} \hfill
\hspace{-5em}
\subfloat{\includegraphics[width=8.2cm]{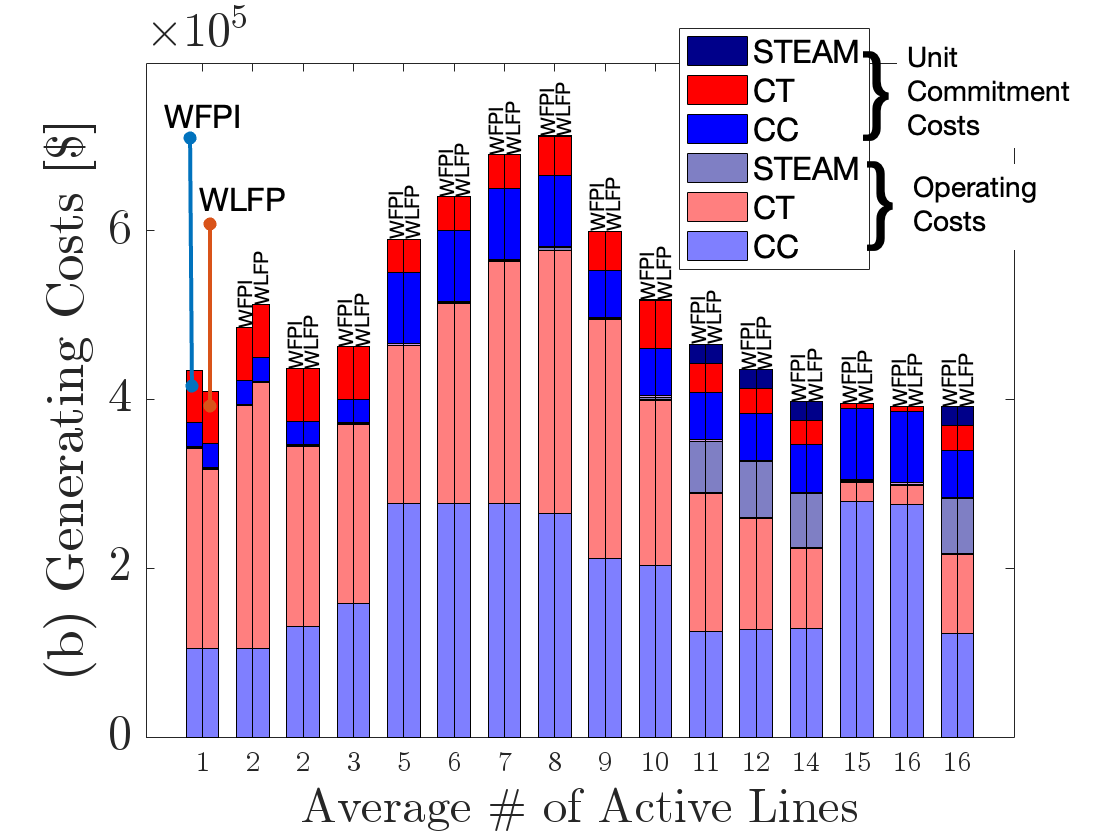}} \hfill
\caption{PSPS (a) generation and (b) total costs (excluding VoLL), optimized for the IEEE RTS 24-bus system.}
\label{fig:IEEE24WFPIvWLFPPgCompare_bygen_1st_DemandRT}
% Folder: H:\My Drive\MAE 207 Grid Integration Project\Project Code\Figures for Journal Paper Spring 2024 PostTSG
% File: PlotsCompareRTSWFPIvWLFP.m
% Set: Grid='24B';   
% Lines: Run whole script
% Figure(9),(10), (6), and (7)
% \vspace{-2em}
\end{figure}

Simulation results for the DA S-PSPS problem are shown in Fig.~\ref{fig:IEEE24WFPIvWLFPPgCompare_bygen_1st_DemandRT} for both WIP-WFPI and WIP-WLFP. In the IEEE RTS 24-bus system, there are 38 lines, and 11 of those lines have zero risk. Different values of $\pi_{\text{tol}}$ lead to a different set of active lines out of the remaining 23 non-zero risk (NZR) lines. Fig.~\ref{fig:IEEE24WFPIvWLFPPgCompare_bygen_1st_DemandRT} compares the total generation and costs in the cases of WIP-WFPI and WIP-WLFP for 1 to 23 NZR active lines obtained by solving the DA S-PSPS problem for consecutively increasing values of $\pi_{\tx{tol}}$. Total generation or demand served is similar in both cases, which increases monotonically until 10 NZR lines are active and is constant afterwards. The reason is that due to the network structure, only 10 active lines are needed to fulfill the maximum demand. But increasing active lines beyond 10 impacts the generation costs.  

Generation costs are not strictly increasing as shown in Fig.~\ref{fig:IEEE24WFPIvWLFPPgCompare_bygen_1st_DemandRT}(b). Costs are maximum for the case of 8 active lines and decrease with increasing number of active lines. This is due to the decrease in the proportion of demand met by the more expensive CC and CT units, and more of the production is supplied by renewable sources. Between WIP-WFPI and WIP-WLFP, the costs differ for the cases of 1 and 2 active lines and are similar for all other cases. While the generation and costs are similar, the combination of line shutoff decisions $z_{ij,t}$ differs between WIP-WFPI and WIP-WLFP, leading to higher RT costs and commitment of high-risk lines in the case of WIP-WFPI compared to WIP-WLFP, as discussed next.

\subsection{Comparing Line Shutoff Decisions}
\label{subsection:Comparing Line Shutoff from WIP-WFPI and WIP-WLFP models}
\begin{figure}
\centering

\begin{overpic} [width=\columnwidth]{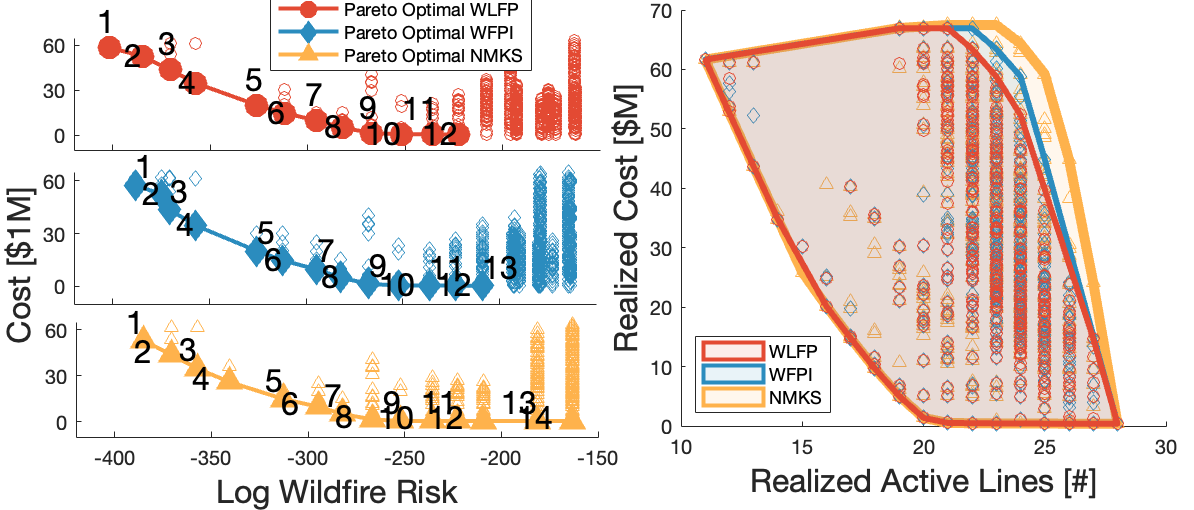}
    \put (10,32) {(a)}
    \put (10,20) {(b)}
    \put (10,8) {(c)}
    \put (60,20) {(d)}
\end{overpic}

% \vspace{-1em}
\caption{Optimized DA costs vs. DA cumulative wildfire risk under permutations of the ten riskiest lines for WLFP (a), WFPI (b), and NMKS (c), with Pareto-optimal strategies marked with lines. (d) Optimized realized (out-of-sample) costs vs. realized number of active lines. The lower boundary shows the Pareto front; the upper boundary shows the worst-case realized costs across active line settings.}
%\vspace{-2em}
\label{fig:WFPIWLFPNMKSDistributionsBoundaryRTS}
\end{figure}

Each of the line shutoff strategies from the DA S-PSPS problem (Fig.~\ref{fig:IEEE24WFPIvWLFPPgCompare_bygen_1st_DemandRT}) is validated against 1,024 RT line damage scenarios outlined in Section~\ref{subsubsection:Second Stage Formulation} and obtained by solving~\eqref{eq:RTobj}. Fig.~\ref{fig:WFPIWLFPNMKSDistributionsBoundaryRTS}(a)-(c) shows the total realized costs plotted against the total realized system wildfire log probability using WIP-WLFP, WIP-WFPI, and $N-k$ strategy (NMKS) from~\cite{Greenough}, respectively. The log wildfire probability is computed from the left-hand side of~\eqref{eq:logWLFP_Nmk} with $z_{ij,t}$ and realized line damages $\xi_{ij,t}$. It can be seen that the solutions marked 1-12 for WIP-WLFP, 1-13 for WIP-WFPI, and 1-14 for NMKS form a Pareto-optimal front that results in the lowest cost and wildfire probability. 

\begin{figure*}[htbp]
\centering
\subfloat[]{\includegraphics[width=6.0cm]{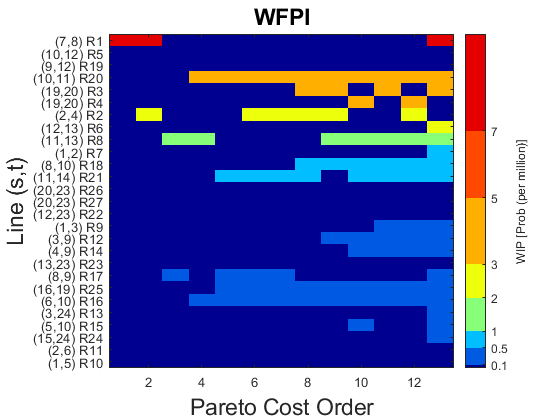}} \hfill
\hspace{-3em}
\subfloat[]{\includegraphics[width=6.0cm]{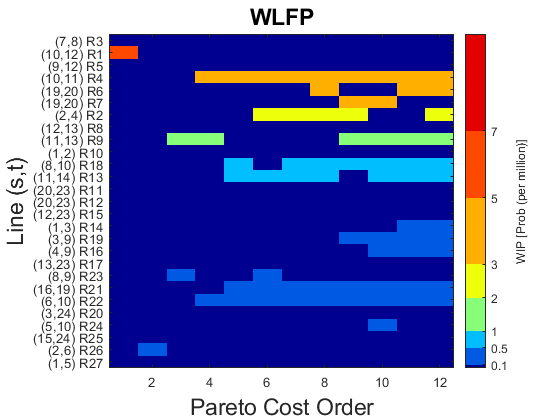}}\hfill
\hspace{-3em}
\subfloat[]{\includegraphics[width=6.0cm]{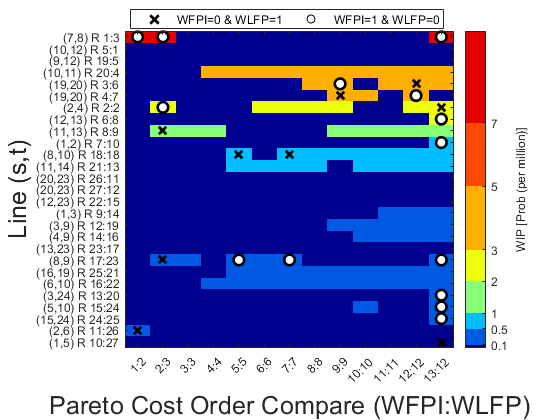}}

\vspace{-0.8em}
\caption{Pareto-optimal shutoff decisions from WIP-WFPI and WIP-WLFP for the IEEE RTS 24-bus system. X’s and O’s highlight differences from WIP-WFPI over the WIP-WLFP heatmap. WIP-WFPI results in more high-risk line shut-offs. Rows are in ascending RT WIP order, and the $y$-axis shows the rank (R:) according to the (a) WFPI, (b) WLFP, and (c) WFPI \& WLFP.}
%\vspace{-1.3em}
\label{fig:IEEE24BWFPIvWLFPLineShutoffComparison}

% Folder: H:\My Drive\MAE 207 Grid Integration Project\Project Code\Figures for Journal Paper Spring 2024 PostTSG
% File: Out_of_Sample_Test_EEV.m
% Set: Grid='24B';   
% Lines: Load all data Line 661
% Figure(15),Figure(16),Figure(18)

\end{figure*}

The line shutoff decisions for each of the numbered Pareto-optimal solutions ($x$-axis) are compared in the heatmap plot of Fig.~\ref{fig:IEEE24BWFPIvWLFPLineShutoffComparison}. The $y$-axis shows line indices $(i,j)$ arranged in the descending order of WIP-WFPI values in Fig.~\ref{fig:IEEE24BWFPIvWLFPLineShutoffComparison}(a) and of WIP-WLFP values in Fig.~\ref{fig:IEEE24BWFPIvWLFPLineShutoffComparison}(b), whereas Fig.~\ref{fig:IEEE24BWFPIvWLFPLineShutoffComparison}(c) shows the differences in line shutoff strategies. There are three instances of the WIP-WFPI energizing the highest risk line (according to the RT WIP forecast) along Pareto optimal de-energization strategies when the WIP-WLFP model never energizes that line. These instances are marked with `$\circ$' along the top row of Fig.~\ref{fig:IEEE24BWFPIvWLFPLineShutoffComparison}(c). Finally, the cumulative log probability of active lines across Pareto optimal solutions is about 4.6 \% lower (or 11.4 \% if you sum up probabilities) across all Pareto optimal de-energization strategies.

The line WIP-WFPI values are generally smaller in magnitude than WIP-WLFP and are more uniformly distributed among the lines, as shown earlier in Figures~\ref{fig:WFPIvWLFP2020HeatmapRTS}(c) and  \ref{fig:WFPIvWLFP2020HeatmapRTS}(d), which results in the underestimation of the line risk. This gives the operator greater flexibility in which lines to energize, but can lead to energizing high-risk lines. An extreme example of uniformity in line risk would be the naive $N$-$k$ approach (all lines are assigned the same amount of risk) discussed in \cite{Greenough}. Both NMKS and WIP-WFPI can lead to higher realized costs due to energizing high-risk lines, as discussed next.

\subsection{Distribution of Real-Time (Out-of-Sample) Optimal Costs}
\label{subsection:Distribution of Real-Time Optimal Costs Across Different Models}
\subsubsection{IEEE RTS 24-bus Results}
\label{subsubsection:IEEE RTS 24-bus Results}
The realized costs from RT simulation are plotted against the realized number of active lines in Fig.~\ref{fig:WFPIWLFPNMKSDistributionsBoundaryRTS}(d). The realized number of active lines incorporates additional line damages from RT scenarios $\Psi$ and is either less than or equal to the active line strategy obtained from the S-PSPS problem. The convex hull of all the RT solutions shows that the worst-case realized costs (upper boundary) are largest for the NMKS strategy and lowest for WIP-WLFP since NMKS assumes the same risk for all lines, which leads to energizing high-risk lines. Damages to high-risk lines then lead to greater generation and VoLL costs. For example, in the Pareto optimal solution no. 9, both WIP-WLFP and WIP-WFPI models lead to roughly a 75\% reduction in the load served at bus 18, but the increase in costs in the WIP-WFPI is from the blackout of the demand at bus 20, which was a loss of 117 MWh at 1500~h to 1600~h and an average load loss of 96 MW for the day.

\begin{figure}[h]
\centering
\begin{overpic}[width=\columnwidth]{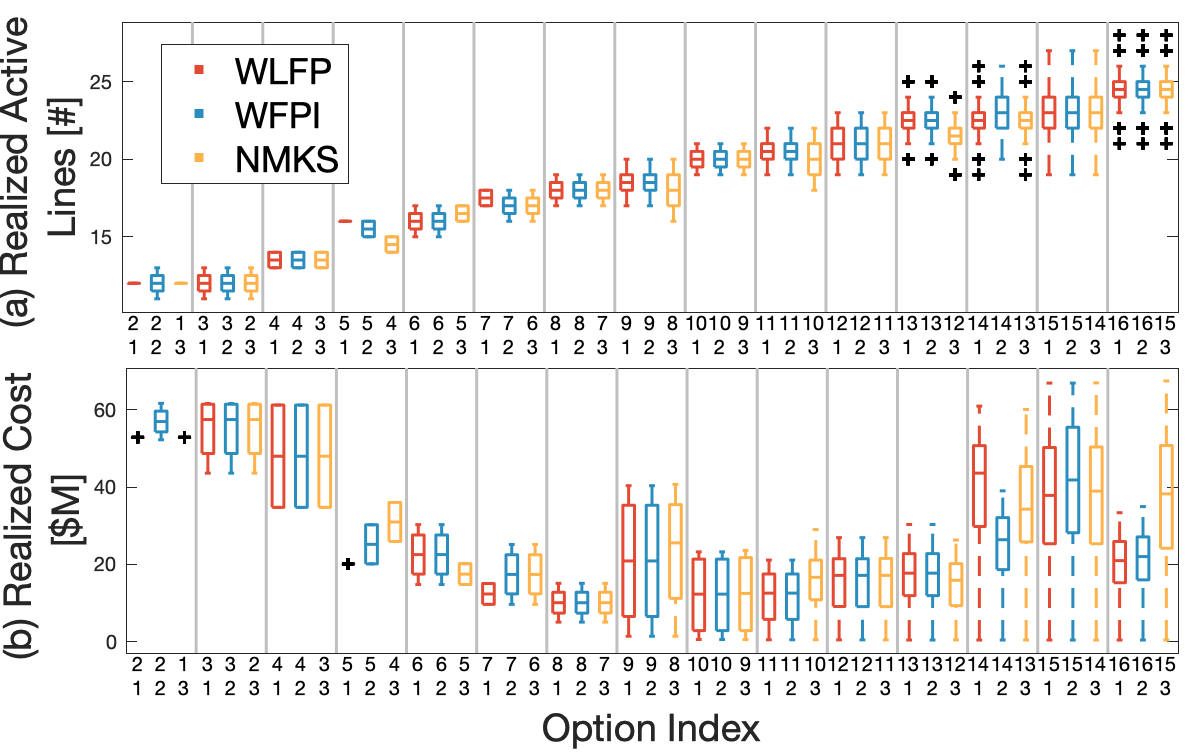}
    % \put (27,68) {(a)}
    % \put (75,68) {(b)}
    % \put (27,33) {(c)}
    % \put (75,33) {(d)}
\end{overpic}

%\vspace{-2em}
\caption{Comparison of realized active line ranges (a) and optimized realized costs (b) under permutations of the ten riskiest lines for WIP-WLFP, WIP-WFPI, and NMKS models along their Pareto fronts. The top row of the option index denotes the Pareto point (from Fig.~\ref{fig:WFPIWLFPNMKSDistributionsBoundaryRTS}(a–b)); the bottom row indicates the forecast model (WLFP=1, WFPI=2, NMKS=3).}
%\vspace{-1em}
\label{fig:IEEE24BWFPIWLFPNMKSboxplots}
\end{figure}

We further compare the distribution of realized active lines and costs in Fig.~\ref{fig:IEEE24BWFPIWLFPNMKSboxplots} for each of the Pareto-optimal solutions. The results show that in most cases, WIP-WLFP results in a lower variability in realized cost when compared to WIP-WFPI and NMKS. Exceptions include the 6-6-7 case in which NMKS performs better than the other two strategies and the 14-14-13 case in which WIP-WFPI out-performs the other strategies. This difference can be attributed to the different combination of line shutoff strategies for the same number of active lines in all three strategies, as discussed earlier.

\subsubsection{RTS-GMLC Results}
\label{subsubsection:IEEE RTS-GMLC Results}
\begin{figure}
%\vspace{-2em}
%\centerline{
\subfloat[]{\includegraphics[width=8.2cm]{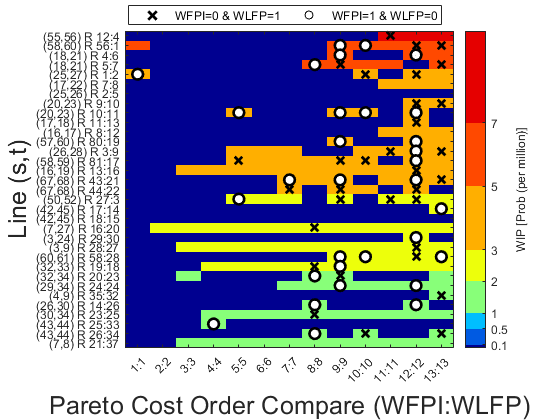}}
\subfloat[]{\includegraphics[width=8.2cm]{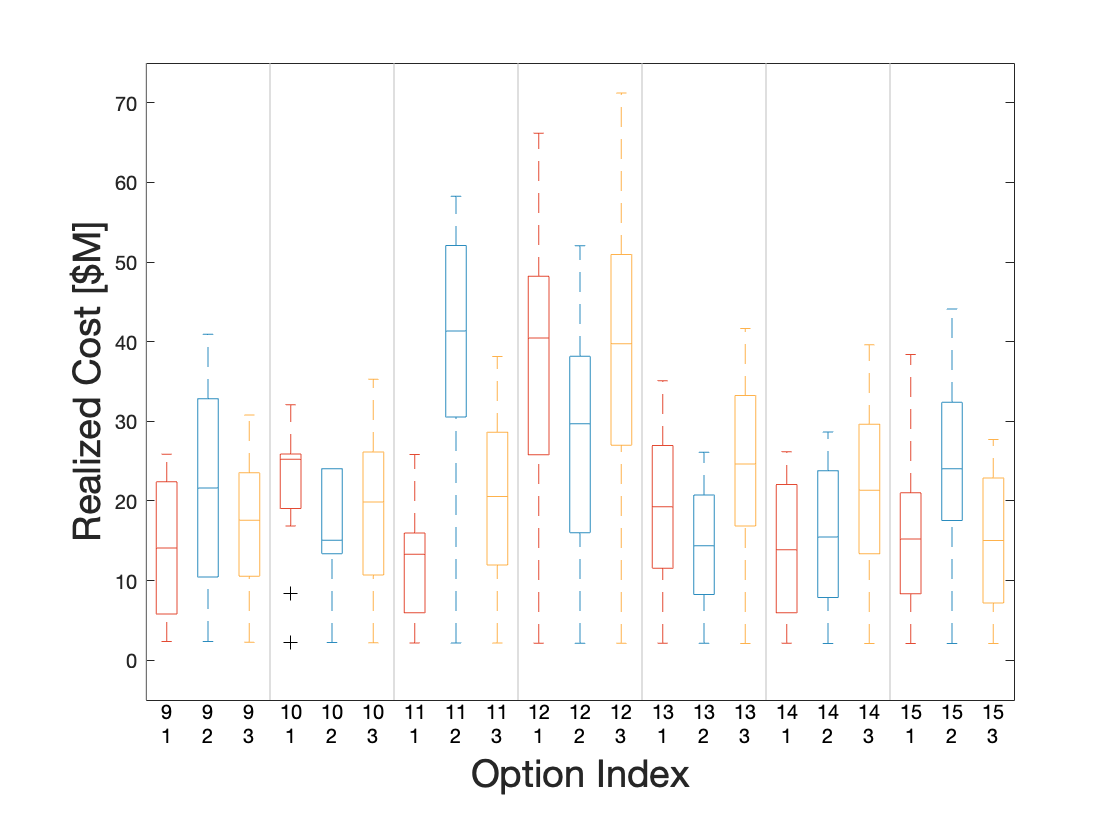}}
%\vspace{-0.8em}
\caption{(a) Comparison of Pareto-optimal shutoff decisions from WIP-WFPI and WIP-WLFP for RTS-GMLC. X’s and O’s indicate differences from WIP-WFPI over the WIP-WLFP heatmap. The top 35 riskiest lines are shown. (b) Realized costs when these decisions face permutations of the ten riskiest lines. Pareto points 1-8 in were removed in (b) due to space restrictions.}
%\vspace{-1.5em}
\label{fig:IEEERTbWFPIvWLFPLineShutoffComparison_RTb2}
\end{figure}

A similar comparison of the range of realized costs is performed for the full RTS-GMLC system and shown in Fig.~\ref{fig:IEEERTbWFPIvWLFPLineShutoffComparison_RTb2}. There are 7 out of 15 instances when the WIP-WLFP has more favorable realized cost distributions (locations 4, 7, 8, 9, 11, 14, 15), as opposed to 2 for the WIP-WFPI (10, 13), and 2 for the $N$-$k$ model (9, 12). The specific DA optimized line shutoff decisions and their differences are depicted in heatmaps in Fig.~\ref{fig:IEEERTbWFPIvWLFPLineShutoffComparison_RTb2}.

Regional loadshed analysis on the RTS-GMLC is omitted due to space limitations. However, in \ref{section:Depth of Outage Analysis RTS-GMLC}, we examine how differences in WIP forecasts translate into realized regional outage severity. While aggregate regional loadshed may appear similar across WIP-WLFP, WIP-WFPI, and NMKS models (Figs.~\ref{fig:IEEE24BWFPIWLFPNMKSboxplotsLoadLossR1R2}–\ref{fig:IEEE24BWFPIWLFPNMKSboxplotsLoadLossR3}), regional analysis reveals different blackout patterns driven by model-dependent shutoff and commitment decisions (Fig.~\ref{fig:WFPIvWLFPLoadLossComparisonRTS}). 

\subsubsection{Run Times and Scalability}

All simulations were conducted using CVX Version 2.2 and YALMIP (R20230609) with Gurobi 9.12 on a desktop PC equipped with an Intel Core i9 Processor (3.00 GHz), 64 GB of RAM, and running Windows 11 Professional. Gurobi's MIP gap and feasibility tolerance were set to 0.01\% and $10^{-6}$, respectively. 

\begin{figure}[H]
\centering
\subfloat[]{\includegraphics[width=8.2cm]{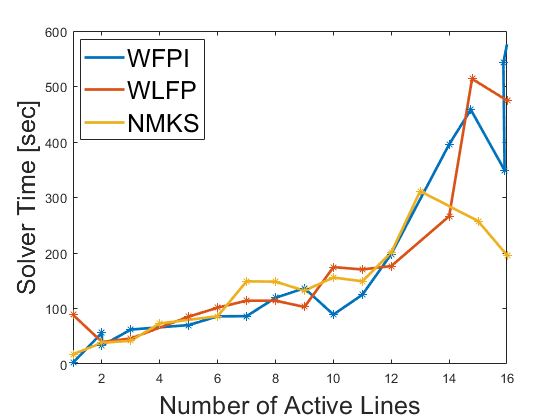}}
\subfloat[]{\includegraphics[width=8.2cm]{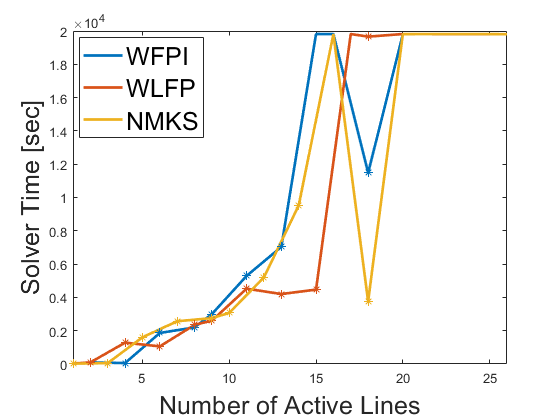}}

\caption{Solution run times along the Pareto front for the WIP-WLFP, WIP-WFPI, and NMKS models on the RTS 24-bus system (a) and RTS-GMLC zero-load loss solutions (b). Gurobi solves to a 0.01\% MIP gap and $10^{-6}$ feasibility tolerance, with time limits of 15 min (24-bus) and 6 hr (RTS-GMLC). Kinks with asterisks indicate optimal solutions. }
%\vspace{-2em}
\label{fig:IEEERTb_SolveTimeNoVoLL}
\end{figure}

The minimum, median, maximum, and standard deviation solve times in seconds are 2.44, 169.9, 543.0, 148.1 for the WIP-WFPI model, 2.06, 114.1, 532.5, 161.9 for the WIP-WLFP model, and 17.5, 149.0, 337.8, 95.5 for the NMK for the RTS bus system with a maximum solve time limit of 15 minutes. A similar analysis of solve times was conducted for Pareto points on the RTS-GMLC (Fig. \ref{fig:IEEERTb_SolveTimeNoVoLL}(b)), but the maximum solve time limit was set to 6 hours.

%\vspace{-1em}

\section{Conclusion}
\label{section:Conclusion}
This paper demonstrates how the choice of wildfire risk metric influences optimal unit commitment decisions during PSPS events. We adapt the S-PSPS framework from \cite{Greenough} and apply it to the IEEE RTS 24-bus and RTS-GMLC (73-bus) systems using two wildfire impact prediction models based on WFPI and WLFP. Using expected demand and WIP forecasts over 51 representative days, we solve a DA unit commitment problem and then evaluate the resulting commitments under RT uncertainty via an out-of-sample Monte Carlo simulation. WFPI-based forecasts tend to underestimate RT transmission line risk, leading to less robust DA unit commitments and higher RT costs. We stress that these models differ in such a way that a WIP-WFPI model cannot just recover a WIP-WLFP through input scaling or tuning of risk-averseness parameters. This study focuses on two widely used wildfire risk metrics from the USGS. Future work could assess how optimized PSPS is affected by alternative forecasting methods, including those developed by utilities or proposed in \cite{Umunnakwe, Bayani, Bayani2}.

\section{Declaration of Generative AI and AI-assisted technologies}
During the preparation of this work, the authors used Open-AI to improve the readability and language of the manuscript. After using this tool, the authors reviewed and edited the content as needed and take full responsibility for the content of the published article.

\section{Funding}
This research received no specific grant from funding agencies in the public, commercial, or not-for-profit sectors.

\section{Declaration of competing interest}
The authors declare that they have no known competing financial interests or personal relationships that could have appeared to influence the work reported in this paper

\pagebreak
\pagebreak
\clearpage

\textbf{SUPPLEMENTAL MATERIAL} 
\appendix
\printnomenclature

\section{Extra WIP vs. OWIP Analysis for IEEE 24-bus and RTS-GMLC systems}
\label{section:Extra WIP vs. OWIP Analysis for IEEE 24-bus and RTS-GMLC systems}

In the following section, we give readers additional insight into how WIP is calculated and the ranges of monthly WIP and OWIP within the various regional clusters of the IEEE RTS 24-bus and RTS-GMLC systems.

The range of WIP based on historical WFPI and WLFP is reported in Table~\ref{table:CompareWIPWFPI} and Table~\ref{table:CompareWIPWLFP} (data taken from \cite{WFPI_Reliability}).

\begin{table*}[h]
\setlength{\tabcolsep}{1pt} % Reduce cell padding
\centering
\caption{Mapping of ranges of WFPI $[-]$ to the distribution of WIP $[10^{-6}]$ based on data from the reliability diagram from \cite{Short}}
\begin{tabular}{l@{\hskip 2pt} c@{\hskip 2pt} c@{\hskip 2pt} c@{\hskip 2pt} c@{\hskip 2pt} c@{\hskip 2pt} c@{\hskip 2pt} c@{\hskip 2pt} c@{\hskip 2pt} c@{\hskip 2pt} c@{\hskip 2pt} c@{\hskip 2pt} c@{\hskip 2pt}}
WFPI & [0,10) & [10,20) & [20,30) & [30,40) & [40,50) & [50,60) & [60,70) & [70,80) & [80,90) & [90,100) & [100,120) & [120,147) \\
\midrule
min  & 0.030 & 0.07 & 0.12 & 0.15 & 0.20 & 0.25 & 0.30 & 0.40 & 0.50 & 0.60 & 0.70 & 1.10 \\
q25  & 0.035 & 0.075 & 0.18 & 0.25 & 0.33 & 0.42 & 0.55 & 0.80 & 0.95 & 1.20 & 1.50 & 1.70 \\
q50  & 0.040 & 0.08 & 0.28 & 0.35 & 0.50 & 0.65 & 0.80 & 1.10 & 1.45 & 1.90 & 2.30 & 2.40 \\
q75  & 0.045 & 0.085 & 0.40 & 0.55 & 0.70 & 0.90 & 1.05 & 1.40 & 1.90 & 2.50 & 3.10 & 5.00 \\
max  & 0.050 & 0.09 & 0.60 & 0.80 & 0.90 & 1.10 & 1.30 & 1.80 & 2.40 & 4.10 & 4.00 & 7.00 \\
mean & 0.040 & 0.08 & 0.20 & 0.28 & 0.47 & 0.61 & 0.82 & 1.12 & 1.57 & 2.12 & 2.88 & 8.26 \\
\end{tabular}
\label{table:CompareWIPWFPI}
\vspace{-1em}
\end{table*}

\begin{table}[h]
\setlength{\tabcolsep}{1pt} % Reduce cell padding
\centering
\caption{Mapping of ranges of WLFP $[10^{-6}]$ to the distribution of WIP $[10^{-6}]$ based on data from reliability diagram from \cite{Short}}
\begin{tabular}{l@{\hskip 3pt} c@{\hskip 3pt} c@{\hskip 3pt} c@{\hskip 3pt} c@{\hskip 3pt} c@{\hskip 3pt} c@{\hskip 3pt} c@{\hskip 3pt} c@{\hskip 3pt}}
%\toprule
WLFP & [0, 0.1) & [0.1, 0.5) & [0.5,1) & [1,2) & [2,3) & [3,5) & [5,7) & $>$7 \\
\midrule
min & 0.017	& 0.12	& 0.25	& 0.53	& 0.86	& 1.97	& 2.39	& 3.84 \\
q25 & 0.027	& 0.12	& 0.53	& 1.05	& 1.70	& 3.09	& 4.40	& 7.63 \\
q50 & 0.041	& 0.25	& 0.72	& 1.33	& 2.26	& 3.75	& 4.87	& 9.32 \\
q75 & 0.055	& 0.48	& 0.90	& 1.61	& 3.06	& 5.10	& 8.10	& 12.32 \\
max & 0.064	& 0.57	& 1.09	& 2.31	& 4.37	& 6.60	& 10.21	& 18.17 \\
mean& 0.040	& 0.26	& 0.72	& 1.38	& 2.40	& 3.88	& 5.57	& 9.78 \\
\end{tabular}
\label{table:CompareWIPWLFP}
\vspace{-1em}
% % Folder: H:\My Drive\MAE 207 Grid Integration Project\Project Code
% % File: WLFPdatascrape.m
% % Set: Rtype='WLFP';   
% % Data: WLFPminq1medq3maxnooutliers_table
\end{table}

\begin{figure}[h]
\centering
\subfloat{\includegraphics[width=8.2cm]{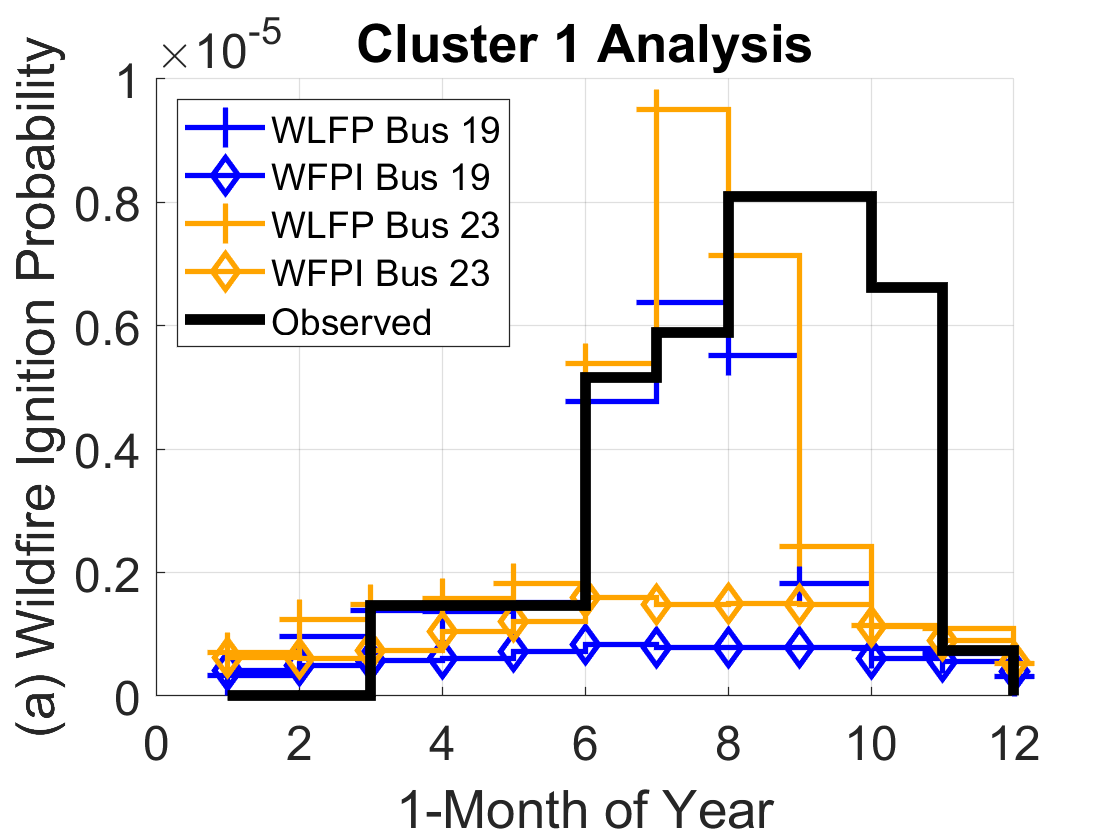}} \hfill
\subfloat{\includegraphics[width=8.2cm]{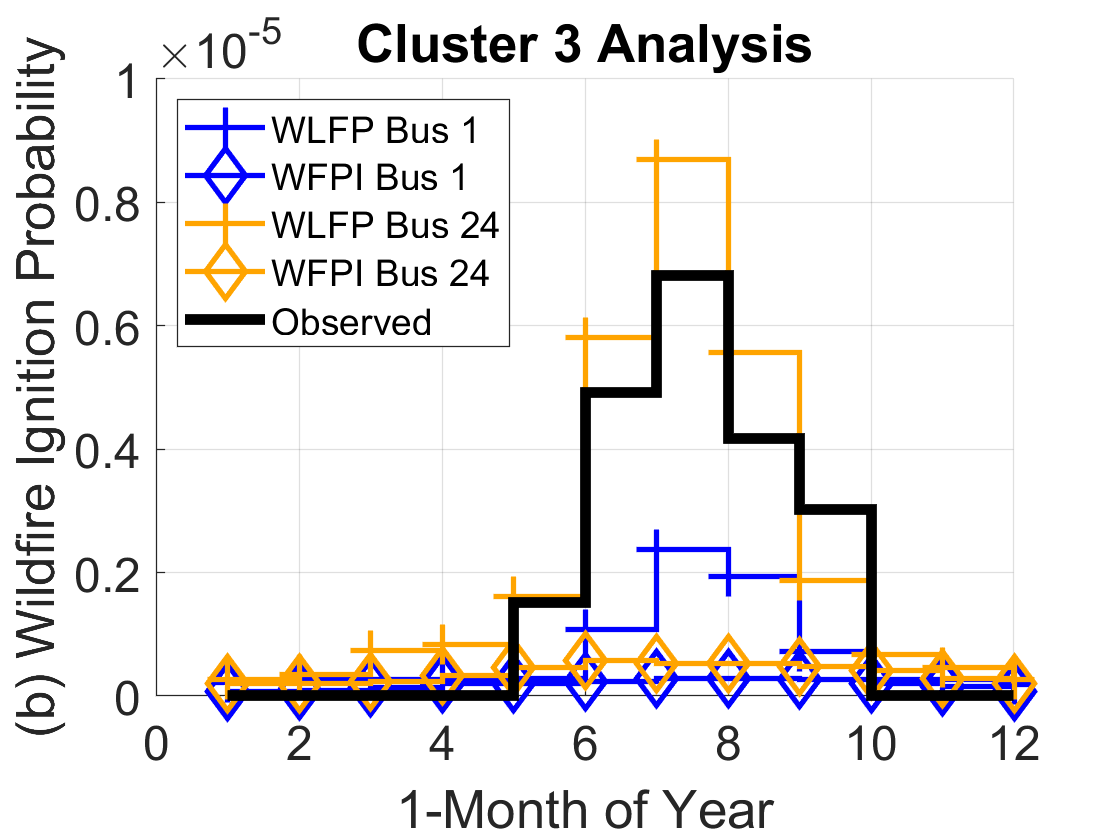}} \hfill
\subfloat{\includegraphics[width=8.2cm]{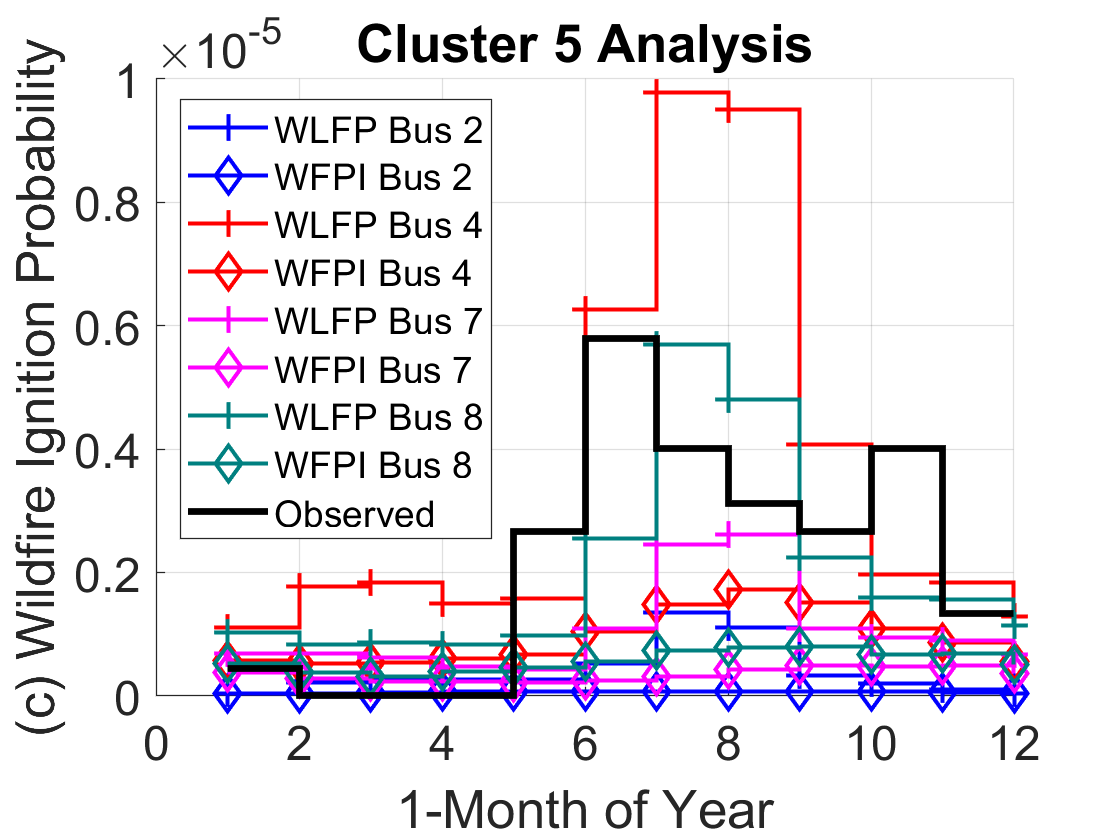}} \hfill
\subfloat{\includegraphics[width=8.2cm]{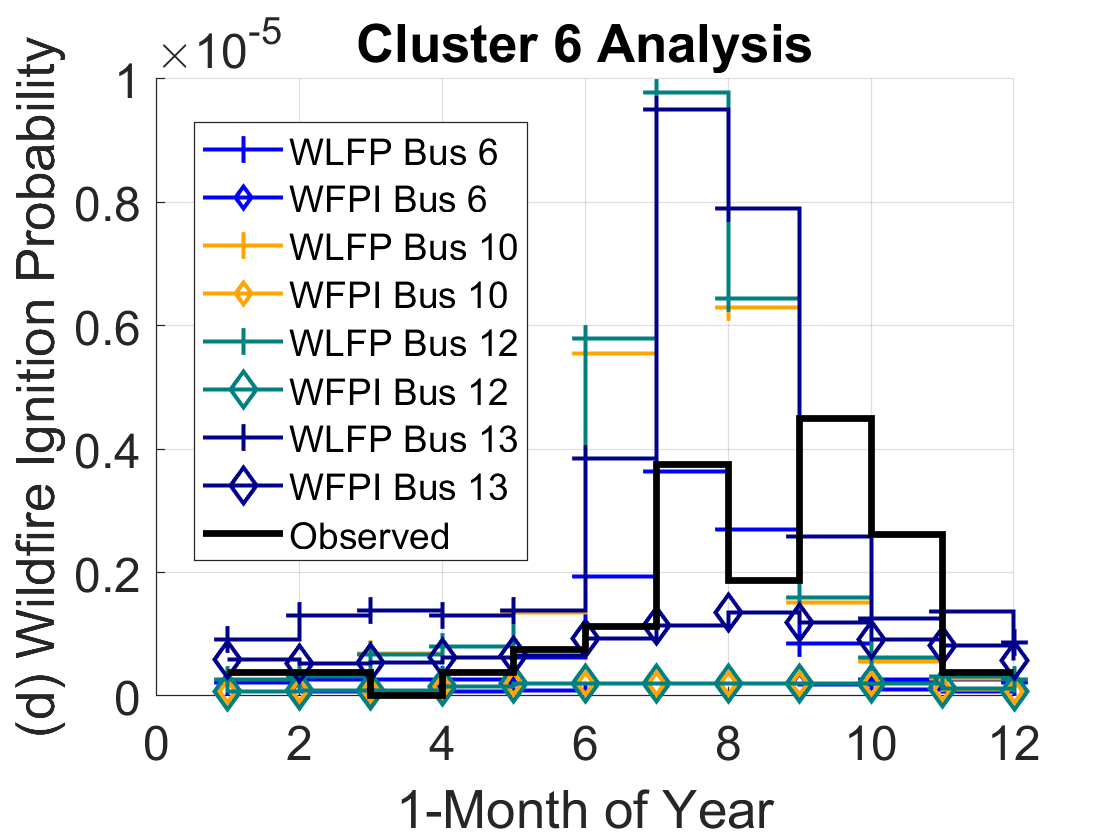}} 
\caption{Comparison of discrete distributions averaged from 2001 to 2020 and with a bin width of 1 month for WIP-WFPI, WIP-WLFP, and OWIP from \cite{Short}. Bus locations are from the IEEE 24-bus system.}
\label{fig:IEEE24ClusterAnalysis1month}
% Folder: H:\My Drive\MAE 207 Grid Integration Project\Project Code
% File: CompareWIPWLFPWFPI.m
% Set: Grid='24B';   
% Lines: Run whole script
% Figure(6), Figure(8), Figure(10), Figure(11)
\end{figure}

For 8 out of the 12 buses there is an improvement in MAE using WIP-WLFP. At buses 4, 10, 12, and 13, the elevated levels of WIP in June, July, and August contributed to a higher MAE from WIP-WLFP than WIP-WFPI, as shown in Fig.~\ref{fig:IEEE24ClusterAnalysis1month}. 

\begin{figure}[h]
\centering
% \vspace{-1em}
\includegraphics[width=0.9\columnwidth]{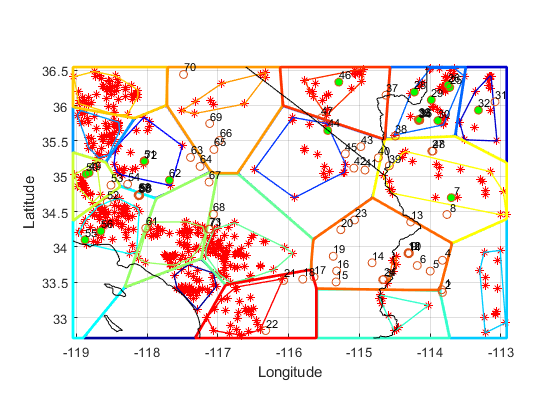}
\caption{K-Means clustering (K=18) of wildfire data from \cite{Short} to compute regional wildfire differences around the RTS-GMLC system.}
\label{fig:WFPIvWLFPDistributionsRTS}
% Folder: H:\My Drive\MAE 207 Grid Integration Project\Project Code
% File: CompareWIPWLFPWFPI.m
% Set: Grid='RTS';   
% Lines: Run whole script
% Figure(3)
\end{figure}

\begin{figure}[h]
\centering
\subfloat{\includegraphics[width=8.2cm]{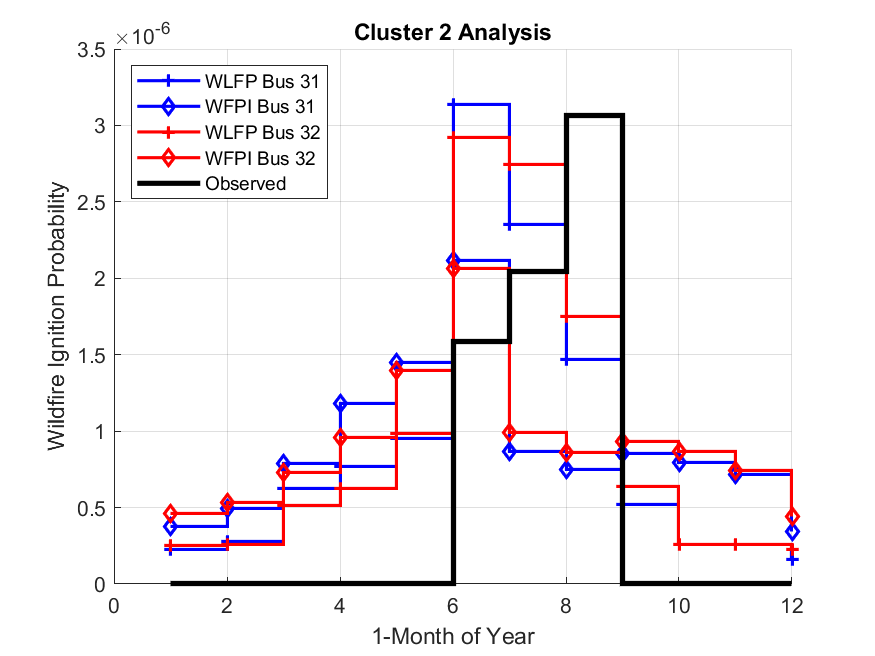}} \hfill
\subfloat{\includegraphics[width=8.2cm]{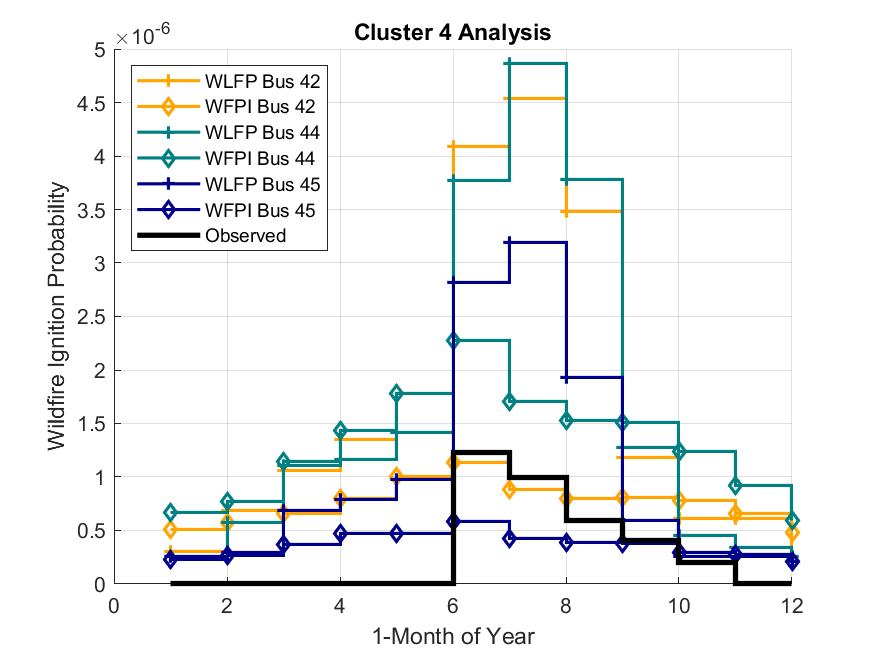}} \hfill
\subfloat{\includegraphics[width=8.2cm]{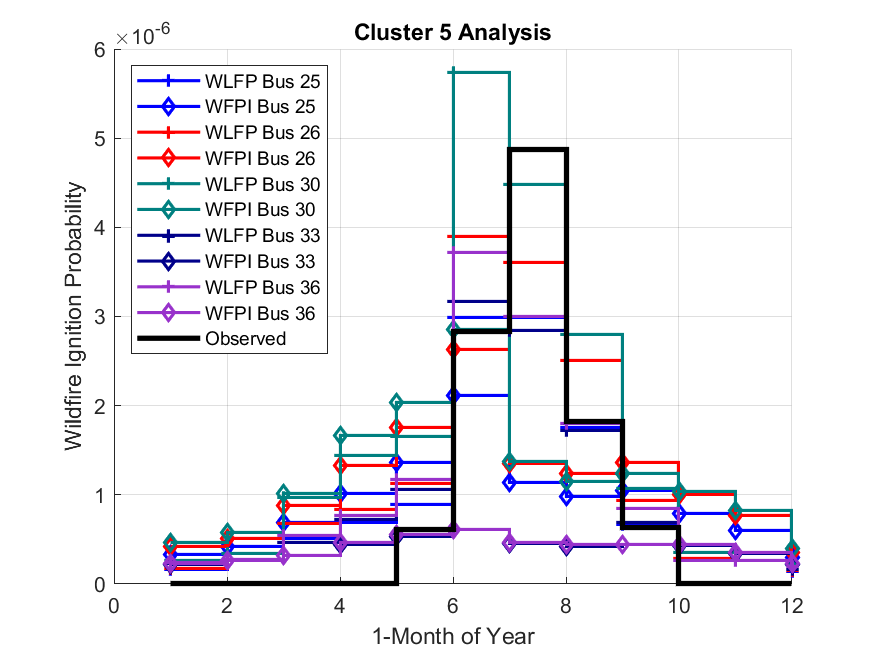}} \hfill
\subfloat{\includegraphics[width=8.2cm]{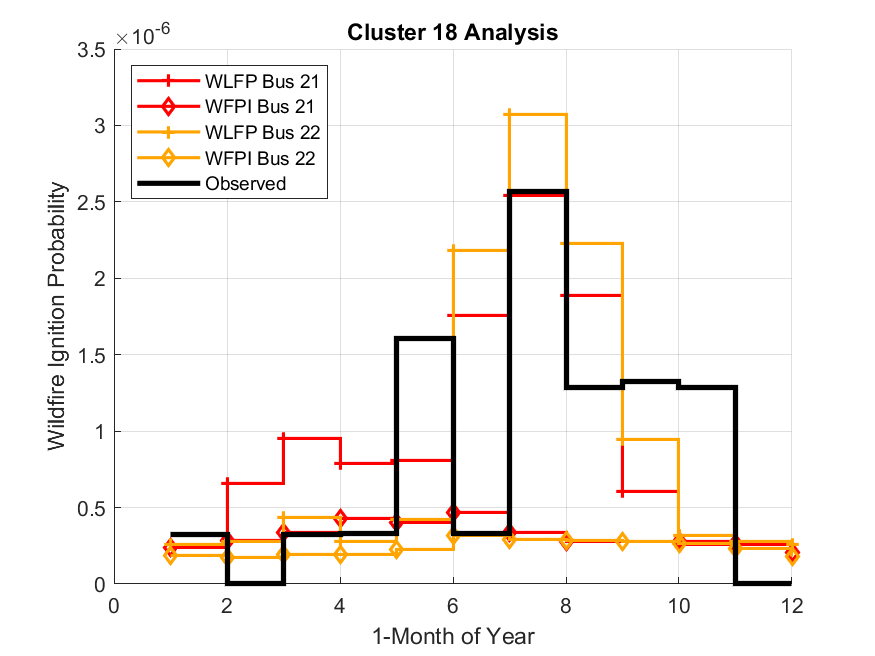}} 
\caption{Comparison of discrete distributions averaged from 2001 to 2020 and with a bin width of 1 month for WIP-WFPI, WIP-WLFP, and OWIP from \cite{Short}. Bus locations are from regions 1 and 2 from the RTS-GMLC system.}
\label{fig:IEEERTSClusterAnalysis1month}
\end{figure}

In Fig. \ref{fig:WFPIvWLFPDistributionsRTS}, we depict how K-means clustering is used to construct 18 area clusters for the OWIP computation around the RTS-GMLC. Four of the most notable regions are analyzed in Fig. \ref{fig:IEEERTSClusterAnalysis1month}.

\section{Depth of Outage Analysis RTS-GMLC}
\label{section:Depth of Outage Analysis RTS-GMLC}

\subsection{Regional Out-of-Sample Loadshed and Hours with Loadshed Analysis}

\begin{figure}
\vspace{-2em}
\centering
\subfloat{\includegraphics[width=8.2cm]{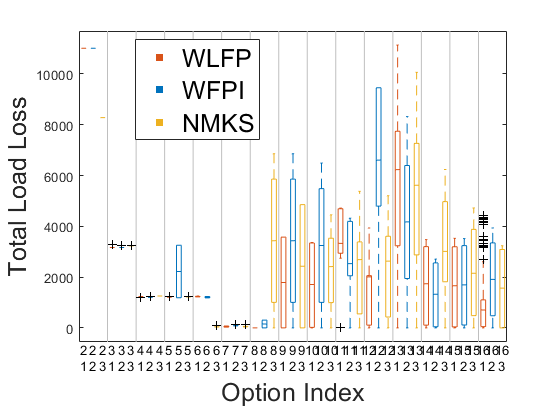}} \hfill
\subfloat{\includegraphics[width=8.2cm]{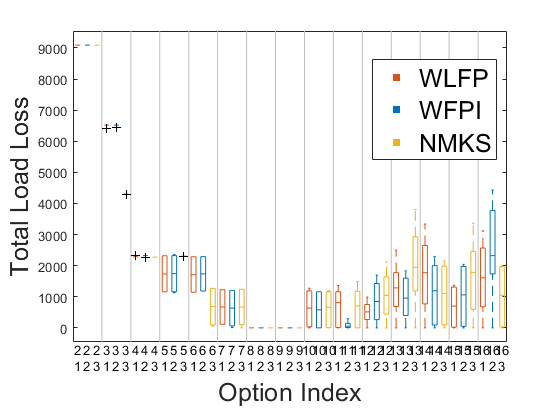}} \hfill
\subfloat{\includegraphics[width=8.2cm]{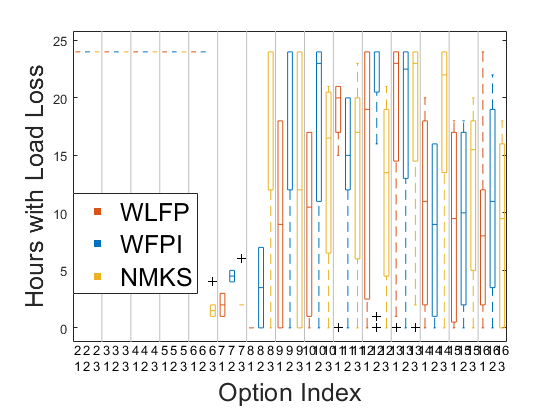}} \hfill
\subfloat{\includegraphics[width=8.2cm]{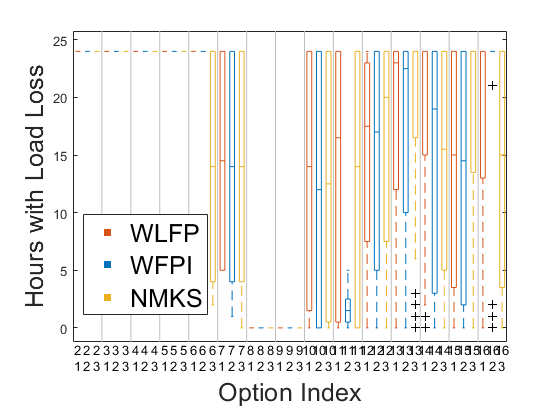}} \hfill
\vspace{-0.5em}
\caption{Comparison of the realized total load loss (a),(b) for the RTS-GMLC and realized hours with any load loss (c),(d) for Region 1 (left) and Region 2 (right)  under permutations of the ten riskiest lines for WIP-WLFP, WIP-WFPI, and NMKS models along their Pareto fronts. The top row of the option index denotes the Pareto point (from Fig.~\ref{fig:WFPIWLFPNMKSDistributionsBoundaryRTS}(a–b)); the bottom row indicates the forecast model (WLFP=1, WFPI=2, NMKS=3). There are no significant differences in the realized loadshed between the three models in Region 3.}

\label{fig:IEEE24BWFPIWLFPNMKSboxplotsLoadLossR1R2}

% Folder: H:\My Drive\MAE 207 Grid Integration Project\Project Code
% File: CompareWIPWLFPWFPI.m
% Set: Grid='RTb';   
% Lines: Run whole script
% Figure (3)

\end{figure}

\begin{figure}
\centering
\subfloat{\includegraphics[width=8.2cm]{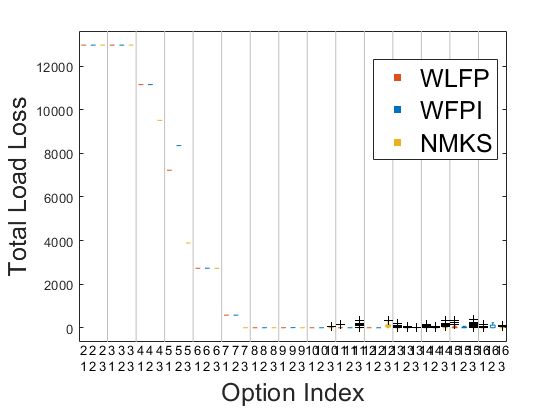}} \hfill
\subfloat{\includegraphics[width=8.2cm]{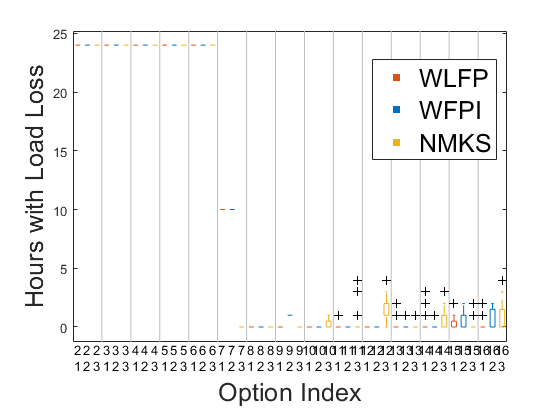}} \hfill
\vspace{-0.5em}
\caption{Comparison of the realized total load loss (a)for the RTS-GMLC and realized hours with any load loss (b) for Region 3 under permutations of the ten riskiest lines for WIP-WLFP, WIP-WFPI, and NMKS models along their Pareto fronts. The top row of the option index denotes the Pareto point (from Fig.~\ref{fig:WFPIWLFPNMKSDistributionsBoundaryRTS}(a–b)); the bottom row indicates the forecast model (WLFP=1, WFPI=2, NMKS=3).}

\label{fig:IEEE24BWFPIWLFPNMKSboxplotsLoadLossR3}

% Folder: H:\My Drive\MAE 207 Grid Integration Project\Project Code
% File: CompareWIPWLFPWFPI.m
% Set: Grid='RTb';   
% Lines: Run whole script
% Figure (3)

\end{figure}

To study how optimal shutoff and commitment decisions affect the severity and impact of a service disruption due to WF-related damages, we investigate the regional differences in realized regional loadshed and the number of hours with any loadshed within the region. These regional differences in loadshed and hours of loadshed based on WIP-WLFP, WIP-WFPI, and NMKS WF risk models are depicted in Fig.~\ref{fig:IEEE24BWFPIWLFPNMKSboxplotsLoadLossR1R2} and Fig~\ref{fig:IEEE24BWFPIWLFPNMKSboxplotsLoadLossR3}.

\begin{figure}[h]
\centering
\subfloat{\includegraphics[width=8.2cm]{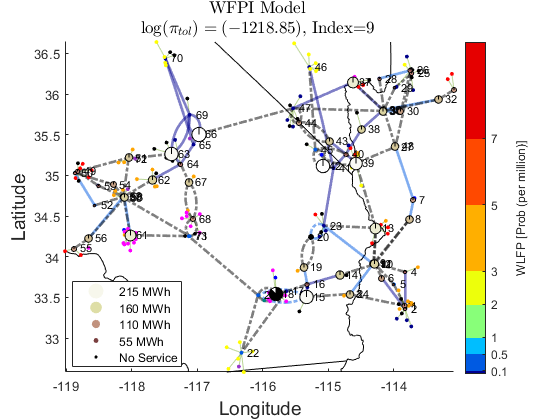}} \hfill
\subfloat{\includegraphics[width=8.2cm]{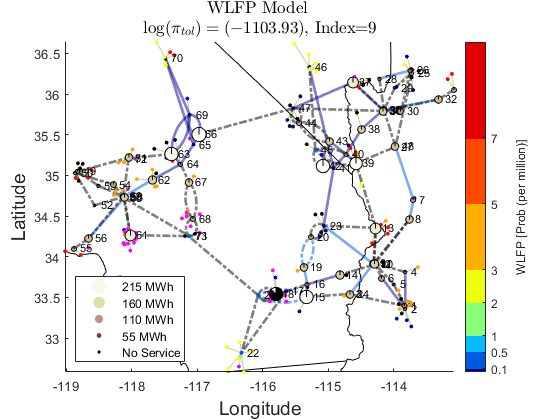}} \hfill

\vspace{-0.5em}
\caption{Effect of the Pareto-optimal shutoff decisions (Index 9 from Fig. \ref{fig:IEEERTbWFPIvWLFPLineShutoffComparison_RTb2}(b)) on the load loss experienced at the 1500~h on the RTS-GMLC based on the WIP-WFPI(a) and WIP-WLFP(b) models. The difference in optimal decisions lead to a blackout of node 20 (for gen colors labels see Fig. \ref{fig:IEEERTScolored})}

\label{fig:WFPIvWLFPLoadLossComparisonRTS}

% Folder: H:\My Drive\MAE 207 Grid Integration Project\Project Code
% File: CompareWIPWLFPWFPI.m
% Set: Grid='24B';   
% Lines: Run whole script
% Figure (3)

\end{figure}

To expand our understanding of the differences in loadshed to the bus level, we depict the Pareto optimal commitment, optimal shutoff, and worst-case realized loadshed results from solution 9 in Fig.~\ref{fig:WFPIWLFPNMKSDistributionsBoundaryRTS}(a–b)) in Fig. \ref{fig:WFPIvWLFPLoadLossComparisonRTS}. Both WIP-WLFP and WIP-WFPI models lead to roughly a 75\% reduction in the load served at bus 18, but the increase in costs in the WIP-WFPI is from the blackout of the demand at bus 20, which was a loss of 117 MWh at 1500~h and an average load loss of 96 MW for the day.

\subsection{Appropriate Riskiest Lines Set Size for Out-of-Sample Test}
Next, we conduct a sensitivity analysis with the number of the riskiest lines used for the out-of-sample test for the RTS-GMLC. The goal is to determine an appropriate set size for the riskiest lines that is large enough to reveal regional load loss differences, yet small enough to keep the resulting number of damage scenarios computationally tractable. We again examine the realized total load loss at the regional level to gain insight into how physical regions of the grid are affected.  

\begin{figure}[h]
\centering
\subfloat{\includegraphics[width=0.48\textwidth]{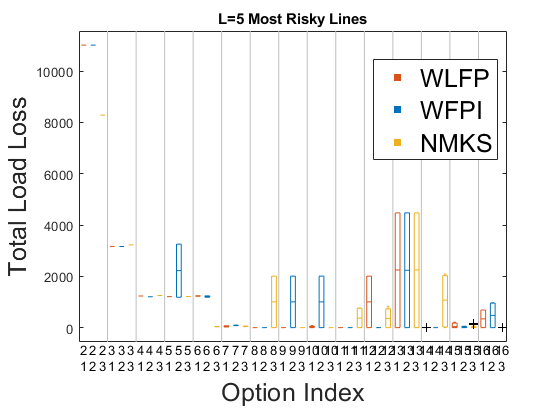}} \hfill
\subfloat{\includegraphics[width=0.48\textwidth]{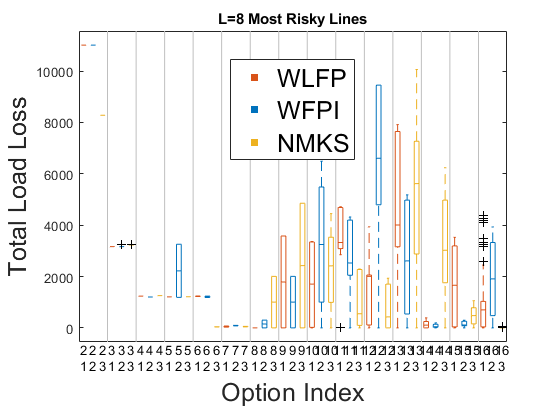}} \hfill
\subfloat{\includegraphics[width=0.48\textwidth]{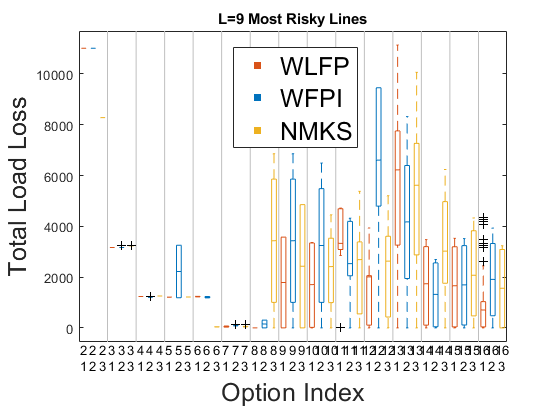}} \hfill
\subfloat{\includegraphics[width=0.48\textwidth]{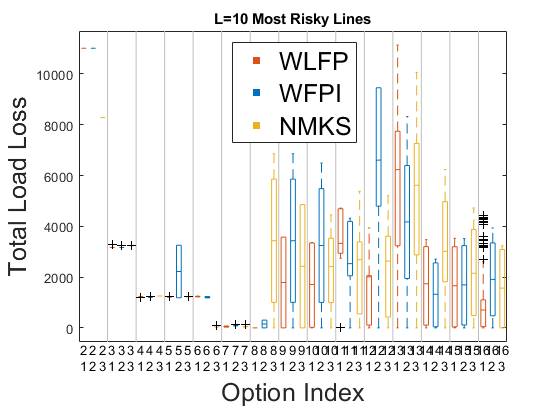}} \hfill
\vspace{-0.5em}
\caption{Comparison of the realized total load loss for the RTS-GMLC for Region 1 under permutations of the five, eight, nine, and ten riskiest lines for WIP-WLFP, WIP-WFPI, and NMKS models along their Pareto fronts. The top row of the option index denotes the Pareto point (from Fig.~\ref{fig:WFPIWLFPNMKSDistributionsBoundaryRTS}(a–b)); the bottom row indicates the forecast model (WLFP=1, WFPI=2, NMKS=3).)}

\label{fig:WFPIvWLFPLoadLossComparisonRTSR1}

% Folder: H:\My Drive\MAE 207 Grid Integration Project\Project Code
% File: CompareWIPWLFPWFPI.m
% Set: Grid='24B';   
% Lines: Run whole script

\end{figure}

\begin{figure}[h]
\centering
\subfloat{\includegraphics[width=8.2cm]{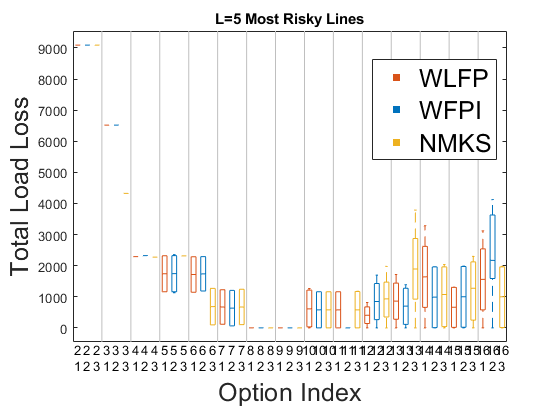}} \hfill
\subfloat{\includegraphics[width=8.2cm]{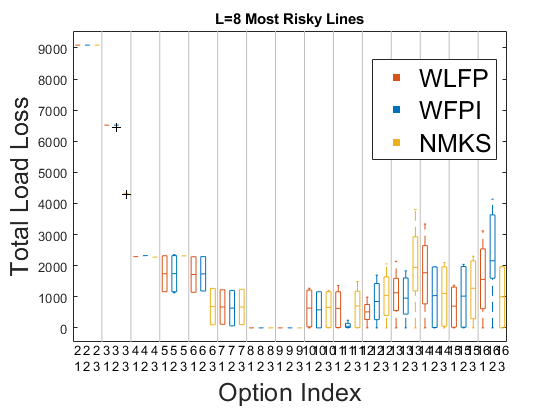}} \hfill
\subfloat{\includegraphics[width=8.2cm]{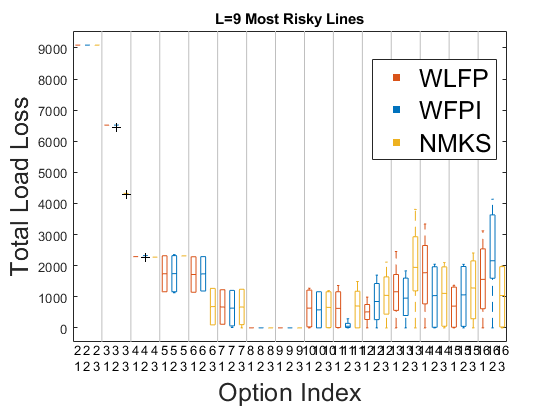}} \hfill
\subfloat{\includegraphics[width=8.2cm]{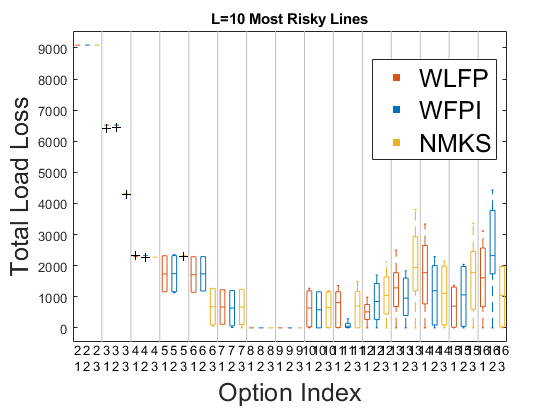}} \hfill
\vspace{-0.5em}
\caption{Comparison of the realized total load loss for the RTS-GMLC for Region 2 under permutations of the five, eight, nine, and ten riskiest lines for WIP-WLFP, WIP-WFPI, and NMKS models along their Pareto fronts. The top row of the option index denotes the Pareto point (from Fig.~\ref{fig:WFPIWLFPNMKSDistributionsBoundaryRTS}(a–b)); the bottom row indicates the forecast model (WLFP=1, WFPI=2, NMKS=3).)}

\label{fig:WFPIvWLFPLoadLossComparisonRTSR2}

% Folder: H:\My Drive\MAE 207 Grid Integration Project\Project Code
% File: CompareWIPWLFPWFPI.m
% Set: Grid='24B';   
% Lines: Run whole script

\end{figure}

\begin{figure}[h]
\vspace{-1em}
\centering
\subfloat{\includegraphics[width=8.2cm]{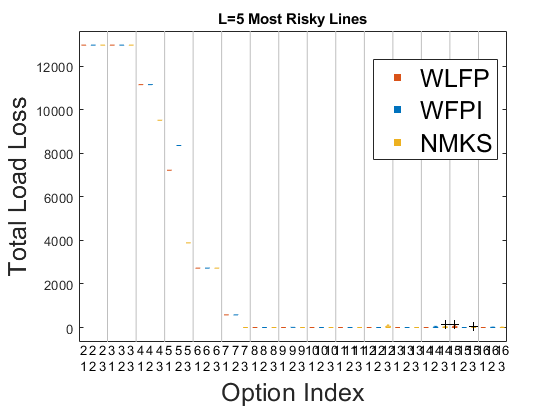}} \hfill
\subfloat{\includegraphics[width=8.2cm]{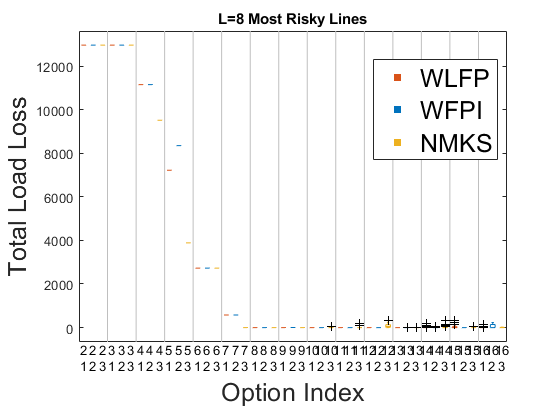}} \hfill
\subfloat{\includegraphics[width=8.2cm]{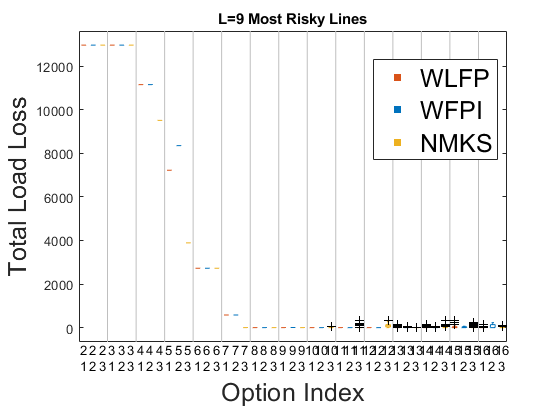}} \hfill
\subfloat{\includegraphics[width=8.2cm]{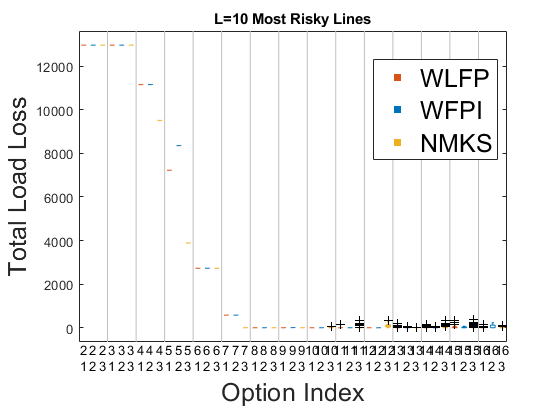}} \hfill
\vspace{-0.5em}
\caption{Comparison of the realized total load loss for the RTS-GMLC for Region 3 under permutations of the five, eight, nine, and ten riskiest lines for WIP-WLFP, WIP-WFPI, and NMKS models along their Pareto fronts. The top row of the option index denotes the Pareto point (from Fig.~\ref{fig:WFPIWLFPNMKSDistributionsBoundaryRTS}(a–b)); the bottom row indicates the forecast model (WLFP=1, WFPI=2, NMKS=3).)}

\label{fig:WFPIvWLFPLoadLossComparisonRTSR3}

% Folder: H:\My Drive\MAE 207 Grid Integration Project\Project Code
% File: CompareWIPWLFPWFPI.m
% Set: Grid='24B';   
% Lines: Run whole script

\end{figure}

\begin{figure}[h]

\centering
\includegraphics[width=8.2cm]{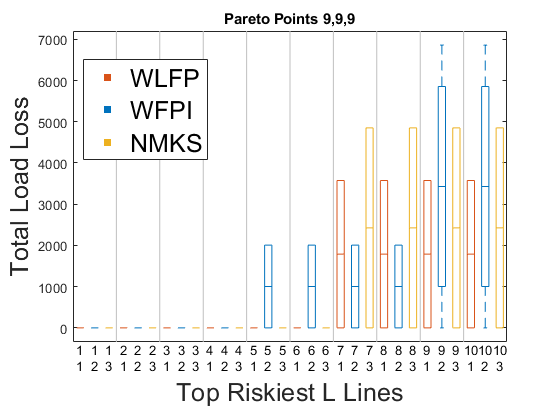}

\vspace{-0.5em}
\caption{Comparison of the realized total load loss for the RTS-GMLC for the 9th Pareto Optimal Point for Region 1 under permutations of under an expanding collection of risky lines from the riskiest to the top 10 riskiest according the the WIP on July 31st for WIP-WLFP, WIP-WFPI, and NMKS models along their Pareto fronts. The top row of the option index denotes the Pareto point (from Fig.~\ref{fig:WFPIWLFPNMKSDistributionsBoundaryRTS}(a–b)); the bottom row indicates the forecast model (WLFP=1, WFPI=2, NMKS=3).)}

\label{fig:WFPIvWLFPLoadLossComparisonRTSR1VaryL}

% Folder: H:\My Drive\MAE 207 Grid Integration Project\Project Code
% File: CompareWIPWLFPWFPI.m
% Set: Grid='24B';   
% Lines: Run whole script

\end{figure}

As depicted in Figures \ref{fig:WFPIvWLFPLoadLossComparisonRTSR2} and  \ref{fig:WFPIvWLFPLoadLossComparisonRTSR3}, the out-of-sample distributions of realized load lost along the Pareto curve are relatively stable within Region 2 and 3 of the RTS-GMLC. In contrast to Region 1, it takes a set size containing at least 10 of the riskiest lines for the out-of-sample distributions of realized load lost along the Pareto curve \ref{fig:WFPIvWLFPLoadLossComparisonRTSR1} to settle. See Fig. \ref{fig:WFPIvWLFPLoadLossComparisonRTSR1VaryL} and Table \ref{table:pp9_convergence} for a more granular snapshot of the out-of-sample realized total load loss from the 9th Pareto optimal (from Fig.~\ref{fig:IEEERTbWFPIvWLFPLineShutoffComparison_RTb2} (b)) versus risky line set size.

\begin{table}[h]
\centering
\caption{Statistical convergence of Total Load Shedding for Region 1 at Pareto Point 9, for the top 5,8,9, and 10 most risky lines. This table includes mean, standard deviation, 95th quartile, inter-quartile range, and KL divergence relative to the distribution with the 10 most risky lines.}
\begin{tabular}{l@{\hskip 6pt} c@{\hskip 6pt} c@{\hskip 6pt} c@{\hskip 6pt} c@{\hskip 6pt} c@{\hskip 6pt} c@{\hskip 2pt} }
\hline
$L$ & Model & Mean & Std & $Q_{0.95}$ & IQR & $D_{KL}(P_L||P_{10})$ \\
\hline
5 & WLFP & 0 & 0 & 0 & 0 & 0.693 \\
5 & WFPI & 1002.9 & 1018.9 & 2005.8 & 2005.8 & 0.693 \\
5 & NMKS & 0 & 0 & 0 & 0 & 0.693 \\
\hline
8 & WLFP & 1787.8 & 1791.3 & 3575.7 & 3575.7 & 0 \\
8 & WFPI & 1002.9 & 1004.9 & 2005.8 & 2005.8 & 0.693 \\
8 & NMKS & 2426.3 & 2431.1 & 4852.7 & 4852.7 & 0 \\
\hline
9 & WLFP & 1787.8 & 1789.6 & 3575.7 & 3575.7 & 0 \\
9 & WFPI & 3429.2 & 2628.0 & 6858.5 & 4852.7 & 0 \\
9 & NMKS & 2426.3 & 2428.7 & 4852.7 & 4852.7 & 0 \\
\hline
10 & WLFP & 1787.8 & 1787.8 & 3575.7 & 3575.7 & 0 \\
10 & WFPI & 3429.2 & 2626.7 & 6858.5 & 4852.7 & 0 \\
10 & NMKS & 2426.3 & 2427.5 & 4852.7 & 4852.7 & 0 \\
\hline
\end{tabular}
\label{table:pp9_convergence}
\end{table}

\section{Independence of Wildfire Events}
\label{section:Independence of Wildfire Events}

\begin{figure}[h]
%\vspace{-3em}
\centering
\includegraphics[width=8.2cm]{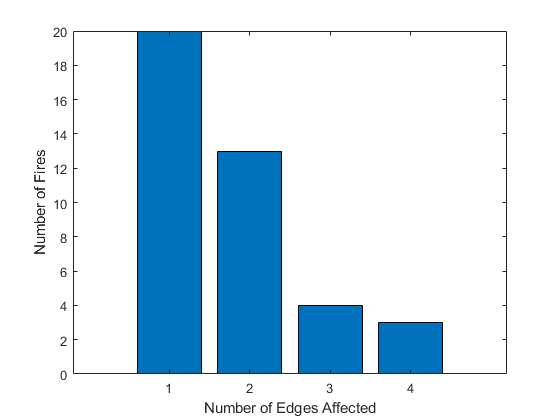}
\caption{A histogram of number of transmission edges affected by individual wildfire events (IEEE-24 test case). Most events affect only a single edge; multi-edge events are rare.} 
\label{fig:histWF24B}
%data: 
%script:
\end{figure}

\begin{figure}[h]
\centering
\includegraphics[width=8.2cm]{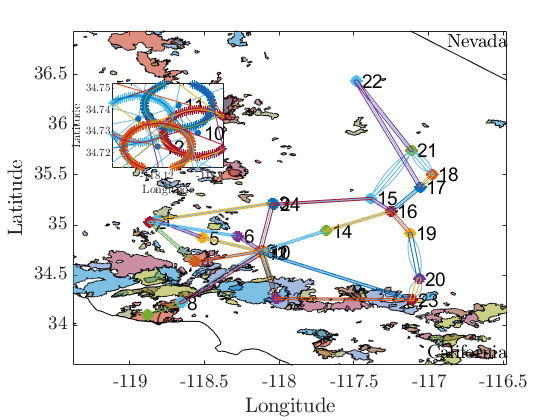}
\caption{Wildfire perimeters overlaid with the IEEE-24 network. Colored perimeters show historical fire footprints; colored lines show the network and the edges affected by events. This spatial view highlights that perimeters and line routes overlap only in a few localized corridors. 3/4 mile buffer distance on either side of the line and around buses applied based on recommendation from \cite{Trakas2}} 
\label{fig:WFperimeters24B}
%data: 
%script:
\end{figure}

\begin{figure}[h]
\centering
\includegraphics[width=8.2cm]{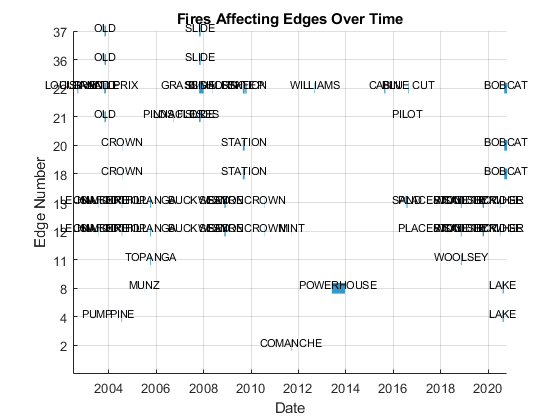}
\caption{Timeline of historical fire events that impacted network edges (edge index on y-axis). Repeated labels indicate multiple events involving the same edge; clustering of events in time is uncommon in this dataset.} 
\label{fig:WFgantt24B}
%data: 
%script:
\end{figure}

Recent catastrophic wildfires (for example, the Camp and Dixie Fires) indicate that ignition tends to concentrate on a small subset of transmission assets rather than being uniformly distributed across the network. In our case studies of superimposing large wildfire perimeters (greater than 500 acres) over the last 25 years onto the RTS and RTS-GMLC test systems, a majority of wildfire ignitions affect a single transmission element, and multi-line impacts are rare. To maintain tractability, we assume independence among ignition events and model each line’s WIP separately. This approximation is supported by our data: only 17\% (7/40) and 11.5\% (6/52) of observed fire events that have occurred on the IEEE 24 and RTS-GMLC affect more than two lines and the median number of lines affected per event is 1.5 and 1, respectively (see Fig. \ref{fig:histWF24B}-\ref{fig:WFgantt24B} for details). Often, fires affecting multiple lines are due to a fire happening near a bus, and the total number of lines affected reflects those connected to that bus (e.g., buses 8, 13, and 23). Furthermore, of the 20 wildfire events affecting more than one transmission line, the median length of wildfire events was 10 days. There were 11 instances of overlapping fire periods, with the median gap between ignitions being 4 days (mean of 5.9 days). A median of 3 lines and a maximum of 6 lines were simultaneously affected by overlapping fires (there are 38 lines in the network and 27 NZR lines). We stress that this assumption is conservative for dispatch-level decision making. But, simultaneous spatially-correlated multi-line failures can be treated as high-impact tail events and handled in a separate, contingency-focused analysis. CPUC situational reports and regional wildfire risk studies support the claim that wildfire damage is often spatially localized.\cite{CPUC,Short}%~\cite{CPUC,Westerling2,Ghassemian}. 

\section{Intuition for Scenario generation from joint distribution of demand and WFPI/WLFP}
\label{section:Intuition for Scenario generation from joint distribution of demand and WFPI/WLFP}

\begin{figure}[t]
\centering
\subfloat{\includegraphics[width=8.2cm]{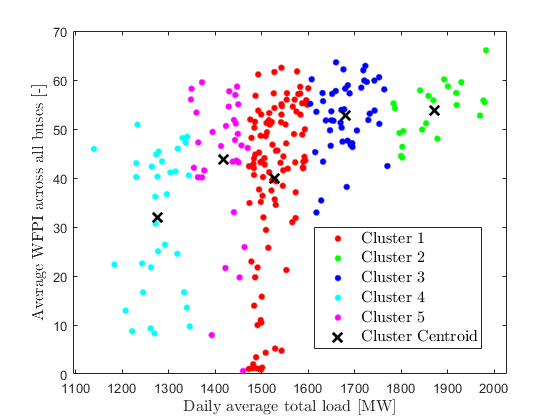}} \hfill
\subfloat{\includegraphics[width=8.2cm]{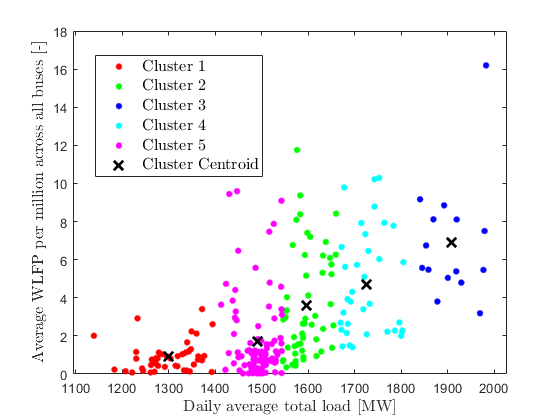}} \hfill
\caption{K-means analysis with five scenarios (for illustration) for the RTS derived from the tree reduction based on either WFPI (a) or WLFP (b) and load data from 2020. 
}
\label{fig:IEEE24BusWFPIWLFPkmeans}
\end{figure}

Fig. \ref{fig:Load_Profiles_IEEE24_WFPI_WLFP} and Tables \ref{table:scenDemWFPI} and \ref{table:scenDemWLFP} illustrate the five representative daily demand and daily bus wildfire risk scenarios obtained through the tree reduction algorithm \cite{Gröwe-Kuska}. Five scenarios are picked for illustration purposes to show how the clustering works with a small number of scenarios.  Scenarios are intended to capture the joint relationship between the electric grid demand and wildfire ignition probability throughout the year. In general, California demand exhibits strong seasonal trends, with peak electricity demand occurring during the summer months that also coincide with peak summer wildfire season (from May to September). However, there is a substantially high wildfire risk during the low demand months in the Fall and Winter that coincide with strong Santa Anna wind gusts. However, from a demand and wildfire clustering perspective, these low demand-high WF risk Santa Ana events are rare. Consequently, some demand and WIP combinations occur frequently, such as high demand with high wildfire risk, while others are comparatively rare, such as low demand with high wildfire risk (e.g., winter storm or Santa Ana wind events) or high demand with low wildfire risk (e.g., post-rainfall summer conditions).
The results further depict an important distinction between the WLFP- and WFPI-based wildfire risk representations. Daily bus WLFP values exhibit stronger seasonal clustering with electricity demand than daily bus WFPI values, producing more distinct scenario clusters across both the demand and wildfire risk dimensions. As a result, the reduced scenarios more clearly represent diverse operating conditions, such as low demand/low WF risk, medium demand/high WF risk, and high demand/high WF risk periods. In contrast, WFPI predictions tend to be more uniform throughout much of the wildfire season and generally underestimate WIP relative to observed wildfire-prone conditions. Consequently, the clustering process is driven more strongly by demand variations than by wildfire risk variations, resulting primarily in low- to moderate-risk wildfire scenarios with less separation across wildfire risk levels. Fig. \ref{fig:IEEE24BusWFPIWLFPkmeans} shows larger variance in WFPI predictions in the 5 clusters obtained by using K-means clustering.

\begin{table}[ht]
\centering
\caption{Interpretation of Demand versus Wildfire Risk Scenario Combinations}
\renewcommand{\arraystretch}{1.2}
\begin{tabular}{|c|c|p{5.2cm}|c|}
\hline
\textbf{Demand Level} & \textbf{Wildfire Risk Level} & \textbf{Typical Conditions} & \textbf{Representation} \\ 
\hline
Low & Low & Winter or early spring conditions with reduced electrical demand and lower wildfire activity & WLFP and WFPI \\ 
\hline
Low & Medium & Late spring or transitional seasonal conditions with moderate wildfire potential & WLFP and WFPI \\ 
\hline
Low & High & Relatively rare conditions typically associated with Santa Ana wind-driven wildfire events outside peak summer demand periods & Rare occurrence \\ 
\hline

Medium & Low & Early spring or late fall conditions with moderate demand and limited wildfire activity & WLFP and WFPI \\ 
\hline
Medium & Medium & Early summer conditions with increasing temperatures and moderate wildfire potential & Primarily WFPI \\ 
\hline
Medium & High & Late summer or early fall wildfire conditions with elevated ignition potential & Primarily WLFP \\ 
\hline
High & Low & Rare post-rainfall summer conditions with elevated cooling demand but temporarily reduced wildfire risk & Rare occurrence \\ 
\hline
High & Medium & Spring or transitional heat conditions with elevated demand and moderate wildfire risk & WLFP and WFPI \\ 
\hline
High & High & Peak summer heat-wave and wildfire season conditions with simultaneous high demand and elevated wildfire ignition risk & Primarily WLFP \\ 
\hline
\end{tabular}
\label{tab:scenario_interpretation}
\end{table}

\begin{table}
\begin{center}
\caption{Five WFPI based input scenarios selected from the scenario tree reduction \cite{Gröwe-Kuska} with each respective scenario probability, daily average load, average WFPI across all buses, and its closest representative day to each scenario from a nearest neighbor search.}
\begin{tabular}{l m{1.5cm} m{1.6cm} m{1.75cm} m{7.0cm}}
\hline
 Date&        Scenario Prob.&    Avg. Total Load [MWh]&    Avg. Bus WFPI [-] & Scenario Type\\
 \hline
25-Mar-2020&        0.125&          1483.7&          13.938 &  Low Demand, Low WF Risk\\  
25-Jun-2020&      0.18333&          1593.7&          43.583 &  Moderate Demand, Moderate WF Risk \\  
06-May-2020&      0.20833&          1671.9&          50.271 &  High Demand, Moderate WF Risk\\
08-Jul-2020&      0.31667&          1543.2&          54.938 &  Moderate Demand, Moderate WF Risk \\  
11-Oct-2020&      0.16667&          1314.8&          41.354 &  Low Demand, Moderate WF Risk\\  
\hline
\end{tabular}
\label{table:scenDemWFPI}
\end{center}
\end{table}

\begin{table}
\begin{center}
\caption{Five WLFP input scenarios selected from the scenario tree reduction \cite{Gröwe-Kuska} with each respective scenario probability, daily average load, average WLFP across all buses, and its closest representative day to each scenario from a nearest neighbor search}
\begin{tabular}{l m{1.5cm} m{1.6cm} m{1.7cm} m{7.0cm}}
\hline
Date &        Scenario Prob. &   Avg. Total Load [MWh] &   Avg. Bus WFLP [$\times 10^{-6}$] & Scenario Type \\
\hline
25-Mar-2020 &      0.23333 &         1483.7 &         0.18279 & Low Demand, Low WF Risk \\ 
08-May-2020 &     0.31667  &        1478.2 &          1.0967  & Low Demand, Moderate WF Risk\\ 
23-Jul-2020 &        0.15  &        1671.9 &          6.6639  & High Demand, High WF Risk\\ 
16-Aug-2020 &    0.058333  &        1486.8 &          5.5706  & Moderate Demand, High WF Risk\\ 
15-Sep-2020 &     0.24167  &        1718.5 &          3.3991  & High Demand, Moderate WF Risk\\ 
\hline
\end{tabular}
\label{table:scenDemWLFP}
\end{center}
\end{table}

\section{Conditional Cross-entropy-based Constraint to Capture Interdependent Failures for Future Work}
\label{section:Conditional Cross-entropy-based Constraint to Capture Interdependent Failures for Future Work}

To account for interdependent failure probabilities, we introduce the likelihood that a wildfire-related damage at line $(i,j)$ could also be due to an incoming wildfire started by a nearby transmission line. $\bar{\pi} _{ij,t}$ is the probability of damage due to the union of independent wildfire events at $(i,j)$, and we denote this wildfire event as $\bar{\xi}_{ij,t}=1$. The damage could be due to a local fire event calculated through the WIP ($\tilde{\xi}_{ij,t}=1$) or the spread of at least one fire originating at a neighboring line to $(i,j)$ ($\sum_{n\in \mathcal{N}_{ij}}\tilde{\xi}_{n,t}\geq1$). $\xi_{n,t}=1$ indicates a wildfire started at neighbor $n$ within the set of neighbors to line $(i,j)$, $\mathcal{N}_{ij}$, and spread to line $(i,j)$. The probability of damage conditioned on this new wildfire spread model is 
% summarized in equation \eqref{Pxijbar}. 
\begin{align} 
\mathbb{P}_{\bar{\xi}} \left[\bar{\xi}_{ij,t}=1\right]=\mathbb{P} \left[(\tilde{\xi}_{ij,t}=1) \cup (\sum_{n\in \mathcal{N}_{ij}}\tilde{\xi}_{n,t}\geq1) \right]= \bar{\pi}_{ij,t} \label{Pxijbar}
\end{align}
If there is no damage due to the wildfire spread from the neighbors and there is only local fire damage, we recover the local wildfire probability $\tilde{\pi}_{ij,t}$ as %in Equation \eqref{Pxijbarlocal}. 
\begin{align}
\mathbb{P}_{\bar{\xi}} \left[\bar{\xi}_{ij,t}=1| \sum_{n\in \mathcal{N}_{ij}}\tilde{\xi}_{n,t}=0\right]&=\tilde{\pi}_{ij,t} \label{Pxijbarlocal}
\end{align}
If there is no local fire but fire damage at $(i,j)$ is due to the spread from neighboring lines, we recover the probability of fire damage as %given in Equation \eqref{Pxijbarspread}. 
\begin{align}
\mathbb{P}_{\bar{\xi}} \left[\bar{\xi}_{ij,t}=1|\tilde{\xi}_{ij,t}=0|\right]&=\mathbb{P} \left[ (\sum_{n\in \mathcal{N}_{ij}}\tilde{\xi}_{n,t}\geq1) \right]=1-\mathbb{P} \left[ (\sum_{n\in \mathcal{N}_{ij}}\tilde{\xi}_{n,t}=0)\right] =1-\mathop \prod \limits_{n \in \mathcal{N}_{ij}} (1-\pi_{n \rightarrow ij,t}) \label{Pxijbarspread} 
\end{align}
Again, the probability of at least one ignition due to a fire spread from a neighbor is given by $1-\Pi_{n\in N_{ij}}(1-\pi_{n \rightarrow ij})$ or the complement of the event in which there is no fire spread from neighbors. Therefore, the joint probability of a damage event considering local and neighboring wildfire spread is given as 
\begin{align} 
\mathbb{P}_{\bar{\xi}}\left[\bar{\psi}\right]= \mathop \prod \limits_{(i,j) \in \mathcal{K}_{\bar{\psi}}}  \bar{\pi}_{ij,t} \mathop \prod \limits_{(i,j) \in \mathcal{K}_{\bar{\psi}}^{\tx{c}}}(1 - \bar{\pi}_{ij,t} ).\label{eq:Nmklinefailures_spread}
\end{align}
The impact of line damaged by wildfire is incorporated in~\eqref{eq:Nmklinefailures_spread} by including $z_{ij,t}$ and $z_{ij,t}^{\textrm{dn}}$ as 
\begin{align} 
\mathop \prod \limits_{(i,j) \in \mathcal{K}_{\bar{\psi}}}  \left(\bar{\pi}_{ij,t}\right)^{z^{\tx{dn}}_{ij,t}}\mathop \prod \limits_{(i,j) \in \mathcal{K}^{\tx{c}}_{\bar{\psi}}}(1 - \bar{\pi}_{ij,t})^{z_{ij,t}} \leq {{\pi _{\tx{tol}}}}, \label{eq:WLFP_Nmk_spread}
\end{align}
where $\pi_{\tx{tol}}$ is an upper bound on the probabilities of line damage scenarios. For simplicity, the same tolerance level is used across all wildfire scenarios in $\bar{\Psi}$.
And its associated conditional cross-entropy constraint (conditioned on the new wildfire spread model) is given as 
\begin{align} 
\mathop \sum \limits_{\left({i,j} \right) \in \mathcal{K}_{\bar{\psi}}} {z^{\tx{dn}}_{ij,t}}\log\left({{\bar{\pi} _{ij,t}}} \right) + \mathop \sum \limits_{\left({i,j} \right) \in \mathcal{K_{\bar{\psi}}^{\tx{c}}}} {z_{ij,t}} {\log}\left({1 - {\bar{\pi} _{ij,t}}} \right) %\nonumber \\
\leq {\log}\left({{\pi _{\tx{tol}}}} \right),\quad \forall \bar{\psi} \in \bar{\Psi}, \, \forall t \in \mathcal{H}. \tag{\ref{eq:logWLFP_Nmk_spread}} \end{align}
To maintain the linearity of constraint \eqref{eq:logWLFP_Nmk_spread} and avoid introducing any endogenous uncertainty to the spread modeling probability, we assume the wildfire spread model parameters are generated offline. 
Using standard probability rules, we can develop the probability of a damage event ($\bar{\pi}_{ij,t}$) based on the probability of a local fire $\tilde{\pi}_{ij,t}$ at $(i,j)$ or the probability of the fire starting at a neighboring line $n$ and reaching line $(i,j)$ ($\pi_{n \rightarrow ij}$):
\begin{align}
\bar{\pi} _{ij,t}=1-(1-\tilde{\pi}_{ij,t})(\Pi_{n\in \mathcal{N}_{ij}}(1-\pi_{n \rightarrow ij,t})) \tag{\ref{piijbar}}
\end{align}
The spread probabilities, $\pi_{n \rightarrow ij}$, can be roughly derived from the USGS WFPI-based Fire Spread Probabilities (WFSP) (the probability a fire spreads at least 500 acres) or from a higher resolution fire spread model. When fire spread terms are deemed negligible (i.e., $\pi_{n \rightarrow ij}=0 \forall n\in\mathcal{N}_{ij}$), the original cross entropy constraint \eqref{eq:logWLFP_Nmk} of independent ignition events is recovered.
For completeness, we show that $\bar{\pi} _{ij,t}$ could be calculated using probability rules about the union of two independent probability events: 
($P(A \cup B)=P(A)+P(B)-P(A \cap B)$) or ($P(A \cup B)=1-P(A^c \cap B^c)$)
\begin{align}
\bar{\pi} _{ij,t}&=\tilde{\pi}_{ij,t}+(1-\Pi_{n\in \mathcal{N}_{ij}}(1-\pi_{n \rightarrow ij,t}))-\tilde{\pi}_{ij,t}(1-\Pi_{n\in \mathcal{N}_{ij}}(1-\pi_{n \rightarrow ij,t})) \label{piijbarA} \\
\bar{\pi} _{ij,t}&=1-(1-\tilde{\pi}_{ij,t})(\Pi_{n\in \mathcal{N}_{ij}}(1-\pi_{n \rightarrow ij,t})) \tag{\ref{piijbar}}
\end{align}

\end{document}